\documentclass[12pt]{JHEP3}

\usepackage{amsmath}
\usepackage{amssymb}
\usepackage{graphicx}
\usepackage{slashed}
\usepackage{booktabs}

\newcommand{\be}{\begin{eqnarray}}
\newcommand{\ee}{\end{eqnarray}}
\newcommand{\nn}{\nonumber}
\newcommand{\bn}{\begin{enumerate}}
\newcommand{\en}{\end{enumerate}}

\def\IR{\mathbb{R}}

\def\CN{{\cal N}}

\def\D{\Delta}

\def\half{\frac{1}{2}}

\def\p{\partial}

\def\Tr{{\rm Tr}}

\title{Line Operator  Index on $S^1\times S^3$ }

\author{ Dongmin Gang,  Eunkyung Koh, and Kimyeong Lee
\\
\\
Korea Institute for Advanced Study, Seoul 130-012, Korea
\\
\\
E-mail: \email{ arima275@kias.re.kr, ekoh@kias.re.kr, klee@kias.re.kr } } 

\abstract{ We derive a general formula of an index for ${\cal N}=2$ superconformal field theories on $S^1\times S^3$ with  insertions  of BPS Wilson line or 't Hooft line operator at the north pole and their anti-counterpart at the south pole of $S^3$.  One-loop and monopole bubbling effects are taken into account in the computation. As examples, we calculate the indices for  $\CN=4$ theories  and $ {\cal N}=2$ $SU(2)$ theory with $N_f=4$, and find good agreements between indices of line operators related by S-duality.  The relation between  Verlinde loop operators and the indices is explored. The holographic correspondence between  the fundamental (anti-symmetric) Wilson line operator   and  the fundamental string (D5 brane) in  $AdS_5\times S^5$ is confirmed by the index comparison.}

\preprint{KIAS-P12009}

\begin{document}

\section{Introduction and  concluding remarks}

Exact field theory results are useful  to probe non-perturbative physics such as S-duality in 
four dimensional gauge theories. Recently, many exact results have been obtained using the 
localization technique, after the seminal work \cite{Pestun:2007rz} on the partition 
function and Wilson loop expectation value of ${\cal N}=2$ theories on 
$S^4$. In \cite{Gomis:2011pf}, the exact 't Hooft loop expectation value on $S^4$ of 
${\cal N}=2$ theories is also obtained. 
Furthermore, the technique is applied to three dimensional theories and the exact   partition function  on $S^3$ is calculated in \cite{Kapustin:2009kz}.  
These exact calculations of sphere partition  functions allow  quantitive studies
of S-duality, AdS/CFT correspondence \cite{Maldacena:1997re}\cite{Aharony:2008ug}, 2d/4d correspondence \cite{Alday:2009aq} and etc.

Another exactly calculable  quantity is the superconformal index (SCI)\cite{Romelsberger:2005eg}\cite{ hep-th/0510251}. 
The index counts gauge invariant BPS local operators and  can be interpreted as a (twisted) supersymmetric partition function on $S^1 \times S^{d-1}$
via radial quantization \cite{Nawata:2011un}. The index is exactly calculable for various four   and three dimensional gauge theories \cite{Kim:2009wb}\cite{Imamura:2011su}.  A natural extension is to calculate the SCI with insertion of BPS defects \cite{Nakayama:2011pa}. In 
this work, we study the index for four dimensional  $\CN=2$ superconformal  theories 
in the presence of a Wilson line or 't Hooft line operator \cite{hep-th/0501015}\cite{Gomis:2009ir}\cite{Giombi:2009ek}. While the calculation of the index with a
Wilson line is straightforward, that with a 't Hooft line needs more 
analysis. 
Our approach is a hybrid of the path-integral and state counting. 
The index gets contributions from 1-loop  and  nonperturbative effects called `monopole bubbling' \cite{Lee:1996vz}\cite{Kapustin:2006pk}.
The most difficult part is to take care of monopole bubbling effects.

We start with the $\CN=4$ supersymmetric Yang-Mills theory (SYM) on $\IR\times S^3$ to define a
superconformal index  for a supercharge 
$Q$ which is compatible with the 1/2 BPS line operators which present at 
the north and south pole of the $S^3$. We do obtain one-loop  contributions to the 
index by calculating the spectrum of the fields in the presence of the 
Dirac magnetic monopole at the north pole and its anti-monopole at 
the south pole and by adding only the contributions from the modes saturating 
$\{Q,Q^\dagger\}=\D=0$. We extend  the 1-loop analysis to $\CN=2$ superconformal theories with 
various compatible chemical potentials. 

For the case where magnetic charge of line operator corresponds to a minuscule 
representation, there is no monopole bubbling and the 1-loop index is exact. 
In the case, we provide an explicit  formula for the 't Hooft line index   
and find a match with corresponding S-dual Wilson line index for many examples. 

However, the index for a 't Hooft line has additional complications 
due to the monopole bubbling when the magnetic charge is larger than 
that of a minuscule representation. The magnetic bubbling can be 
regarded as massless monopoles surrounding the infinitely massive 't Hooft monopole which is  the Dirac monopole. 
Fortunately, its contribution to the index has been sorted out in \cite{Gomis:2011pf} on $S^4$ and the work 
has been extended to calculate the line operator index on $S^1\times 
\IR^3$ recently \cite{Ito:2011ea}. We employ the result obtained there to write down the bubbling 
index on $\IR\times S^3$. The index is glued by multiplying the 
contributions from the north and south pole. The contribution from  each pole is 
turned out to be same with the monopole bubbling index on $S^1\times \mathbb{R}^3$. 
 We employ this method to set up a formula for  't Hooft line index  including monopole bubbling effects 
 and perform a consistency check using S-duality. Especially, 
 we find that for ${\cal N}=4$  $SU(2)$ theory, the index of 't Hooft line with 
non-minimal magnetic charge corrected with the magnetic bubbling corresponds to 
that of the product of the Wilson line in the fundamental 
representation, not in an irreducible representation.  The S-dual 
Wilson line index calculation shows in detail how the decomposition of 
indices to irreducible representation should be carried out. It is well known that magnetic monopoles 
with unbroken non-abelian gauge group have the global color problem 
which prevents the color rotation due to the infinite inertia \cite{Nelson:1983fn}. Our 
setting is done on $S^3$ where the total magnetic flux vanishes and so 
there is no such problem. The massless monopoles appear naturally and 
contribute to the index via the monopole bubbling mechanism \cite{Lee:1996vz}.

More interestingly, we consider line operator indices for $\mathcal{N}=2$ $SU(2)$ theory with four flavors, which is superconformal and known to have S-duality. 
Since the theory contains fundamental matters, the minimal charge of a 't Hooft line does not correspond to a minuscule representation, thus the monopole bubbling occurs. 
Taking into account the monopole bubbling, we find that the index of the minimal 't Hooft line matches
the index of the fundamental Wilson line. The mass parameters in the theory transform under the S-duality in a way permuting $SU(2)$ subgroups of the $SO(8)$ flavor symmetry.

 In the recent paper \cite{Dimofte:2011py}, 
 line operator index for $\mathcal{N}=4$ $SU(2)$ theory  
 is indirectly calculated by adopting   a similar method  used in calculation of line operator 
 partition function on $S^4$ from a Liouville theory \cite{Alday:2009fs}\cite{Drukker:2009id}.
 They introduce a ``half-index", $\Pi$, which is the index of the 4d gauge theory on half of $S^3$.
 The superconformal index for the 4d gauge theory can be obtained by combining the half-indices for two hemispheres,
 \begin{align}
 I= \langle \Pi_N | \Pi_S \rangle \;.
 \end{align}
Then, assuming the existence of the dictionary between line operators in the half sphere and   operators $\hat{O}_L$ acting 
on the half-index,   the index with insertion of line operator $L$ can be computed as 
\begin{align}
I_L = \langle \Pi_N | \hat{O}_L |\Pi_S\rangle \;.
\end{align}
The explicit form of operator $\hat{O}_L$ can be constructed from a Verlinde loop operator in 2d Liouville theory. 
The  index  computed in this method  for minimally charged 't Hooft line operator in $\mathcal{N}=4$ $SU(2)$ theory exactly match 
with  our 1-loop results. Extending the calculation to non-minimally charged  line operators,  
we can rederive the monopole bubbling index.

We check the AdS/CFT correspondence between Wilson line operators in ${\cal N}=4$ $U(N)$ SYM theory
and macroscopic objects (string or D-branes) in $AdS_5\times S^5$\cite{Rey:1998ik}\cite{Maldacena:1998im}
\cite{Drukker:2005kx}\cite{Yamaguchi:2006tq}. Explicitly, we calculate 
the large $N$ index for a Wilson line operator in the fundamental representation and find that the index can be 
factorized into two factors. The first factor matches the index of gravity spectrum on $AdS_5 \times S^5$ and the second factor matches the index of   fluctuations around fundamental string wrapping $AdS_2$ part. In a similar way, confirm the correspondence between a Wilson line operator in totally anti-symmetric representation and a D5 brane wrapping $AdS_2 \times S^4$.


There are several directions to take from this point. The most 
interesting one seems to extend the index calculation using  Verlinde loop operators  to 
$\mathcal{N}=4$ theories with general gauge group or to   $\mathcal{N}=2$ theories and 
confirm the results obtained here. Instead of laborious sum over the 
contributions of colored Young diagrams, the resulting generalization 
 would take care of the monopole bubbling algebraically. 
 There is  an interesting proposal of relating  4d superconformal indices to topological  correlations  in a concrete 2d model \cite{Gadde:2009kb}\cite{Gadde:2011ik}\cite{Gadde:2011uv}.
 Using the 2d/4d correspondence they propose the exact indices for non-Lagangian $T_N$ theories \cite{Gaiotto:2009we}, which is difficult to obtain from  conventional field theory techniques. It would be  nice if one can extend their  results to  indices with  line operators. One may also consider extensions to superconformal indices with other BPS defects, such as surface operators or domain walls. Index for a  surface operator in $\mathcal{N}=4$ SYM theory is  obtained in \cite{Nakayama:2011pa} by analyzing  the defect field theory living on the  surface operator.  It would be interesting to see whether one can obtain these indices either from direct field theory calculations or from a 2d/4d correspondence. In section 4.1, we propose several mathematical identities obtained by  equating an Wilson line index with the corresponding 't Hooft line index in $\mathcal{N}=4$ SYM theories with $U(N), O(2N),O(2N+1)$ and $Sp(2N)$ gauge groups. These identities are confirmed only up to a few lowest order in $x$. It is worthy to give analytic proof of these identities using properties of Hypergeomtric functions or other mathematical tricks \cite{Dolan:2008qi}\cite{Spiridonov:2009za}\cite{Spiridonov:2010qv}\cite{Spiridonov:2011hf}.


The plan of this work is as follows. In Sec 2. we introduce   operators and define its index on $S^3$. We calculate the index in the small coupling limit which captures the 1-loop contributions to the index in  Sec 3.
We include monopole bubbling effect in Sec.4 and check some of the S-duality.
We use Verlinde loop operators for  index calculation and  recapture the bubbling effect in Sec.5.
The AdS/CFT correspondence  for the index is checked in Sec. 6. We add the Lagrangian for $\mathcal{N}=4$ SYM on $\IR\times S^3$ and the spectrum of differential operators  in appendices.

\section{Line operator superconformal index in $\mathcal{N}=4$ SYM }

In this section, we will consider line operators in four dimensional maximally 
suspersymmetric Yang-Mills theory (SYM). The theory contains a  gauge field $A_\mu$, 
four Weyl spinors $\{ \psi^A \}_{A=1,2,3,4}$, and six real scalars $X_4, \ldots, X_9$. In appendix A,  Lagrangian of the theory on $\mathbb{R}\times S^3$ is given. We will then define a superconformal index compatible with these line operators.

\subsection{Supersymmetric 't Hooft/Wilson line operators }

\begin{figure}[h!]
  \begin{center}
    \includegraphics[width=5.5cm]{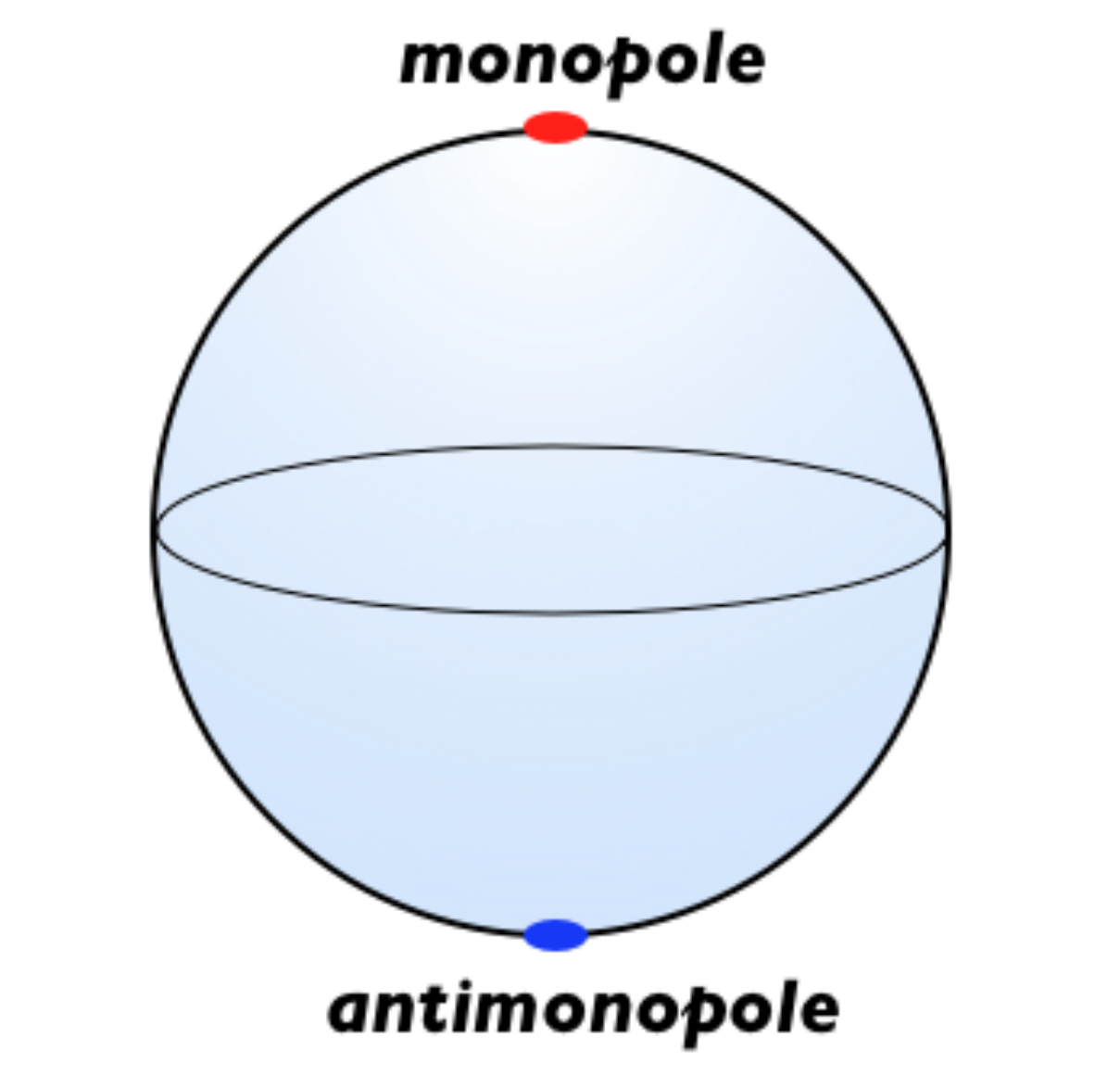}
\caption{A monopole and an antimonopole on $S^3$}\label{monopole-antimonopole}
  \end{center}
\end{figure}

A 't Hooft line operator in the $\mathcal{N}=4$ SYM on  $\mathbb{R}^4 = (x_0, \vec{x}) $ can be  defined by the path-integral integrating over fluctuations of the fields around the following  singular fields configuration \cite{hep-th/0501015},
\begin{align}
&F = \frac{B}4 \epsilon_{ijk} \frac{x^i}{|\vec{x}|^3} dx^k \wedge dx^j \; , \nonumber
\\
& X_6 = \frac{B}{ 2 |\vec{x}|} \; .  \label{tHooft in R4}
\end{align}
The line operator is located at the singular point $\vec{x}=0$. 
Here $B$ denotes the  magnetic charge and it takes values in Cartan subalgebra $\mathbf{h}$ of the gauge group $G$. For the case $G=U(N)$ ,
\begin{align}
B = n_i H^i = \textrm{diag}(n_1, \ldots , n_N) \;, 
\end{align}
where $H^i$s are basis of $\mathbf{h}$. Using a conformal map, the Euclidean space $\mathbb{R}^4$ can be mapped to $\mathbb{R}\times S^3$ in the following way, 
\begin{align}
(x_0 , \vec{x}) = e^{- \tau} (\cos \chi , \sin \chi  \Omega_i), \quad \Omega_i  = (\sin \theta \cos \varphi, \sin \theta \sin \varphi, \cos \theta)  . \nonumber
\end{align}
 $\tau$ denotes the Euclidean time  and $(\chi, \theta, \varphi)$ denote a coordinate system of the three sphere in $\mathbb{R}\times S^3$. 
 Under the conformal transformation, the above 't Hooft line operator is  mapped  to 
\begin{align}
&F = -\frac{B}2 \sin \theta d\theta \wedge d \varphi \; , \nonumber
\\
& X_{12} := X_{34}^\dagger = \frac{1}2 (X_6 + i X_9)  = \frac{B}{4 \sin \chi} \; .  \label{tHooft line operator}
\end{align}
Here we introduce complex scalars $X_{AB}$ (see Appendix A for the relation between $X_{AB}$ and $X_m$). In the coordinate system, the line operator is located at  the north pole ($\chi =0$)  and the south pole ($\chi =\pi$). The line operator can be interpreted as an infinitely heavy magnetic monopole at the north pole (with magnetic charge $B$) and an antimonopole (with charge $-B$) at the south pole (see Figure~\ref{monopole-antimonopole}). 
The line operator preserves the subgroup $SO(1,2) \times SO(3)\times SO(5)$ of  bosonic conformal symmetries, $SO(2,4)\times SO(6)$, and 1/2 of the 32 supercharges of the theory.  The supersymmetry variation of $\mathcal{N}=4$ SYM theory on $\mathbb{R}\times S^3$ is given in \cite{hep-th/0605163} ($\sigma^a := (-\mathbb{I}, \vec{\sigma}) $, $a=0,1,2,3$)\footnote{After Wick rotation, we will consider the theory on $\mathbb{R}\times S^3$ with Minkowski time $t = i \tau$.}
\begin{align}
&\delta_\eta A_a =  i (- \psi_A^\dagger \sigma_a \eta^A + \eta^\dagger_A \sigma_a \psi^A)\; , \nonumber
\\
&\delta_\eta X^{AB} = i (- \eta^{AT} \sigma^2 \psi^B + \eta^{BT} \sigma^2 \psi^A - \epsilon^{ABCD}\psi^\dagger_C \sigma^2 \eta^*_D) \; , \nonumber
\\
&\delta \psi^{A} = \frac{1}2 F_{ab } \sigma^a \sigma^b \eta^A+ 2 D_a X^{AB} \sigma^a (\sigma^2 \eta_B^*) +X^{AB}\sigma^a \nabla_a (\sigma^2 \eta^*_B) + 2i [X^{AC}, X_{CB}] \eta^B \; .
\end{align}
The Killing spinors $\eta^A$ on $\mathbb{R}\times S^3$ are ($A=1,2,3,4$)
\begin{align}
&\eta_A = \eta^A_+  + \eta^A_-  \;, \nonumber
\\
&\eta_+^A =  e^{\frac{i}2 t }e^{ i/2 \chi \sigma_1}e^{i/2 \theta \sigma_3} e^{i/2 \varphi \sigma_1}  \epsilon^A_+ \;,  \quad \nonumber
\\
& \eta^A_- = e^{- \frac{i}2 t }e^{ -i/2 \chi \sigma_1}e^{i/2 \theta \sigma_3} e^{i/2 \varphi \sigma_1} \epsilon^A_- \; . \label{Killing spinor}
\end{align}
For the choice of vielbein basis, see appendix B.2. Here $\epsilon^A_{\pm}$ are constant two component spinors.  They parametrize the 32 real super and superconformal symmetries.  They satisfy the Killing spinor equations $\nabla_a \eta_\pm = \pm \frac{i}{2} \sigma_a \eta_\pm$.   These Killing spinors and  32 supercharges $(Q, \bar{Q}, S,\bar{S})$ in a flat space-time are related in the following way (based on the notation used in appendix A in \cite{hep-th/0510251}), 
\begin{align}
\delta (\eta^A)= \delta \big{(}\epsilon_+^A \bar{Q}_A + \epsilon_-^A S_A + (i \sigma^2 \epsilon^*_{+A}) \bar{S}^A + (i \sigma^2 \epsilon^*_{-A}) Q^A \big{)} \;.
\end{align} 
$\delta(\ldots)$ denotes the fermionic variation generated by supercharges (or Killing spinor) in the argument. We ignore spinor indices $\alpha$ and $\dot{\alpha}$ which should be properly contracted. 
The 't Hooft line operator  \eqref{tHooft line operator} preserves 16 fermionic symmetries, parametrized by the $\epsilon^A_\pm $ satisfying the following conditions
\begin{align}
\epsilon^{1}_\pm =  i \sigma^2 \epsilon^*_{\mp,2},\quad \epsilon^{3}_\pm =   i \sigma^2  \epsilon^*_{\mp,4} \;.  \label{SUSY projection for tHooft}
\end{align}
This implies that
\begin{align}
&\eta^1_\pm =  i e^{ \pm i \chi \sigma_1} \sigma^2  \eta^*_{\mp,2}\;, \; \eta^3_\pm =  i e^{ \pm i \chi \sigma_1} \sigma^2  \eta^*_{\mp,4} \; , \nonumber
\end{align}
and thus for the line operator \eqref{tHooft line operator},
\begin{align}
&\delta \psi^1 =\big{\{}- F_{\hat{\theta} \hat{\phi}}\sigma_1. e^{i \chi \sigma_1}  +2  \sigma_1 D_1 X^{12}    -2  i X^{12}  \big{\}} \sigma_2   (\eta^2_-)^* -\big{\{}\sigma_1 \leftrightarrow - \sigma_1\big{\}} \sigma_2   (\eta^2_+)^*   \nonumber
\\
& =\big{ \{}  \frac{B}{ 2 \sin^2 \chi} \sigma_1 . e^{i \chi \sigma_1} - \frac{\cos \chi B}{2 \sin^2 \chi}\sigma_1  - \frac{i B}{2 \sin \chi} \big{\}} \sigma_2  (\eta^2_-)^* -\big{\{}\sigma_1 \leftrightarrow - \sigma_1\big{\}} \sigma_2  (\eta^2_+)^*    \nonumber
\\
&= 0 \;.
\end{align}
It can be shown that $\delta \psi^A =0$ for $A=2,3,4$ in a similar way.    Thus, we check that the line operators \eqref{tHooft line operator} is invariant under the 16 fermionic symmetries given by the projection conditions \eqref{SUSY projection for tHooft}.

One may also consider Wilson line operators in the $\mathcal{N}=4$ SYM theory. A Wilson line operator in flat $\mathbb{R}^4$ can be written as 
\begin{align}
W_R =\textrm{tr}_R P \exp [\int_C  dx^0 (-i A_0 +  X_9  )]  \; ,
\end{align}
where the curve $C$ is supported on the $x^0$ direction. Here `$\textrm{tr}_R$' denote a trace in representation $R$ of gauge group. It preserves 16 supercharges and 8 among them are common to the supercharges preserved by the 't Hooft line operator considered above.  The electric charge of the Wilson line is given by the representation $R$ of the gauge group $G$. Note that both Wilson-line and 't Hooft line can present at the same time. 
\\

\subsection{Superconformal index compatible with line operators}

In this subsection, we will define a superconformal index (SCI) which is compatible with line operators  introduced in the previous subsection.

Among 16+16 supercharges in the $\mathcal{N}=4$ SYM, line operators preserve 16 supercharges. To define the superconformal index, we choose a supercharge $Q$ as (based on the notation used in appendix A in \cite{hep-th/0510251})
\begin{align}
Q = Q^{\alpha=1 , A=1} + \bar{Q}_{A=2}^{\dot{\alpha}=1} \; , 
\end{align}
$(\alpha, \dot{\alpha})$ are $\mathbf{(2,0)}$ and $\mathbf{(0,2)}$ indices for $SU(2)_L\times SU(2)_R = SO(4)$. $A=1,\ldots 4$ are for ${\bf 4}$ indices of $SU(4)$ R-symmetry.  Supercharges $Q^{\alpha=1,A=2}$ and $\bar{Q}_{A=2}^{\dot{\alpha}=1}$ are related to the Killing spinors $\eta^A$  \eqref{Killing spinor} on $\mathbb{R}\times S^3$ in the following way,
\begin{align}
&Q^{\alpha=1, A=1} \; \leftrightarrow \; \epsilon^*_{-,A=1} = \left(\begin{array}{c} 2^{-\frac{1}2}\\ 2^{-\frac{1}2}\end{array}\right) \;, \nonumber
\\
&\bar{Q}_{A=2}^{\dot{\alpha}=1} \; \leftrightarrow \; \epsilon^{A=2}_+ = \left(\begin{array}{c} 2^{-\frac{1}2}\\ -2^{-\frac{1} 2}\end{array}\right) \;.
\end{align}
Useful (anti-)commutation relations are 
\begin{align}
&\Delta :=\{ Q, Q^\dagger\} = \epsilon- j_L - j_R - r_1, \quad [Q, \epsilon + j_L + j_R] =0 \; . \label{property of Q}
\end{align}
$\epsilon$  is the energy associated with the time direction in $\mathbb{R}\times S^3$ and also the conformal dimension of operators in $\mathbb{R}^4$. $j_L, j_R$ denote the Cartan generators of $SU(2)_L, SU(2)_R$ respectively. Our 't Hooft line  preserves only the diagonal subgroup $SU(2)_{\textrm{diag}}$ of  $SU(2)_L \times SU(2)_R$, which is the rotational group of $S^2$ defined by $\theta,\varphi$ in $S^3$. $r_1$ denotes one of three Cartans of $SU(4)$,
\begin{align}
r_1 = \textrm{diag} (1,-1,0,0) \;.
\end{align} 
$r_1$ charge for scalar $X^{AB}$ is $\delta^{A,1}- \delta^{A,2} + \delta^{B,1}- \delta^{B,2}$ and the charge for fermion $\psi^A$ is $\delta^{A,1}-\delta^{A,2}$.

We define a superconformal index as
\begin{align}
I_L (x, \eta_a):= \textrm{Tr}_{\mathcal{H}_L} (-1)^F x^{\epsilon + j_L + j_R}  \prod_a \eta_a^{h_a}\;. \label{def of index}
\end{align}
Here $\mathcal{H}_L$ denotes the Hilbert space on $S^3$ in the presence of line operator $L$. We introduce chemical potentials $\{ \eta_a \}$ for Cartan charges $\{h_a\}$ of global symmetry group $H$ which commutes with the chosen supercharge $Q$, 
\begin{align}
[Q,H] =0 \;. \label{property of Q2}
\end{align}
When all chemical potentials except $x$ are turned off, the index  is sometimes called Schur index which is first introduced in \cite{Gadde:2011ik}\cite{Gadde:2011uv}.\footnote{In \cite{Gadde:2011ik}\cite{Gadde:2011uv}, the Schur index is defined as $I(q):=\textrm{Tr}(-1)^F q^{\epsilon-\frac{1}2  r_1}$. Using the BPS bound $\Delta=0$, our index can be rewritten as $I(x)= \textrm{Tr}(-1)^{F}x^{2\epsilon -r_1} $. Thus under the identification $q=x^2$, our index is same with the Shur index. }  For $\mathcal{N}=4$ SYM theory, we can turn on the chemical potential $\eta$ for $r_3$ which is another Cartan of $SU(4)$,
\begin{align}
r_3 =\textrm{diag}(0,0,1,-1) \;.
\end{align}
The index is independent of continuous ($Q$-preserving) deformations in the theory. 
One simple such deformation is changing the coupling constant of the theory  and thus the index is independent of the coupling constant. 

There are two approaches to calculate the index. 
First, one can explicitly construct the Hilbert space $\mathcal{H}_L$ from canonical quantization and calculate the index by taking  the trace over $\mathcal{H}_L$ . Due to the properties in eq.\eqref{property of Q} and \eqref{property of Q2}, only states which saturate the BPS bound  ($\Delta=0$) contribute to the index, thus we only need to take trace over  these BPS states. To extract contributions from gauge invariant states, we need to integrate the multi-particle index with Haar measure for subgroup of $G$ unbroken by line operators. 
The second approach is using the path integral representation of the index, 
\begin{align}
I_L (x, \eta_a) = \int D\Phi|_{L}  e^{-S_E[\Phi;(x,\eta_a )]}
\end{align}
After Wick rotation ($\tau = - i t$) and compactification of the Euclidean time ($\tau \sim \tau + \beta$),  we put the theory on $S^1\times S^3$. The size of the circle $\beta$ is related to the chemical potential $x$ as $x = e^{- \beta}$.\footnote{In general, one can consider more general  SCI by including new chemical potential $e^{-\beta^{\prime}}$, $I (\beta^\prime, x= e^{-\beta}, \eta_a) :=\textrm{Tr}(-1)^F e^{- \beta^\prime \Delta} e^{-\beta (\epsilon+j_L+ j_R ) } \prod_a \eta_a^{h_a}$. The path-integral representation of  index will be changed accordingly. For example, the size of the thermal circle will be $(\beta +\beta^\prime)$. However, we know that the index does not depend on $\beta^\prime$ since  only states with $\Delta=0$ contributes to the index. Thus we  set $\beta^\prime =0$  for simplicity.  } $\Phi$ denotes all  fields in the theory and the periodic boundary condition on the $S^1$ direction is imposed. The Euclidean action $S_E[\Phi]$ is twisted by chemical potentials in the following way
\begin{align}
\partial_\tau \rightarrow \partial_\tau - (j_L + j_R) + h_a \frac{\ln \eta_a}{\beta} \; . \label{twisting by chemicals}
\end{align}
 In the path-integral approach we have to sum over all the field configurations around the singular background \eqref{tHooft line operator}, which defines the  line operator $L$.
  As the superconformal index is independent of the coupling constant, we can  obtain the exact index  in the free theory limit. In the  limit, we only need to take into account of the one-loop corrections around the background, which needs the harmonic expansion around the  background. Due to the appearance of the thermal circle $S^1$, we can turn on holonomy $U = e^{i \lambda}$ of gauge fields along the circle direction, $A_0 = \frac{\lambda}{\beta}$. After the one-loop computation, the path integral can be written as  an integration of the holonomy variable $U$. These two approaches are, of course, equivalent and will give the same answer. 
We mainly use the first approach (canonical quantization) and comment on the relation to the second approach (path-integral) when it is necessary.  One advantage of the first approach is that it is more intuitive and  concrete since we are actually constructing states and counting them.  Another advantage is that we do not need to treat horribly complicated multi-particle index directly. We only need to count index contribution from single particle BPS states which saturate the bound $\Delta =0$. Then, the multi-particle index can be obtained simply by  taking Plethystic exponential of the single particle index. On the other hand, there is an advantage of  the second approach over the first one.  
Using the path-integral representation of the index, one can  relate the index to  partition functions on other manifolds. For example,  
taking small thermal circle $S^1$ limit, the index calculation can be reduced to  the calculation of partition function on $S^3$ \cite{Gadde:2011ia}\cite{Imamura:2011uw}\cite{Dolan:2011rp} (see also \cite{Benini:2011nc}). 

\section{Index calculation : classical and  1-loop contributions}

Since the index defined in the previous section is invariant of the coupling constant $g_{YM}$, the index calculated in the free theory limit ($g_{YM}\rightarrow 0 $) is exact. In the limit, only classical and 1-loop effect are relevant in the perturbative index computation. We will calculate these contributions in this section. We mainly focus on   the index with insertion of a 't Hooft line operator. The index formula with a Wilson line operator can be obtained   easily  as we will see in the section 3.5.

\subsection{Classical contribution}

Let us first  calculate the classical  value of $\epsilon+j_L +j_R$ of the 't Hooft line operator  \eqref{tHooft line operator}. Since the 't Hooft line operator is spherically symmetric and time independent, the classical value of $j_L + j_R$ vanishes.
\begin{align}
(j_L +j_R)^{(cl)} =0.
\end{align}
Classical energy $\epsilon^{(cl)}$ for the line operator is ($X^{12} \rightarrow X$ for simplicity and suppressing other irrelevant fields)
\begin{align}
\epsilon^{(cl)}_{bulk} = \frac{1}{g_{YM}^2}  \int d \chi (4 \pi) \sin^2 \chi \textrm{Tr} (\frac{1}2 F_{\hat{\theta}\hat{\phi}}^2  + 2 (D_\chi X)^2 + 2 X^2) := \int d\chi  \textrm{Tr}(\mathcal{E})\;. 
\end{align}
The unregulated energy is clearly divergent, as it measures the infinite self-energy of point-like monopole. The divergence can be regulated by introducing  a cutoff $\omega\ll 1$ and integrate $\chi$ on the interval $(\omega, \pi - \omega)$. We also need a boundary term for the energy which is supported on the boundaries,  $\chi =\omega$  and $\pi - \omega$.
\begin{align}
\epsilon^{(cl)}_{bdy} = -\textrm{Tr}( X P_X )  |^{\pi -\omega}_{\omega} = -\frac{16 \pi }{g_{YM}^2} \sin^2 \chi \textrm{Tr}(X \partial_\chi X) |^{\pi -\omega}_{\omega} \;,  \label{bdy term}
\end{align}
where $P_X := \frac{\partial \mathcal{E}}{\partial (\partial_\chi X)}$. See the section 2.2 in \cite{Drukker:2005kx} for similar boundary terms for DBI action.  We will  why the  boundary term is neccesary.
Under the variation of fields ($\delta A_i$, $\delta X$), the bulk energy functional varies as
\begin{align}
\delta \epsilon^{(cl)}_{bulk} = \int d \chi \big{[}(\textrm{e.o.m for $A_i$})\delta A_i +(\textrm{e.o.m for $X$})\delta X \big{]} + \textrm{Tr}(P_{X}\delta X )|^{\pi -\omega}_{\omega} \;.
\end{align}
Since $\lim_{\chi \rightarrow 0, \pi} P_X= - \frac{4\pi }{g_{YM}^2}  B$, the $P_X$ near the boundaries measure the magnetic charge of a line operator, which should be a fixed value for the given line operator. 
Thus we should impose the following boundary condition on field $X$,
\begin{align}
\delta P_X =0\;, \quad \textrm{at $\chi = \omega$ and $\pi - \omega$} \;.
\end{align}
This boundary condition should  be considered as a part of definition of a line operator. 
With the boundary condition, the boundary variation term  in $\delta \epsilon_{bulk}^{(cl)}$ does not vanish.  
To cancel the boundary variation term, we introduced the boundary term in eq.~\eqref{bdy term}.
Let $ \epsilon^{(cl)} := \epsilon^{(cl)}_{bulk} +\epsilon^{(cl)}_{bdy}$, then
\begin{align}
\delta \epsilon^{(cl)} &=  \int d \chi \big{[}(\textrm{e.o.m for $A_i$})\delta A_i +(\textrm{e.o.m for $X$})\delta X \big{]}  - \textrm{Tr} ( \delta P_X X)|^{\pi - \omega}_{\omega} \nonumber
\\
&= \int d \chi \big{[}(\textrm{e.o.m for $A_i$})\delta A_i +(\textrm{e.o.m for $X$})\delta X \big{]}  .
\end{align}
The energy is extremized  by saddle points satisfying the equations of motion only after introducing the boundary term \eqref{bdy term}.  With bulk and boundary energy term, the classical energy for the line operator \eqref{tHooft line operator} vanishes. 
%
\begin{align}
\epsilon^{(cl)} &= \epsilon^{(cl)}_{bulk} +\epsilon^{(cl)}_{bdy}\; , \nonumber
\\
 &=\frac{\pi \textrm{Tr}(B^2)}{g_{YM}^2} \big{(} \int^{\pi -\omega}_\omega \csc^2 \chi d \chi + \cot \chi|^{\pi -\omega}_{\omega} \big{)} =0\;. 
\end{align}
Thus the classical contribution to the index is $1$, 
\begin{align}
x^{(\epsilon+j_L+j_R)^{(cl)}} =1 \;.
\end{align}
It's expected because if  the classical value  is non-zero, it will appear in  the index as an overall factor in the form of $\exp (...{\frac{1}{g_{YM}^2}})$
 which contradicts   the fact that the index is independent of the coupling constant $g_{YM}$.
 The quadratic fluctuations around the  line operator \eqref{tHooft line operator}  can be decomposed into four pieces.
\begin{align}
\mathcal{L}_{quad} &= \mathcal{L}_1 (X_{AB},X^{\dagger}_{AB})|_{(A,B)=(1,3),(1,4)} + \mathcal{L}_2 (\psi^{A}, \psi_A^\dagger)|_{A=1,2} \nonumber
\\
&+\mathcal{L}_3 (\psi^A, \psi_A^\dagger)|_{A=3,4} + \mathcal{L}_4 (X_{12},X^\dagger_{12}, A_\mu) \;. 
\end{align}
In the below, we will construct the Hilbert space on $S^3$ by quantizing each quadratic actions and calculate their contributions to the index. 
\\

\subsection{Single particle index contribution from $\mathcal{L}_1$}

The Lagrangian $\mathcal{L}_1$ is given by (we absorbed $g_{YM}$ by field redefinition)
\begin{align}
 \mathcal{L}_1 &= \frac{1}2 \sum_{(A,B)=(1,3),(1,4)}\textrm{Tr}\big{(} \dot{X}_{AB}^\dagger\dot{X}_{AB} - X_{AB}^\dagger (-D_{S^3}^2 + 1 ) X_{AB} + \frac{1}{4\sin^2 \chi}[X_{AB}, B][X_{AB}^\dagger, B] \big{)}\;.
\end{align}
Let us first expand $X_{AB}$ in a basis of Lie algebra $t_\alpha$s, $X = X^\alpha t_\alpha$,  where the generators satisfy
\begin{align}
&t_\alpha^\dagger = t_{-\alpha}, \quad \textrm{Tr} (t_\alpha t_\beta) = \delta_{\alpha, - \beta}, \quad [  B, t_\alpha ] = \alpha (B) t_\alpha \;.\nn 
\end{align}
$\alpha=1, \ldots , \textrm{dim}(G)$ label the roots (including zero roots) of gauge group $G$. Then the action becomes
\begin{align}
& \mathcal{L}_1 = \frac{1}2 \sum_{(A,B)} \sum_\alpha (\dot{X}^{\alpha}_{AB})^* \dot{X}_{AB}^{\alpha}-( X_{AB}^{\alpha})^*   M_{q_\alpha}^2 \cdot X_{AB}^{\alpha}\;, 
\\
&\textrm{where, } M_{q_\alpha}^2 = - D^2_{q_\alpha} + 1 + \frac{q_\alpha^2}{\sin ^2\chi}\;. 
\end{align}
$D_q$ denote the covariant derivative on $S^3$ with magnetic fluxes of monopole charge $q$  turned on $S^2$ directions in $S^3$, $F = q  \textrm{vol ($S^2$)}$. $q_\alpha$ is given by $q_\alpha =\frac{1}2 \alpha (B)$. Spectrum of the differential operator $M_{q}^2$ is analyzed in Appendix B. Expanding $X_{13}$ and $X_{14}$
in terms of eigenfunctions of $M_q^2$,
\begin{align}
X_{AB} (t, \Omega_3) = \sum c_{AB}^{q,n,J,m} (t)Y^{q}_{n,J,m} (\Omega_3)\;,
\end{align}
the quadratic action describes  infinitely many decoupled quantum mechanical (QM) harmonic oscillators. Each mode $c_{AB}^{q,n,J,m}$  has following quantum numbers,\footnote{$n$-particles state $|n\rangle$ in a harmonic oscillator of mass $\epsilon$ has energy $n \epsilon$  ignoring the zero-point energy $\frac{\epsilon}2$. For more discussion on the zero-point energy, see the section 3.7.}
\begin{align}
\begin{tabular}{c|c|c|c|c|}  & $\epsilon$ (mass) & $j_L + j_R$ & $r_1$ & $r_3$  \\\hline $c_{(13),(14)}^{q,n,J,m}$ & $n+1$ & $m$ & $1$ &  $(1,-1)$  \end{tabular}
\end{align}
In the table, we only list quantum numbers for modes $c$ but not for its conjugation $c^*$, which can be obtained simply by flipping signs of quantum numbers $(j_L+j_R), r_1$ and $r_3$ for $c$.  The range of quantum numbers is given as
\begin{align}
J=|q|,|q|+1, \ldots, \; |m| \leq  J, \; \; n= J, J+1, \ldots
\end{align} 
Modes  $c^{q,n,J=n,m=n}_{(13),(14)}$ ($n= |q|, |q|+1,\ldots$) saturate the BPS bound $\Delta= 0$, and their contribution to the index is $x^{2 n+1 }$. Summing all contributions from these harmonic modes, we obtain the single particle  index from $\mathcal{L}_1$
\begin{align}
I_{sp;\mathcal{L}_1} ( e^{i \lambda_i},x, \eta) = (\eta +\eta^{-1})\sum_\alpha \sum_{n=|q_\alpha|} x^{2n+1} e^{i \alpha (\lambda)}= (\eta +\eta^{-1}) \sum_\alpha \frac{x \cdot x^{ |\alpha (B)|}e^{i \alpha (\lambda)}}{1-x^2} \;.
\end{align}
Here we include chemical potentials $\{e^{i \lambda_i}\}|_{i=1,\ldots, \textrm{rank}(G)}$ for Cartan algebra basis $\{H_i\}$ of gauge group $G$. Actually, non-BPS states from $\mathcal{L}_1$   contribute to the index, but  they will be canceled by  non-BPS states from fermionic modes (in $\mathcal{L}_2$ or $\mathcal{L}_3$) and we  ignore these contributions at this stage.

\subsection{Single particle index contribution from $\mathcal{L}_2$  and $\mathcal{L}_3$}
The action $S_2$ from quadratic interaction of $\psi^{1,2}$ is given by
\begin{align}
\mathcal{L}_2 =&\sum_{A=1,2} ( i \psi_A^\dagger  \dot{\psi}^A+i \psi^\dagger_A \sigma^i D_i \psi^A)+ \frac{1}{2\sin \chi} \psi_1^\dagger \sigma^2 [B, (\psi^\dagger_2)^T]- \frac{1}{2\sin \chi} {\psi^{1T}} \sigma^2 [B,  \psi^2] \; .
\end{align}
Expanding fields in terms of basis of Lie algebra $\psi = \psi^\alpha t_\alpha$, the action become
\begin{align}
 \sum_{\alpha}  (\psi^{\dagger\alpha} ,\zeta^{\dagger\alpha}) \left(\begin{array}{cc} i   \partial_t +i \slashed  D_{q_\alpha} & \frac{   q_\alpha }{\sin \chi} \\    \frac{ q_\alpha }{\sin \chi}  &  i   \partial_t -i \slashed  D_{q_\alpha}\end{array}\right) \left(\begin{array}{c} \psi^\alpha \\ \zeta^\alpha \end{array}\right)
\end{align}
Here $\psi = \psi_1, \zeta^\alpha = \sigma^2 \cdot (\psi_2^{-\alpha})^\dagger$. $i \slashed D_q$ denote the Dirac operator on $S^3$ with monopole flux turned on $S^2$ direction ($F  = q \textrm{vol}(S^2)$). One needs to analyze the spectrum of the following operator
\begin{align}
\slashed M_q := \left(\begin{array}{cc} i \slashed  D_{q} & \frac{   q }{\sin \chi} \\    \frac{ q }{\sin \chi}  & -i \slashed  D_{q}\end{array}\right) 
\end{align}
The spectrum of the operator is analyzed in Appendix B. Expanding $(\psi, \zeta)$ in terms of the eigen-spinor $ \Psi^{q;\pm , \kappa}_{n,J,m}$ of $\slashed M_q$, 
\begin{align}
\left(\begin{array}{c}\psi \\ \zeta\end{array}\right) (t, \Omega_3) = \sum c^{q;\pm , \kappa}_{n,J,m} (t) \Psi^{q;\pm , \kappa}_{n,J,m} (\Omega_3)
\end{align}
the action describe the infinitely many decoupled fermionic (QM) harmonic oscillators. Each pair of $(c^+, c^-)$ form a harmonic oscillator. Each modes have following quantum numbers
\begin{align}
\begin{tabular}{c|c|c|c|c|c}  & $\epsilon$ (mass) &  $(j_L + j_R)$&  $r_1$ & $r_3$   \\\hline $(c^{q;+,\kappa}_{n,J,m},c^{q;-,\kappa}_{n,J,m})$ & $n+1$ & $m$  & 1 & 0   \end{tabular}
\end{align}
The range is given by
\begin{align}
\textrm{Range : } & J= |q|- \frac{1}2 \; (\textrm{exist for $|q|\neq 0$})  , |q| + \frac{1}{2} , |q|+ \frac{3}2,  \; \\
&  |m|\le J, \ n=J, J+1 \ldots
\end{align}
Modes with $n=J, \ m=J$ saturate the BPS bound and they give $- x^{2J+1}$ to the index. Summing index contribution from all these modes, one gets (note that $\kappa=1$ for $J= |q|- \frac{1}2$ and $\kappa=1,2$ otherwise)
\begin{align}
&I_{sp;\mathcal{L}_2} ( e^{i \lambda_i}, x, \eta)  = \sum_\alpha \big{[} \sum_{J= |q_\alpha|+\frac{1}2}^\infty (-2 x^{2J+1} e^{i \alpha (\lambda)}) - (1- \delta_{q_\alpha,0}) x^{2|q_\alpha|}e^{i \alpha (\lambda)} \big{]} \; , \nonumber
\\
&= \sum_\alpha \big{(} \frac{-2 x^2 \cdot x^{|\alpha (B)|}}{1-x^2} -(1- \delta_{\alpha(B),0})x^{|\alpha(B)|} \big{)} e^{i \alpha (\lambda)}\; .
\end{align}
Quadratic interaction action $\mathcal{L}_3 (\psi^3, \psi^{\dagger}_4)$ is identical to the action  $\mathcal{L}_2 (\psi^1, \psi^\dagger_2)$. Unlike $\psi^1$ and $\psi^\dagger_2$, however,  $\psi^3$ and $\psi^\dagger_4$ have $r_1$-charge 0 and thus modes from them can't saturate the BPS bound.  Thus 
\begin{align}
I_{sp;\mathcal{L}_3} =0 \; . 
\end{align}

\subsection{Single particle index contribution from $\mathcal{L}_4$}
First consider the $B=0$ case. In the case, the index can be calculated  just by counting gauge invariant operators in the theory on $\mathbb{R}^4$. One can easily see that `letters' made of fields in $\mathcal{L}_4$ can not saturate the BPS bound and thus one can conclude that there is no index contribution from $\mathcal{L}_4$ when $B=0$. Even after introducing 't Hooft line operator with charge $B \neq 0$, we expect that fluctuation modes from $\mathcal{L}_4$ can not contribute to the index.  It is unphysical  that non-BPS states become  BPS states after interacting with  BPS defects. 
Thus we can conclude  that 
\begin{align}
I_{sp;\mathcal{L}_4} =0\; . 
\end{align} 
It is worth checking it explicitly by honestly analyzing the fluctuation of fields in $\mathcal{L}_4$.

\subsection{Summary}

From the above analysis, one gets  the (1-loop) single particle index  as follow: 
\begin{align}
&\tilde{I}_{sp} (e^{i \lambda_i},x, \eta) = \sum_{i=1}^4 I_{sp;\mathcal{L}_i}   \nn
\\
&= \sum_\alpha \big{(} \frac{(\eta+\eta^{-1})x \cdot x^{|\alpha (B)|}}{1-x^2} -\frac{2x^2 \cdot x^{|\alpha (B)|}}{1-x^2} \big{)}e^{i \alpha (\lambda)} - \sum_{\alpha(B)  \neq 0 } e^{i \alpha (\lambda)} x^{|\alpha(B)|} \;.
\end{align}
Multi-particle index can be obtained by taking Plethystic exponential (P.E) of the single particle index. 
\begin{align}
I_{\textrm{multi}} (e^{i \lambda_i},x, \eta) = \textrm{P.E} [\tilde{I}_{sp} (e^{i \lambda_i},x, \eta) ]\;,
\end{align}
where the action of $\textrm{P.E}$ is defined by 
\begin{align}
\textrm{P.E}[f (e^{i \lambda_i},x, \eta)] = \exp \big{[} \sum_{n=1}^\infty \frac{1}{n}f ( e^{i n \lambda_i},x^n, \eta^n )\big{]} \;. \label{P.E}
\end{align}
The multi-particle index $I_{\textrm{multi}} (x, \eta, e^{i \lambda_i})$ contains contributions from gauge-variant states. To count index from gauge-invariant sates, we  need to integrate the multi-particle index by (normalized) Haar measure of the gauge group $G_B$ unbroken by magnetic charge $B$. More explicitly, the unbroken subgroup $G_B$ is given by
\begin{align}
G_B = \{ g : g \in G \textrm{ and } [g,B]=0\}\;.
\end{align}
Thus, the 1-loop superconformal index with magnetic charge $B$ is given as 
\begin{align}
&I^{1-loop}_B (x,\eta) = \int [\widetilde{dU}]_B  \; \textrm{P.E}[\tilde{I}_{sp} ( e^{i \lambda},x, \eta)] \;, \; \textrm{with} \nn
\\
&  \widetilde{[dU]}_B \equiv \frac{1}{\textrm{sym}(B)}  \big{(} \prod_{i=1}^{\textrm{rank}(G)}\frac{d\lambda_i}{ 2\pi} \big{)}   \prod _{\alpha (B)  = 0, \alpha \neq 0} (1-e^{i \alpha (\lambda)}) \;. \label{Pre : 1-loop index}
\end{align}
Here $\widetilde{[dU]}_B$ denote the Haar measure of unbroken gauge group $G_B$ with normalization $\int \widetilde{[dU]}_B =1$.  The symmetric factor $\textrm{sym}(B)$ is  the order (number of elements) of the Weyl group of $G_B$, $\textrm{sym}(B) =|\textrm{Weyl}(G_B) |$. The integration variables, $\lambda= \lambda_i H^i$,  parametrize the maximal torus $\mathbb{T}^n$ of gauge group $G$ . 
\begin{align}
\mathbb{T}^n =\{ e^{i \lambda_i H^i}  : \lambda_i \sim \lambda_i +2\pi\}\;.
\end{align}
Here we will give a comment on how the integration of $ \widetilde{[dU]}_B$ appears in  the path-integral approach. The integral variables $\lambda$  correspond to holonomy  $U = e^{i \lambda}$ of the gauge field around the thermal  circle, $A_0 = \frac{\lambda}{\beta} $. The Haar measure for the subgroup $G_B$ can be understood as Faddeev-Popov determinant for a gauge fixing of zero mode of $A_0$, $\frac{d}{d\tau} \int_{S^3} A_0 =0$. See  Appendix B.2 in \cite{Kim:2009wb} for the explicit derivation.

For more succinct expression, we rewrite the 1-loop superconformal index \eqref {Pre : 1-loop index}
in the following way:
\begin{align}
&I^{1-loop}_B (x,\eta) =  \int  [dU]_B  Z^{1-loop}_B (e^{i \lambda_i},x, \eta)\;,   \textrm{ where}\nonumber
\\
& [dU]_B \equiv  \frac{1}{\textrm{sym}(B)}  \big{(} \prod_{i=1}^{\textrm{rank}(G)}\frac{d\lambda_i}{ 2\pi} \big{)} \prod _{\alpha \neq 0} (1-e^{i \alpha (\lambda)}x^{|\alpha(B)|}) \; ,  \textrm{ and}\nonumber \;
\\
& Z^{1-loop}_{B} ( e^{i \lambda_i},x, \eta) := \textrm{P.E} [I_{sp} ( e^{i \lambda_i}, x, \eta)]\; , \textrm{ with}\nn
\\
& I_{sp} (e^{i \lambda_i},x,\eta) = \sum_{\alpha=1}^{\textrm{dim}(G)} \big{(} \frac{(\eta+\eta^{-1})x \cdot x^{|\alpha (B)|}}{1-x^2} -\frac{2x^2 \cdot x^{|\alpha (B)|}}{1-x^2} \big{)}e^{i \alpha (\lambda)}\; . \label{1-loop for N=4 SYM}
\end{align}
This is the final expression for the  1-loop superconformal index for $\mathcal{N}=4$ SYM with gauge group $G$ in the presence of 't-Hooft line operator with magnetic charge $B$.

For the 1/2 BPS Wilson line operator in representation $R$ of gauge group $G$ at the north pole and its anti-counterpart at the south pole, the SCI is given by
\begin{align}
&I_{R} (x ,\eta )=  \int [dU]_{B=0} \Big{[}  \chi_R (e^{i \lambda_i}) \chi_{\bar{R}}  (e^{ i \lambda_i}) \Big{]} Z^{1-loop}_{B=0} (x,\eta, e^{i \lambda_i}) \; \label{index for Wilson in N=4 SYM} 
\end{align} 
 $\chi_R$ and $\chi_{\bar{R}}$ denote the character of representation $R$ and $\bar{R}$ respectively, $\chi_{\bar{R}} (e^{i \lambda}) =\chi_{R} (e^{-i \lambda})$.  In the path integral point of view, the factor $\chi_R (e^{i \lambda})\chi_{\bar{R}} (e^{i \lambda})$ is nothing but the classical value for the Wilson line operators when the holonomy of gauge fields, $U= e^{i \lambda}$,  along the thermal circle $S^1$ are turned on. 

  For explicit calculation of the index, let us briefly summarize some group theory facts relevant to  the index formula.
We mainly focus on cases where $G=U(N), Sp(2N), SO(2N+1)$  and $SO(2N)$. We choose basis $\{ H_i\}$ of Cartan algebra $\mathbf{h}$ for each group $G$ as 
\begin{align}
\lambda_i H^i &=  \textrm{diag}(\lambda_1, \ldots, \lambda_N) \;, \quad \textrm{for $U(N)$} \; \nn
\\
&= \sigma_3 \otimes \textrm{diag}(\lambda_1, \ldots, \lambda_N)\;, \quad \textrm{for $Sp(2N)$} \nn
\\
&= \left(\begin{array}{cc}\textrm{diag}(\lambda_1, \ldots, \lambda_N) \otimes \sigma_2  & \mathbf{0}_{1\times 2N} \\ \mathbf{0}_{2N\times 1}  & 0\end{array}\right)\;, \quad \textrm{for $SO(2N+1)$}\nn
\\
&= \textrm{diag}(\lambda_1, \ldots, \lambda_N) \otimes \sigma_2 \;, \quad \textrm{for $SO(2N)$} \;.
\end{align}
We choose a skew-symmetric matrix $\Omega$ used in the definition of  $Sp(2N)$ as
\begin{align}
\Omega = i \sigma_2 \otimes \mathbb{I}_{N}, \;
\end{align} 
where  $\mathbb{I}_N$ denote the unit matrix of size $N$. In these choices, action of  roots $\alpha$ on  $\lambda= \lambda_i H^i$, a element in Cartan subalgebra,   is given by $(i,j=1,2,\ldots,N)$
\begin{align}
\{\alpha(\lambda) \} &= \{ \lambda_i - \lambda_j \} \;,  \quad \textrm{for $U(N)$}\nn
\\
&= \{ \pm (\lambda_i \pm \lambda_j )|_{ i<j } , \pm 2 \lambda_i, 0^{N}\} \;,  \quad \textrm{for $Sp(2N)$}\nn
\\
&= \{ \pm (\lambda_i \pm \lambda_j )|_{i<j } , \pm  \lambda_i, 0^{N}\} \;, \quad \textrm{for $SO(2N+1)$}\nn
\\
&= \{ \pm (\lambda_i \pm \lambda_j )|_{i<j } \} \;, \quad \textrm{for $SO(2N)$} \;. \nn
\end{align}
Here the superscript of $e^N$ represents that the  element $e$ is repeated $N$ times. Order of Weyl group for each gauge group is
\begin{align}
&|\textrm{Weyl} (U(N))| = N! \; , \nn
\\
&|\textrm{Weyl} (Sp(2N))| =|\textrm{Weyl} (SO(2N+1))|=2^N N! \;, \nn
\\
&|\textrm{Weyl} (SO(2N))| = 2^{N-1} N! \;. 
\end{align}
For the case when $G=SU(N)$ the index formula  resembles  $G=U(N)$ case except two major differences, both of which are originated from the absence of diagonal $U(1)_b\subset U(N) $ in $SU(N)$.\footnote{For $N=2$, $SU(2)$ is equivalent to $Sp(2)$. In this case  using the  index formula for $Sp(2N)$ theory, instead of ``quotienting'' the $U(2)$ theory index   by $U(1)_b$, one can obtain the index  for $SU(2)$ theory. } Firstly, $SU(N)$ has $N^2-1$ roots and one root $\alpha$ in $U(N)$ with $\alpha (\lambda)=0$ is absent. Thus for $\mathcal{N}=4$ SYM, 
\begin{align}
I^{SU(N)}_{sp} (e^{i \lambda_i},x, \eta)  = I^{U(N)}_{sp} (e^{i \lambda_i},x, \eta) - \frac{(\eta+\eta^{-1})x -2x^2}{1-x^2} \;.
\end{align}
Secondly, $SU(N)$ has maximal torus $\mathbb{T}^{N-1}$   and we need to integrate over traceless part of Cartan subalgebra $\mathbf{h}$ of $U(N)$. One simple way of restricting the integral range to traceless  part is  multiplying integrand by a delta function,
\begin{align}
[dU]^{SU(N)}_B = [dU]^{U(N)}_B \times \delta (\sum_i \lambda_i) = [dU]^{U(N)}_B  \sum_{q \in \mathbb{Z}} e^{i q (\sum_i \lambda_i)}\;.
\end{align}
Here charge $q$ can be considered as a baryon charge, $U(1)_b$. Since the $U(1)_b$ is not a part of gauge symmetry in $SU(N)$ theory, we should count states charged under the $U(1)_b$. For $\mathcal{N}=4$ SYM, however,  index contributions from nonzero baryon charges vanish since all matters are in adjoint representation (thus $q=0$) and total baryon charge of the line operator is zero due to the cancellation of baryon charges from two poles. Therefore  two indices for  $SU(N)$  and $U(N)$  $\mathcal{N}=4$ SYM theories  are  just simply related by a overall  factor, 
\begin{align}
I^{\textrm{$U(N)$ $\mathcal{N}=4$ SYM}}_{B(R)} (e^{i \lambda_i},x, \eta) = I_{B(R)}^{\textrm{$SU(N)$ $\mathcal{N}=4$ SYM}}(e^{i \lambda_i},x, \eta) \times \textrm{P.E}[\frac{(\eta+\eta^{-1})x -2x^2}{1-x^2}] \;.
\end{align}
In the next section, we will generalize the index formula to general $\mathcal{N}=2$ superconformal theories. For these theories, two indices for $SU(N)$ and $U(N)$ gauge theories can not be  related   in a simple way  since index contribution from non-zero baryon charge does not vanish in general $SU(N)$ theories.
\subsection{Generalization to $\mathcal{N}=2$ theories}
It is straight forward to extend the above results to the general $\mathcal{N}=2$ superconformal field  theories. 
We choose an $\mathcal{N}=2$ subalgebra as algebra generated by  supercharges in $\mathcal{N}=4$ with $SU(4)_R$ index $A=1,2$.  In terms of  multiplets in the $\mathcal{N}=2$ algebra, the $\mathcal{N}=4$ multiplet is  decomposed as 
\begin{align}
&(X_{12}, A_\mu, \psi_1, \psi_2) \; : \; \textrm{Vector multiplet}, \nonumber
\\
&(X_{13}, X_{23}, \psi_3, \psi_4^\dagger) \; : \; \textrm{Hyper-multiplet} \;. 
\end{align}
Under the $SU(2)$ R-symmetry in the $\mathcal{N}=2$ algebra, 
\begin{align}
\left(\begin{array}{c}  X_{13} \\ X_{23} \end{array}\right) , \quad \left(\begin{array}{c}\psi_1 \\ \psi_2\end{array}\right) \; : \; \mathbf{2}
\end{align}
and others are in the singlet. The charge $r_1$ in the BPS bound, $\Delta = \epsilon- j_L -j_R - r_1 =0$, corresponds to the Cartan of the $SU(2)_R$. 
Since the line operators in the $\mathcal{N}=4$ theory is defined using fields  in a vector multiplet of the $\mathcal{N}=2$  subalgera, 
the definition can be easily extended to general $\mathcal{N}=2$ theories. The   index \eqref{def of index} also can be easily generalized to $\mathcal{N}=2$ theories, since we use a supercharge $Q$ in the $\mathcal{N}=2$ algebra to define the index.

To extend the index computation in the $\mathcal{N}=4$ to  general $\mathcal{N}=2$ theories, we only need to replace the index contributions from a hyper-multiplet $(X_{13}, X_{23}, \psi_3, \psi_4^\dagger)$ in the adjoint representation with  index contributions from hyper-multiplets $H_i $ in arbitrary representation  $R_i$.
\begin{align}
\sum_\alpha \frac{ (\eta+\eta^{-1}) x \cdot x^{|\alpha(B)|}}{1-x^2} e^{i \alpha (\lambda)}\; \rightarrow \; \sum_i \sum_{\rho \in R_i} \frac{x \cdot x^{|\rho(B)|}}{1-x^2} ( e^{i \rho(\lambda)} \prod_a \eta_a^{h_{i,a}} +  e^{- i \rho(\lambda)}  \prod_a \eta_a^{-h_{i,a}})\; .
\end{align}
 $\rho \in R_i$ denote the all the weights of representation $R_i$ of gauge group $G$. We introduce chemical potentials $\{ \eta_a \}$ for Cartan charges $\{h_a\}$ of flavor symmetry. Charge $h_a$  for  a hypermultiplet $H_i$ is given by  $h_{i,a}$.
 For general $\mathcal{N}=2$ gauge theory with gauge group $G$ with  hyper-multplets $H_i $ in representation $R_i$, the 't Hooft line superconformal index is given by
\begin{align}
&I^{1-loop}_{B} (x, \eta_a ) =  \int [dU]_B  Z^{1-loop}_B (e^{i \lambda_i},x,\eta_a) \;, \textrm{ where}\nonumber \;
\\
&Z^{1-loop}_B (x,\eta_a, e^{i \lambda_i})   = \textrm{P.E}[I_{sp}(e^{i\lambda_i},x, \eta)]\;,  \;I_{sp} = I^{vec}_{sp} + I^{hyper}_{sp}\;\textrm{ with}\nonumber
\\
&I^{vec}_{sp} = - 2 \sum_\alpha \frac{ x^2 \cdot x^{|\alpha(B)|}}{1-x^2} e^{i \alpha (\lambda)} \textrm{ and}\nn
\\
& I_{sp}^{hyper} = \sum_i \sum_{\rho \in P_i } \frac{x \cdot x^{|\rho(B)|}}{1-x^2} (e^{i \rho (\lambda)} \prod_a \eta_a^{h_{i,a}} +e^{- i \rho(\lambda)}\prod_a \eta_a^{-h_{i,a}})\;. \label{N=2 index} 
\end{align}
Here $B$  denote again the magnetic charge for 't Hooft line operator. For Wilson line operators in representation $R$, the index is 
\begin{align}
&I_{R} = \int [dU]_{B=0} \chi_R (e^{i \lambda}) \chi_{\bar{R}}(e^{i \lambda}) Z^{1-loop}_{B=0}( e^{i \lambda_i},x,\eta_a)  \;.
\end{align} 

One consistency check  can be done by comparing  the conventional superconformal index of $\mathcal{N}=2$ gauge theories with our index when line operator is absent, $B=R=0$. The conventional superconformal index is defined as (see \cite{arXiv:1011.5278})
\begin{align}
I_{usual}(t, y ) = \textrm{Tr} (-1)^F t^{2(E+j_L)} y^{2 j_R}  \prod_a \eta^{h_a}\; .
\end{align}
When $t^2 = x, y^2=x$, this coincide with our definition of index and thus formulae for two indices should be equivalent. The single particle index for $I_{usual} (t, y)$ is given by \cite{hep-th/0510251}\cite{arXiv:1011.5278}
\begin{align}
&I^{vec}_{sp;usual} (e^{i \lambda_i},t,y) = \sum_{\alpha} (\frac{t^2 - t^4}{(1-t^3 y)(1-t^3 y^{-1})} + \frac{2t^6 - t^3 (y+1/y)}{(1-t^3 y)(1-t^3 y^{-1})} )e^{i \alpha (\lambda)} \; , \nonumber
\\
&I^{hyper}_{sp;usual} (e^{i \lambda_i},t,y) = \sum_i \sum_{\rho\in R_i} (\frac{t^2 - t^4}{(1-t^3 y)(1-t^3 y^{-1})}) (e^{i \rho (\lambda)} \prod_a \eta_a^{h_{i,a}} +e^{- i \rho(\lambda)}\prod_a \eta_a^{-h_{i,a}})\; . 
\end{align}
When $t^2=x, y^2=x$, this gives the exactly same expression in eq.~\eqref{N=2 index} with $B=0$.

\subsection{Casimir energy and 1-loop $\beta$-function}
So far we ignore a contribution from  Casimir energy $\epsilon_0 $ (or sometimes called zero-point energy), which is an  important element in  three dimensional  superconformal index \cite{Kim:2009wb}\cite{Imamura:2011su}.  
It is because, unlike 3d superconformal index,  we do not need to sum over indices from various magnetic charges $B$. For fixed magnetic charge $B$, the Casimir energy gives an overall factor $x^{\epsilon_0(B)}$  in the index.  Demanding the index to  start with the zeroth power in $x$, we can fix the $\epsilon_0 (B)$. 

However, the situation changes when we are considering the monopole bubbling effect. In the case (as we will see in the section 4), we need to sum over  1-loop indices  from various magnetic  charges with   proper weight factors. Thus, in this case, the differences between the Casimir energies for various magnetic  charges become relevant.  Now, we will determine the Casimir energy up to an addictive constant. Following the prescription in \cite{Kim:2009wb}\cite{Imamura:2011su}, the Casimir energy can be computed as
\begin{align}
\epsilon_0 &= \frac{1}2 \partial_x \big{(}\tilde{I}_{sp} (e^{i\lambda_i}=1,x, \eta_a =1 ) \big{)}|_{x=1} \;.
\end{align}
Here, 
\begin{align}
&\tilde{I}_{sp} (e^{i\lambda_i}=1,x, \eta_a=1)  \;  \nonumber
\\
&=\sum_i\sum_{\rho \in R_i}  \big{(}\frac{2x^{1 +|\rho (B)|} }{1-x^2} \big{)} - \sum_{\alpha \in G}\big{(}  \frac{2x^{2 +|\alpha (B)|} }{1-x^2}  \big{)} -\sum_{\alpha (B)\neq 0} x^{|\alpha(B)|} \;.
\end{align}
The naive application of the formula for $\epsilon_0$ to the above $\tilde{I}_{sp}$  lead to a divergent results. To regulate the divergence,  we will subtract the Casimir energy for zero magnetic charge, $B=0$, from the divergent quantity. Then, one get following finite Casimir energy
\begin{align} 
\epsilon_0 (B) &= \frac{1}{2} \partial_x  \big{(}\tilde{I}^B_{sp} (x) - \tilde{I}^{B=0}_{sp}(x) \big{)}|_{x\rightarrow 1} \nonumber
\\
&=\frac{1}2 \big{(} \sum_{\alpha \in G} |\alpha (B)|^2 - \sum_i \sum_{\rho \in R_i} |\rho(B)|^2 \big{)}\;. \label{casimir energy}
\end{align}
Note that for any $\mathcal{N}=2$ superconformal field theories, the Casimir energy vanishes. This can be seen by rewriting the Casimir energy in the following form
\begin{align}
\epsilon_0 (B) =\frac{1}2 \textrm{Tr}(B^2)\times \big{(}T(\textrm{adj})- \sum_i T(R_i)\big{)}\;. \label{Casimir 2}
\end{align}
$T(R)$ denote the second Casmir of representation $R$, which is defined as
\begin{align}
\textrm{tr}_R (h_1 h_2) = T(R) \times \textrm{Tr}(h_1 h_2)\;, \; \forall h_1, h_2 \in \textbf{h}\;.
\end{align}
Note that $\textrm{tr}_R$ denote a trace  in a representation $R$  and $\textrm{Tr}$ denote the trace in the defining representation.
Note that the second factor in eq.~\eqref{Casimir 2}, $T(\textrm{adj})- \sum_i T(R_i)$, is nothing but the coefficient of 1-loop $\beta$-function for $\mathcal{N}=2$ theories and it vanishes for superconformal field theories. In  3d  case,  the Casimir energy does not vanish even for  superconformal theories such as ABJM theory.   It would be nice if one can explain the absence of  Casimir energy from the representation theory of 4d superconformal algebra.

\section{Index calculation : Monopole bubbling effect}

\begin{figure}[h!]
  \begin{center}
    \includegraphics[width=8.8cm]{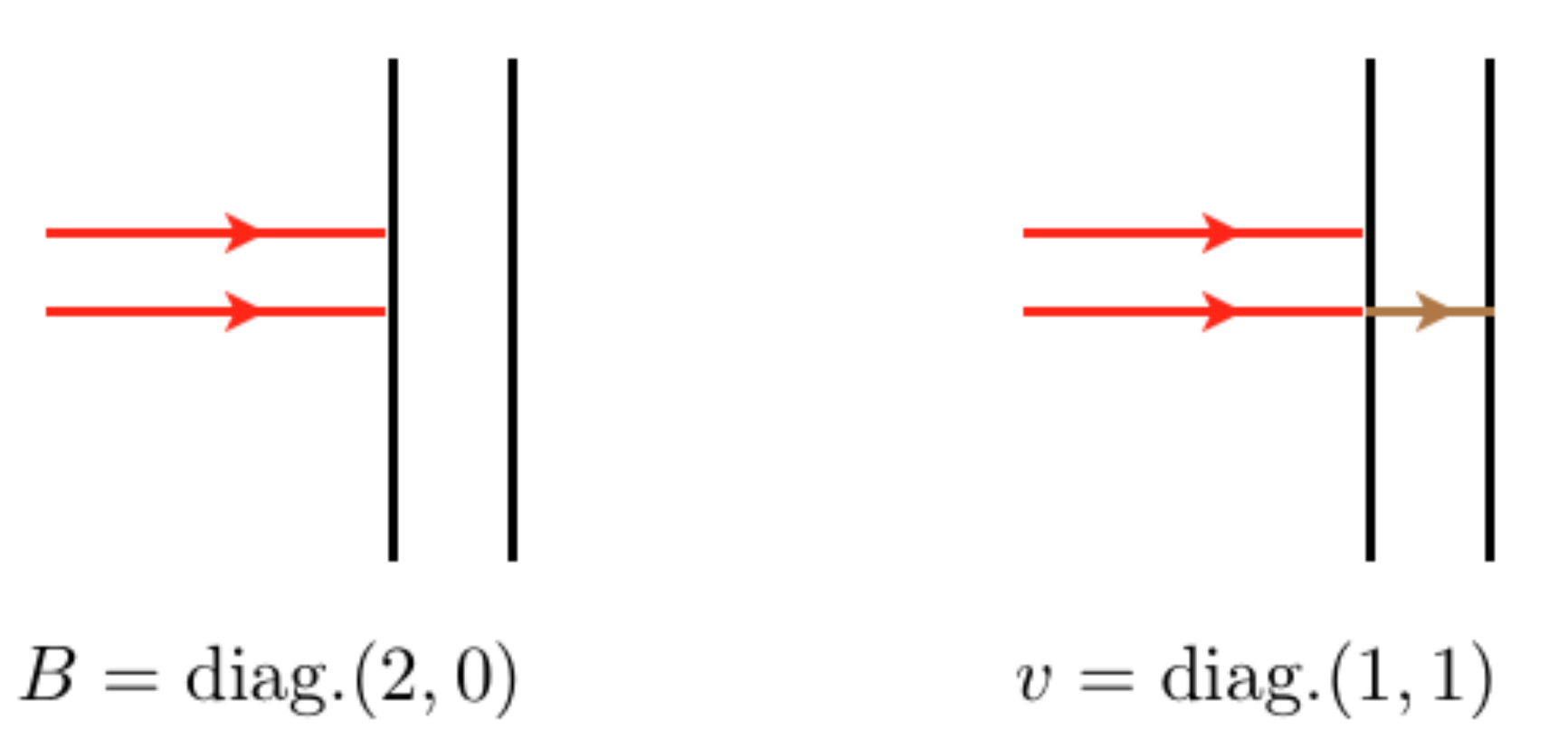}
\caption{A simple brane picture of monopole bubbling. Black lines represent D3-branes (0123) and horizontal lines represent D1-branes (04). Infinitely stretched D1s (red line) ending on the D3s correspond to 't Hooft line operators in the field theory on D3s. A line operator with magnetic charge $B=(2,0)$ (left) can be  screened by a massless  monopole (brown line) and have reduced charge $v=(1,1)$ (right).}\label{monopole bubbling}
  \end{center}
\end{figure}

Our calculation of superconformal index of 't Hooft line operator is not complete since we did not take into account of `monopole bubbling' effect.   Generally, the charge of our singular monopole \eqref{tHooft line operator} can be screened by regular monopoles surrounding the singular monopole. A simple picture of the monopole bubbling is given in Figure~\ref{monopole bubbling}.
The monopole charge is given by the coweight $w=B$. The coweight $B$ can be seen as a weight in the Langland dual group $G^L$. For each $B$, one can assign the representation $\textrm{Rep}(B)$ of $G^L$ whose highest weight is $B$. Then, the possible (screened) asymptotic monopole charges $\{ v \}$ of the singular monopole  are weights in $\textrm{Rep}(B)$, $\{ v\} = \textrm{Rep}(B)$. If all the weights in $\textrm{Rep}(B)$ are related to each other by the action of the Weyl group,  there's no bubbling. In the case, the representation $\textrm{Rep}(B)$ is called the `minuscule representation'. 
\\

\subsection{S-duality check : minuscule representations}

We will consider Wilson line operators in minuscule representations in $\mathcal{N}=4$ SYM theory   and their S-dual 't Hooft line operators. In this case, there is no monopole bubbling effect, thus the 1-loop results \eqref{1-loop for N=4 SYM} are exact. Thus, using the 1-loop results, one can check our index for compatibility with $S$-duality. For  gauge groups $G=U(N), Sp(2N), SO(2N+1)$ and $SO(2N)$, the minuscule representations  of $G$ and their dual magnetic charges of the Langland dual group $G^{L}$ are summarized in table \ref{tab:minuscule}. 
Using the table and formulas in section 3.5,  one can write down index formulas for Wilson line operators in the minuscule representation and that for the corresponding 't Hooft line operators.  In the subsections below, we will explicitly write down these formulas  and check the S-duality. 
%
\begin{table}[htbp]
  \centering
  \begin{tabular}{@{} l|c|c|c|c@{}} 
     \toprule

       \quad \; $G$ & $R$ &  $\chi_R (e^{i \lambda})$& $G^L$ &B \\
     \midrule
     \quad \; $U(N)$ & $A_k$ ($ k \leq N$) & $\sum_{0\leq i_1<\ldots < i_k \leq N} e^{i (\lambda_{i_1} +\ldots +\lambda_{i_k})} $ & $U(N)$ & $(1^k,0^{N-k})$   \\
              \midrule
   \quad $ Sp(2N)$ & $\mathbf{2N}$  & $ \sum_i \sum_\pm  e^{\pm i \lambda_i} $ & $SO(2N+1)$ &$(1,0^{N-1})$  \\
    \midrule
    $SO(2N+1)$& spinor & $\prod_{i=1}^N (e^{\frac{i}2 \lambda_i}+e^{- \frac{i}2 \lambda_i})$ & $ Sp(2N)$ & ($\frac{1}2^N$) \\
    \midrule
     \; \;$SO(2N)$& chiral spinor & $ \frac{1}2\sum_\pm \prod_{i=1}^N (e^{\frac{i}2 \lambda_i}\pm e^{-\frac{i}2 \lambda_i})$ & $SO(2N)$ & ($\frac{1}2^N$) \\
     \cmidrule(r){2-3} \cmidrule(r){5-5}
      & $\mathbf{2N}$ &  $\sum_{i=1}^N \sum_\pm e^{\pm i \lambda_i} $ &  & ($1$,$0^{N-1}$) \\
     \bottomrule
  \end{tabular}
  \caption{Nontrivial minuscule representations and its character of $G=U(N)$, $Sp(2N), SO(2N+1),SO(2N)$ and their corresponding magnetic charges $B$ in $G^L$. $A_k$ denotes the $k$-th totally anti-symmetric representation of $U(N)$.}
  \label{tab:minuscule}.
\end{table}

\subsubsection{$U(N)$ SYM}
The index for the Wilson line operator in the $k$-th antisymmetric representation of $U(N)$  is
\begin{align}
I^{U(N)}_{R=A_k} (x, \eta) &= \frac{1}{N!}\int \prod_{i=1}^N (\frac{d\lambda_i}{2\pi}) \big{(} \prod_{i \neq j}(1-e^{i(\lambda_i - \lambda_j)}) \big{)} \textrm{P.E} \big{[} \frac{(\eta+\eta^{-1})x -2x^2}{1-x^2} \sum_{i,j=1}^N e^{i (\lambda_i - \lambda_j)}  \big{]}
\nn
\\
&\quad \times \prod_{\pm}\big{(} \sum_{1\leq  i_1 < i_2 \ldots < i_k \leq N} e^{\pm i (\lambda_{i_1} +\lambda_{i_2}+\ldots+ \lambda_{i_k})}  \big{)}   \;. \label{Wilson in A_k}
\end{align}
For the corresponding 't Hooft line operator, the index is
\begin{align}
&I^{U(N)}_{B=(1^{k},0^{N-k})} (x, \eta) = \frac{1}{k!(N-k)!} \int \prod_{i=1}^N(\frac{d\lambda_i}{2\pi} ) \nn
\\
& \times \prod_{(i \neq j)=1}^k(1-e^{i(\lambda_i - \lambda_j)}) \prod_{(i \neq j)=k+1}^N(1-e^{i(\lambda_i - \lambda_j)})\prod_{i=1}^k \prod_{j=k+1}^N \prod_\pm (1- x e^{\pm i(\lambda_i - \lambda_j)})   \nn
\\
&\times \textrm{P.E} \bigg{[} \frac{(\eta+\eta^{-1})x-2x^2}{1-x^2}\bigg{(}  (\sum_{i,j=1}^k+ \sum_{i,j=k+1}^N  )e^{i (\lambda_i - \lambda_j)} +\sum_{i=1}^k \sum_{j=k+1}^N \sum_\pm  e^{\pm i (\lambda_i - \lambda_j )}x  \bigg{)} \bigg{]} \;.
\label{tHooft in A_k}
\end{align}
$k! (N-k)!$ is the order of Weyl group for unbroken gauge group, $U(k)\times U(N-k)$. Although they look different, $S$-duality predicts that they should agree to each other for any positive integers $k,N$ satisfying $k \leq N$. For several simple cases, one can check the prediction by comparing two indices in  $x$ expansion. For example,  
\begin{align}
&  I^{U(2)}_{R = A_1} = I^{U(2)}_{B = (1,0)} \nn
\\
&= 1+2(\eta +\eta^{-1})x +(1+3 \eta^2 +3\eta^{-2})x^2 + 4(\eta^3 +\eta^{-3}) x^3 + (1 + 5 \eta^4 +5 \eta^{-4}) x^4 \nn
\\
& \quad + (6 \eta^{-5}+2 \eta^{-1}+ 2 \eta + 6 \eta^{5})x^5 + (7 \eta^{-6}+\eta^{-2}-1+\eta^2+7 \eta^6)x^6 +\ldots, \nn
\\ \nn
\\
& I^{U(4)}_{R=A_2}=I^{U(4)}_{B = (1,1,0,0)} \nn
\\
&= 1+ 2 (\eta+\eta^{-1}) +(3 + 5 \eta^{-2}+5 \eta^2) x^2 + (8 \eta^{-3}+6 \eta^{-1}+ 6 \eta +8 \eta^3) x^3\nn 
\\
&\quad +(14 \eta^{-4}+7 \eta^{-2}+10 + 7\eta^2 + 14 \eta^8)x^4 +10 (2 \eta^{-5}+ \eta^{-3}+\eta^{-1}+\eta +  \eta^{3} + 2 \eta^{5}) x^5 +\ldots,  \nn\end{align}

\subsubsection{$SO(2N+1)/Sp(2N)$ SYM}
For the Wilson line operator in the vector representation ($\mathbf{2N}$) of $Sp(2N)$ theory, the index is
\begin{align}
& I^{Sp(2N)}_{R=\mathbf{2N}} (x, \eta)  \nn
\\
&= \frac{1}{2^N N!} \int  \big{(} \prod_{i=1}^N\frac{d\lambda_i}{2\pi}  \big{)} \big{(}\prod_{i<j}\prod_{\pm}\prod_{\pm}(1-e^{\pm i (\lambda_i \pm \lambda_j)}) \big{)} \big{(}\prod_{i=1}^N \prod_{\pm} (1-e^{\pm 2 i \lambda_i }) \big{)} \big{(}\sum_{i=1}^N \sum_\pm  e^{\pm i \lambda_i} \big{)}^2  \nn
\\
&\quad \quad \times  \textrm{P.E}\bigg{[}  \frac{(\eta+\eta^{-1})x-2x^2}{1-x^2} \bigg{(}N+\sum_{i<j} \sum_\pm \sum_\pm e^{ \pm i (\lambda_i \pm \lambda_j) } + \sum_{i=1}^N \sum_\pm  e^{\pm 2 i \lambda_i} \bigg{)}\bigg{]} \;. \label{Sp(2N) R=2N}
\end{align}
For the corresponding 't Hooft line operator, $B=(1,0,\dots 0)$ in $SO(2N+1)$ theory, the index is
\begin{align}
&I^{SO(2N+1)}_{B=(1,0^{N-1})} (x, \eta) \nn
\\
&= \frac{1}{2^{N-1} (N-1)!} \int \big{(} \prod_{i=1}^N \frac{d\lambda_i}{2\pi} \big{)}\prod_{i<j}\prod_{\pm}\prod_\pm (1-e^{\pm i(\lambda_i \pm \lambda_j)}x^{\delta_{i,1}}) \big{(}\prod_{i=1}^N \prod_{\pm} (1-e^{\pm  i \lambda_i } x^{\delta_{i,1}}) \big{)} \nn
\\
&\quad \times  \textrm{P.E} \bigg{[}  \frac{(\eta+\eta^{-1})x-2x^2}{1-x^2} \big{(} N+ \sum_{(j>i)=1}^N \sum_{\pm} \sum_\pm e^{\pm i (\lambda_i \pm \lambda_j)} x^{\delta_{i,1}} + \sum_{i=1}^N \sum_\pm e^{\pm i \lambda_i}x^{\delta_{i,1}}  \big{)} \bigg{]} \;. \label{SO(2N+1) B=1}
\end{align}
Again they match as expected from S-duality. For example,
\begin{align}
&I^{Sp(2)}_{R=\mathbf{2}} (x, \eta) =  I^{SO(3)}_{B=(1)} (x, \eta)  \nonumber
\\
&=1+ (\eta^{-1} +\eta) x+ (\eta^{-2} +\eta^{2}) x^2 + (\eta^{-3} - \eta^{-1}-\eta +\eta^3 )x^3 + (\eta^{-4}+1+\eta^4) x^4  \nn
\\
&\quad + (\eta^{-5}+\eta^5) x^5 + (\eta^{-6} -1 +\eta^6) x^6 +(\eta^{-7}-\eta^{-3}-\eta^{-1}-\eta -\eta^3 +\eta^7) x^7 +\ldots, \nn
\\ \nn
\\ 
&I^{Sp(4)}_{R=\mathbf{4}} (x, \eta) =  I^{SO(5)}_{B=(1,0)} (x, \eta)  \nonumber
\\
&=1+ (\eta^{-1} +\eta) x+ (2\eta^{-2} +1+2\eta^{2}) x^2 + (2\eta^{-3} +2 \eta^3)x^3 + (3\eta^{-4}+\eta^{-2}+1+\eta^2+3 \eta^4) x^4  \nn
\\
&\quad + (3 \eta^{-5}- \eta^{-3} -2 \eta^{-1}-2 \eta -\eta^3 +3 \eta^5) x^5 + (4 \eta^{-6}+ \eta^{-4}+3\eta^{-2}+5 +3 \eta^2 + \eta^4 +4 \eta^6 ) x^6  +\ldots \;.
\end{align}
For the Wilson line operator in the spinor representation (dim $2^N$) of $SO(2N+1)$ theory, the index is
\begin{align} 
&I^{SO(2N+1)}_{R=\textrm{spinor}} (x, \eta)  \nn
\\
&= \frac{1}{2^N N!} \int  \big{(} \prod_{i=1}^N\frac{d\lambda_i}{2\pi}  \big{)} \big{(}\prod_{i<j} \prod_{\pm}\prod_{\pm}(1-e^{\pm i (\lambda_i \pm \lambda_j)}) \big{)}   \big{(}\prod_{i=1}^N \prod_{\pm} (1-e^{\pm  i \lambda_i }) \big{)}  \nn
\\
&\times \big{(} \prod_i (e^{\frac{i}2 \lambda_i} +e^{-\frac{i}2 \lambda_i})\big{)}^2\textrm{P.E}\bigg{[}  \frac{(\eta+\eta^{-1})x-2x^2}{1-x^2}  \big{(}N+\sum_{i<j}\sum_\pm \sum_\pm e^{ \pm i (\lambda_i \pm \lambda_j) } + \sum_{i=1}^N \sum_\pm e^{\pm i \lambda_i} \big{)} \bigg{]} \;. \label{SO(2N+1) R=spinor}
\end{align}
For the corresponding 't Hooft line operator, $B=(\frac{1}2,\ldots, \frac{1}2) $, in $Sp(2N)$ theory, the index is
\begin{align}
& I^{Sp(2N)}_{B=(\frac{1}2^N )} (x, \eta) \nn
\\
&= \frac{1}{N!} \int  \big{(} \prod_{i=1}^N\frac{d\lambda_i}{2\pi}  \big{)} \big{(}\prod_{i<j} \prod_{\pm}(1-e^{\pm i (\lambda_i + \lambda_j) } x) \big{)}  \big{(}\prod_{i<j} \prod_{\pm}(1-e^{\pm i (\lambda_i - \lambda_j) } ) \big{)}\big{(}\prod_{i=1}^N \prod_{\pm} (1-e^{\pm 2 i \lambda_i } x) \big{)}   \nn
\\
& \quad \times \textrm{P.E}\bigg{[}  \frac{(\eta+\eta^{-1})x-2x^2}{1-x^2} \bigg{(}N+\sum_{i<j} \sum_\pm  e^{ \pm i (\lambda_i + \lambda_j) } x+\sum_{i<j} \sum_\pm e^{ \pm i (\lambda_i - \lambda_j) } + \sum_{i=1}^N \sum_\pm  e^{\pm 2 i \lambda_i} x \bigg{)}\bigg{]} \;. \label{Sp B=1/2^k}
\end{align}
Note that unbroken gauge group is $U(N)$, thus the symmetric factor is $N!$. Again,
\begin{align}
&I^{SO(7)}_{R=\textrm{spinor}} (x,\eta) = I^{Sp(6)}_{B=(\frac{1}2 ,\frac{1}2, \frac{1}2 )} (x, \eta) \nn
\\
&=1+ (\eta^{-1} +\eta) x+ (1+2\eta^{-2} +2\eta^{2}) x^2 + (3\eta^{-3} + \eta^{-1}+\eta +3\eta^3 )x^3  \nn
\\
&\quad +(4\eta^{-4}+2 \eta^{-2}+3+2\eta^2 + 4\eta^4) x^4   + (5\eta^{-5} + \eta^{-3}+\eta^3+5\eta^5) x^5 + \ldots \;. \nn 
\end{align}
Note that for $N=1,2$, the indices in eq~\eqref{SO(2N+1) R=spinor}, \eqref{Sp B=1/2^k} are identical to indices in  eq~\eqref{Sp(2N) R=2N}, ~\eqref{SO(2N+1) B=1} respectively since $SO(3) = Sp(2)$ and $SO(5) = Sp(4)$ up to discrete groups.

\subsubsection{$SO(2N)$ SYM}
For the Wilson line operator in the vector representation $(\mathbf{2N})$ of $SO(2N)$, the index is
\begin{align}
&I^{SO(2N)}_{R=\textbf{2N}} (x, \eta) = \frac{1}{2^{N-1} N!} \int  \big{(} \prod_{i=1}^N\frac{d\lambda_i}{2\pi}  \big{)} \big{(}\prod_{i<j} \prod_{\pm}\prod_{\pm}(1-e^{\pm i (\lambda_i \pm \lambda_j)}) \big{)}    \nn
\\
&\times \big{(} \sum_{i=1}^N (e^{ i \lambda_i}+e^{-i \lambda_i}) \big{)}^2\textrm{P.E}\bigg{[}  \frac{(\eta+\eta^{-1})x-2x^2}{1-x^2}  \big{(}N+\sum_{i<j}\sum_\pm \sum_\pm e^{ \pm i (\lambda_i \pm \lambda_j) } \big{)} \bigg{]} \;.
\end{align}
For the corresponding 't Hooft line operator, $B=(1,0,\ldots,0)$, the index is
\begin{align}
I^{SO(2N)}_{B=(1,0^{N-1})} (x, \eta) &= \frac{1}{2^{N-2} (N-1)!} \int  \big{(} \prod_{i=1}^N\frac{d\lambda_i}{2\pi}  \big{)} \big{(}\prod_{i<j} \prod_{\pm}\prod_{\pm}(1-e^{\pm i (\lambda_i \pm \lambda_j)} x^{\delta_{i,1}}) \big{)}    \nn
\\
&\quad  \times \textrm{P.E}\bigg{[}  \frac{(\eta+\eta^{-1})x-2x^2}{1-x^2}  \big{(}N+\sum_{i<j}\sum_\pm \sum_\pm e^{ \pm i (\lambda_i \pm \lambda_j) } x^{\delta_{i,1}} \big{)} \bigg{]} \;.
\end{align}
Again exact matches can be checked in $x$ expansion. For example,
\begin{align}
&I^{SO(4)}_{R = \textbf{4}} (x, \eta) = I^{SO(4)}_{B = (1,0)} (x, \eta) \nn
\\
&= 1+ 2(\eta^{-1}+\eta) x + (3\eta^{-2}+2 +3\eta^2) x^2 + 4(\eta^{-3}+\eta^3)x^3 +5(\eta^{-4}+\eta^4)x^4  \nn
\\
&\quad + 2(3 \eta^{-5}+\eta^{-1}+\eta +3\eta^5)x^5 +(7 \eta^{-6}+3\eta^{-2} +2 +3\eta^2 +7 \eta^6)x^6+\ldots, \nn
\\ \nn
\\
&I^{SO(6)}_{R = \textbf{6}} (x, \eta) = I^{SO(4)}_{B = (1,0,0)} (x, \eta) \nn
\\
&= 1+ (\eta^{-1}+\eta) x + (3\eta^{-2}+2 +3\eta^2) x^2 + (3\eta^{-3}+\eta^{-1}+\eta+3\eta^3)x^3   \nn
\\
&\quad + (6 \eta^{-4}+3\eta^{-2}+4+3\eta^2 +6\eta^4)x^4 +2 (3 \eta^{-5}-\eta^{-1} -\eta +3\eta^5 )x^5+\ldots \;.
\end{align}
For Wilson line operators in the chiral spinor representation (of dimension $2^{N-1}$) of $SO(2N)$, the index is
\begin{align}
& I^{SO(2N)}_{R=\textrm{chiral spinor}} (x, \eta) = \frac{1}{2^{N-1} N!} \int  \big{(} \prod_{i=1}^N\frac{d\lambda_i}{2\pi}  \big{)}  \big{(}\prod_{i<j} \prod_{\pm}\prod_{\pm}(1-e^{\pm i (\lambda_i \pm \lambda_j)} ) \big{)}\nn
\\
& \qquad \times   \prod_\pm \big{(} \sum_{\vec{\epsilon} = (\pm 1 , \pm  1,\ldots , \pm 1) } \delta_{\textrm{sgn} (\vec{\epsilon}) ,1} e^{ \pm \frac{i}2 ( \epsilon_1 \lambda_1 + \ldots  +\epsilon_N \lambda_N  ) } \big{)}  \nn
\\
& \qquad  \times  \textrm{P.E}\bigg{[}  \frac{(\eta+\eta^{-1})x-2x^2}{1-x^2}  \big{(}N+\sum_{i<j}\sum_\pm \sum_\pm e^{ \pm i (\lambda_i \pm \lambda_j) } \big{)} \bigg{]} \;, \nn
\\
&\textrm{where }\textrm{sgn}(\vec{\epsilon}) : = \prod_{i=1}^N \epsilon_i \;.
\end{align}
For the corresponding 't Hooft line operator, $B=(\frac{1}2,\ldots, \frac{1}2)$, the index is
\begin{align}
& I^{SO(2N)}_{B=(\frac{1}2^N)} (x, \eta) \nn
\\
&= \frac{1}{N!} \int  \big{(} \prod_{i=1}^N\frac{d\lambda_i}{2\pi}  \big{)} \big{(}\prod_{i<j} \prod_{\pm}(1-e^{\pm i (\lambda_i + \lambda_j) } x) \big{)}  \big{(}\prod_{i<j} \prod_{\pm}(1-e^{\pm i (\lambda_i - \lambda_j) } ) \big{)}  \nn
\\
& \quad \times \textrm{P.E}\bigg{[}  \frac{(\eta+\eta^{-1})x-2x^2}{1-x^2} \bigg{(}N+\sum_{i<j} \sum_\pm  e^{ \pm i (\lambda_i + \lambda_j) } x+\sum_{i<j} \sum_\pm e^{ \pm i (\lambda_i - \lambda_j) }  \bigg{)}\bigg{]} \;.
\end{align}
Again they match. For example,
\begin{align}
&I^{SO(4)}_{R=\textrm{chiral spinor}} (x,\eta) = I^{SO(4)}_{B=(\frac{1}2, \frac{1}2 )} (x, \eta) \nn
\\
&= 1+ (\eta^{-1}+\eta) x +(2\eta^{-2}+1 +2\eta^2) x^2 + (2 \eta^{-3}-\eta^{-1}-\eta + 2\eta^3) x^3 \nn
\\
&\quad + (3 \eta^{-4}+\eta^{-2}+2 + \eta^2 +3 \eta^4) x^4 + (3\eta^{-5}-\eta^{-3}-2\eta^{-1}-2 \eta -\eta^3+ 3 \eta^5)x^5 +\ldots \;,\nn
\\ \nn
\\ \nn
&I^{SO(6)}_{R=\textrm{chiral spinor}} (x,\eta) = I^{SO(6)}_{B=(\frac{1}2, \frac{1}2,\frac{1}2 )} (x, \eta) \nn
\\
&= 1+ (\eta^{-1}+\eta)x + (2\eta^{-2}+1 +2\eta^2) x^2 + (3 \eta^{-3}+\eta^{-1}+\eta + 3\eta^3) x^3 \nn
\\
&+(4 \eta^{-4}+ \eta^{-2}+ 2 + \eta^2 +4 \eta^4) x^4 + (5\eta^{-5}+\eta^{-3}+\eta^3 +5 \eta^5) x^5 +\ldots \;.
\end{align}
It would be interesting to find  underlying mathematical identities in the indices agreement.

\subsection{Final expression for the index : 1-loop + monopole bubbling}
Taking into account of monopole bubbling effect, the final index formula can be written as
\begin{align}
I_B (x,\eta_a) = \sum_{v \in \textrm{Rep}( B)} \int [dU]_v Z^{S^3}_{mono} (B,v;e^{i \lambda_i}, x,\eta_a) Z^{1-loop}_v ( e^{i \lambda_i},x, \eta_a)\;. \label{final index formula}
\end{align}
$Z^{1-loop}_v$ is the 1-loop results calculated in the previous section, summarized in eq~\eqref{N=2 index}. The remaining non-trivial problem is to determine the monopole bubbling effect denoted by $Z_{mono} (B,v)$.
Since monopole bubbling  happens near two poles of $S^3$, the `monopole bubbling' index $Z^{S^3}_{mono}(B,v)$ can be factorized as follow
\begin{align}
Z^{S^3}_{mono} (B,v) = Z^S_{mono} (B,v) Z^N_{mono} (B,v) \;, \; \textrm{with } Z^S_{mono} (B,v)=  Z^N_{mono} (B,v)\;. \label{mono bubbling : S3 and N,S}
\end{align}
Here $N$ and $S$ represent the north and south pole respectively. Furthermore, since the monopole bubbling happens at the small region around each pole, which is locally $\mathbb{R}^3$,  one may guess that
\begin{align}
Z^{(N,S)}_{mono} (B;v) =Z^{\mathbb{R}^3 }_{mono} (B,v) \;. \label{mono bubbling : S3 and R3}
\end{align}
Here $Z^{\mathbb{R}^3 }_{mono}$ denotes the index contribution from monopole bubbling effect in theories defined on $\mathbb{R}^3$ (times the thermal circle $S^1$). Actually this quantity is calculated in the recent paper \cite{Ito:2011ea}. Using their results (with a proper identification of variables appearing in the index formula), one can obtain $Z_{mono} (B;v)$ just by taking the square of their results. The quantity calculated in the paper is\footnote{We replace a chemical potential $\lambda$ in their paper with $\rho$. $\lambda$ is used for holonomy variable in our paper.}
\begin{align}
I_L^{\mathbb{R}^3 }(a,b,\rho, m_i) = \textrm{Tr}_{\mathcal{H}^{\mathbb{R}^3}_L} (-1)^F e^{2 \pi i \rho (j_L + j_R + \frac{1}2 r_1 )} \;.
\end{align}
The trace is taken over Hilbert space $\mathcal{H}^{\mathbb{R}^3}_L$ on $\mathbb{R}^3$ in the presence of the line operator $L$. The index  depends on several parameters ($a,b,m_i$) and chemical potential $\rho$.  See  \cite{Ito:2011ea} for the meaning of these parameters. As in our $S^3$ case, the index gets contributions from  1-loop effects and non-perturbative monopole bubbling effect. Schematically, 
\begin{align}
I^{\mathbb{R}^3}_L = \sum_{v \in \textrm{Rep}(B)} e^{-S_0 (v)} Z^{\mathbb{R}^3}_{mono}(B,v) Z^{\mathbb{R}^3}_{1-loop}(v) \; ,
\end{align}
where $S_0$ denotes the classical Euclidean action. Due to the difference of  base manifolds ($\mathbb{R}^3 $ and $S^3$), the 1-loop results in their paper are totally different from ours. However, as mentioned above, the index from monopole bubbling effects on $\mathbb{R}^3$ can be related to that on $S^3$ as in eq.~\eqref{mono bubbling : S3 and R3}. To translate their results on monopole bubbling into our language, we keep only parameters $(\textrm{Re} [a] $, $m_i,\rho)$ and set other parameters ($\textrm{Im}[a],b$) to be zero. Then, the real parameter $a$ is   the asymptotic value of $A_0$ with a factor $(a =  \frac{\beta}{2\pi} A^{(\infty)}_0 $)  and can be identified with the holonomy variable $\lambda = \beta A_0$, which appears as an integral variable in  our index formula. 
\begin{align}
\lambda_i = 2 \pi a_i \;.
\end{align}
The mass parameters $m_i$ for each hypermultiplet $H_i$ can be identified with the chemical potentials $\eta_i$ for the $U(1)$ symmetries rotating phase of $H_i$. 
\begin{align}
\eta_i = \exp (2 \pi i m_i)\;.
\end{align}
This is because the introduction of chemical potentials $\eta_i$ induces the mass term for hypermultiplets $H_i$, see eq~\eqref{twisting by chemicals}.\footnote{Introducing $\eta$ induces a mass term with mass $M_i = \frac{\ln \eta_i}\beta$. In \cite{Ito:2011ea}, they use the convention $m_i   = i R M_i $, where $R$ denotes the radius of the thermal circle ($\beta =2\pi R$). } Using the BPS bound ($\epsilon = j_L + j_R + r_1$), our index can be rewritten as 
\begin{align}
\textrm{Tr} (-1)^F x^{\epsilon+ j_L + j_R} = \textrm{Tr} (-1)^F x^{2 (j_L+j_R) + r_1} \;.
\end{align}
Thus, we can relate their chemical potential $\rho$ with our $x$ in the following way,
\begin{align}
x = e^{\pi i \rho} .
\end{align}
In sum,  $Z^{(N,S)}_{mono}(B,v;e^{i \lambda_i},x,\eta_i )$ can be obtained from $Z^{\mathbb{R}^3 }_{mono} (B,v;a_i, \rho,m_i)$ by 
\begin{align}
&Z^{N,S}_{mono}(B,v;  e^{i \lambda_i}, x,\eta_i)  = Z^{\mathbb{R}^3 }_{mono} (B,v; a_i, \rho,m_i )|_{a_i   \rightarrow \frac{\lambda_i}{2\pi} ,\ \rho \rightarrow \frac{\ln x}{\pi i}, \ m_i \rightarrow \frac{\ln \eta_i}{2 \pi i} }\;. \label{dictionaary}
\end{align}
\subsubsection{Review on monopole bubbling index on $\mathbb{R}^3$}

\begin{figure}[h!]
  \centering
  \includegraphics[width=.5\textwidth]{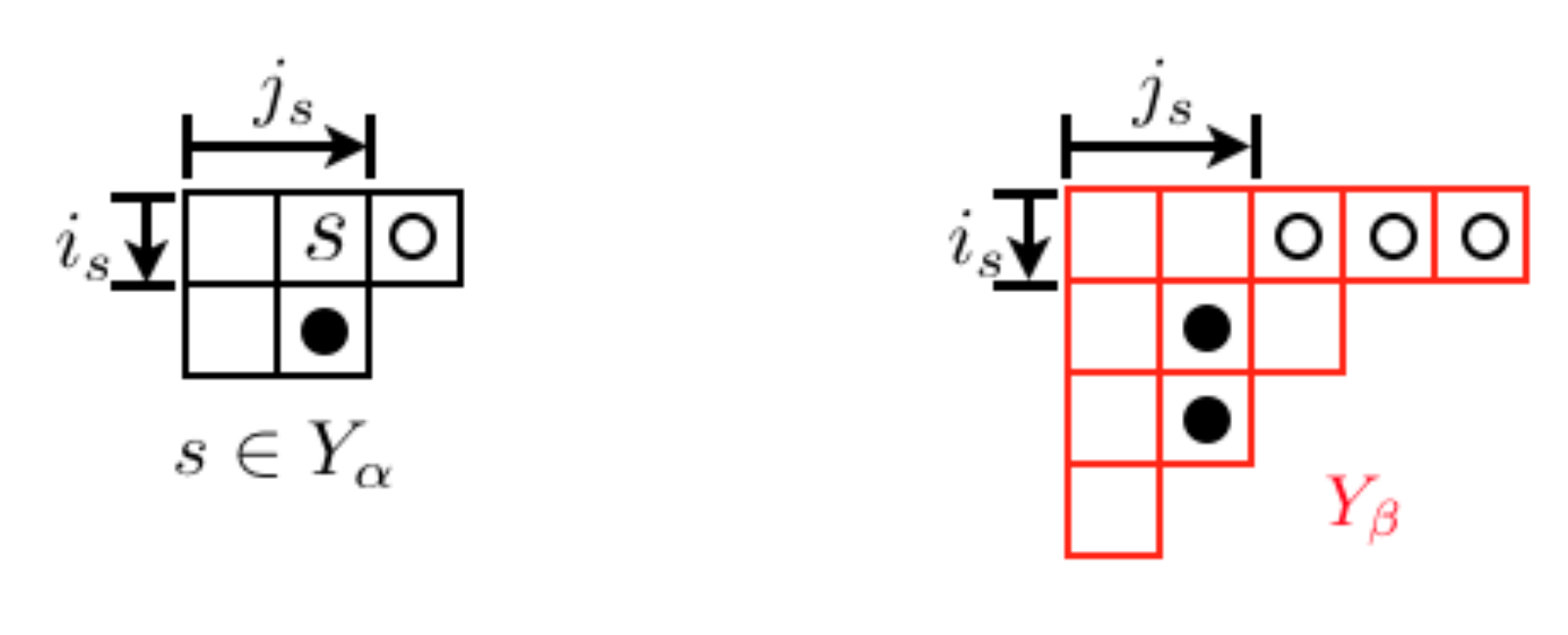} 
  \caption{$(i_s, j_s)$ denote the location of $s$ in $Y_{\alpha}$. In this example, $i_s=1, j_s=2$, and the arm-and leg-length are denoted by the white and black disk, $A_{Y_{\alpha}}(s)=1, L_{Y_{ \alpha}}(s)=1$ (left). The arm- and leg-length can be defined for $Y_{ \beta}$ such that  $s \notin Y_{ \beta}$. 
 In this example,  the arm-length $A_{Y_{ \beta}} (s)= 5-2=3$, and the leg-length $L_{Y_{ \beta}}(s)= 3-1=2$ (right).}
  \label{fig:young}
\end{figure}

Here we will briefly summarize the  formula for the monopole bubbling index $Z^{\mathbb{R}^3 }_{mono} (B,v)$  obtained in  \cite{Ito:2011ea}. For details of the derivation, see section 5 in the paper. They consider   $\mathcal{N}=2$   $U(N)$  gauge theories with   hypermultiplets $H_i$ in representations $R_i$ (the fundamental or adjoint) of the gauge group.   $Z^{\mathbb{R}^3}_{mono}$  can be written as
\begin{align}
Z^{\mathbb{R}^3 }_{mono} (B,v; a, \rho, m_i ) = \sum_{\vec{Y}} z^{\textrm{vec}}_{\vec{Y}} (B,v; a, \rho) \prod_i \prod_{ R_i} z^{R_i}_{\vec{Y}} (B,v;a, \rho, m_i) \;. 
\label{Bubbling index in Okuda}
\end{align}
To understand the formula, first introduce a $k \times k$ diagonal matrix $K = \textrm{diag} (K_1 , K_2,\ldots, K_k)$, which is determined by the following condition
\begin{align}
\textrm{Tr} e^{2 \pi i B \nu} = \textrm{Tr} e^{2\pi i v \nu} + (e^{2\pi i \nu} + e^{-2\pi i \nu}-2)\textrm{Tr} e^{2\pi i K \nu}\;, \quad\forall \nu \;. \label{condition for K}
\end{align}
We will consider $N$ Young diagrams $\{Y_\alpha\}_{\alpha=1}^N$ with $k_\alpha$ boxes in the $\alpha$-th diagram, where $k_\alpha$'s satisfy
\begin{align}
k_1 + k_2+ \ldots + k_N = k\;.
\end{align}
The collection of $N$ Young diagrams are collectively denoted by $\vec{Y}$ and called $N$-colored Young diagram. Only colored Young diagrams satisfying the following condition are summed in the formula  eq.~\eqref{Bubbling index in Okuda},
\begin{align}
K_s = v_{\alpha (s)} + j_{s} - i_{s} \ , \label{monopole bubbling : range 1}
\end{align}
up to a permutation $s \in \{ 1, \ldots , k\}$. There are $k$ boxes in $\vec{Y}$ and we can label each box by an integer $s$, $1\leq s \leq k$. $\alpha(s)$ (between 1 and $N$)  denote the position of Young diagram in $\vec{Y}$ where the $s$-th box is belonging to, $s\in Y_{\alpha(s)}$. $(i_{s}, j_{s})$ denote the location of the $s$-th box  in the $i_s$-th row and the $j_s$-th column of Young diagram $Y_{\alpha(s)}$(see Figure~\ref{fig:young}). $v_\alpha$ denote the $\alpha$-th diagonal elements in the screened monopole charge $v$, $v = \textrm{diag}(v_1, v_2, \ldots,  v_N)$. 
To define functions $z^{vec}_{\vec{Y}}$ and $z^R_{\vec{Y}}$ in \eqref{Bubbling index in Okuda}, we introduce arm- and leg-lengths
\begin{align}
A_{Y_\alpha} (s) = n_{i_s}(Y_\alpha) - j_{s}\;,  \;L_{Y_\alpha} (s) = n^T_{j_s}(Y_\alpha) - i_{s}\;,
\end{align}
where $n_{i}(Y)$ and $n_{i}^T (Y)$ denote the numbers of boxes in the $i$-th row and columm of $Y$ respectively (see Figure~\ref{fig:young}). From vector multiplets and hypermultiplets in the adjoint representation,
\begin{align}
&z^{vec}_{\vec{Y}} = \prod_{(\alpha, \beta, s \in Y_\alpha)} \prod_\pm  \left( \sin \big{[}\pi \big{(}a_\alpha - a_\beta + \frac{1}2 (A_{Y_\alpha}(s)-L_{Y_\beta}(s) \pm 1)\rho \big{)}] \right)^{-1} \;, \label{Bubbling index in Okuda2}
\\
&z^{R=\textrm{adj}}_{\vec{Y}} =  \prod_{(\alpha, \beta, s \in Y_\alpha)} \prod_\pm \sin \big{[}\pi \big{(}a_\alpha - a_\beta + \frac{1}2 (A_{Y_\alpha }(s)-L_{Y_\beta} (s))\rho \pm m\big{)}] \;. \label{Bubbling index in Okuda3}
\end{align}
In both cases, the product is over triples $(\alpha, \beta, s \in Y_\alpha)$ satisfying
\begin{align}
v_\alpha - v_\beta + L_{Y_\beta}(s) +A_{Y_\alpha} (s)+1 =0\;. \label{monopole bubbling : range 2}
\end{align}
From a hypermultiplet in the fundamental representation, one gets
\begin{align}
z^{R=\textrm{fund}}_{\vec{Y}} =  \prod_{(\alpha, s \in Y_\alpha)}  \sin \big{[}\pi \big{(}a_\alpha -m + \frac{1}2 (i_s + j_s -1)\rho \big{)}] \;, \label{Bubbling index in Okuda4}
\end{align}
%
where the product is over the pairs $(\alpha, s \in Y_\alpha)$ satisfying 
\begin{align}
v_\alpha - i_s + j_s =0\;. \label{monopole bubbling : range 3}
\end{align}
The monopole bubbling index formula can be summarized by eq. \eqref{Bubbling index in Okuda},\eqref{Bubbling index in Okuda2},\eqref{Bubbling index in Okuda3}, \eqref{Bubbling index in Okuda4} where the summation or product is over  variables satisfying  eq.~\eqref{monopole bubbling : range 1},\eqref{monopole bubbling : range 2},\eqref{monopole bubbling : range 3}. The formula is rather complicated and it seems difficult to obtain a closed form of monopole bubbling index in full generality.

\subsection{S-duality check : non-minuscule representations}
\subsubsection{$\mathcal{N}=4$ $SU(2)$ theory }
As an simple example,  consider $G=U(2)$  $\mathcal{N}=4$  theory with $B=(2,0)$ and $v=(1,1)$ (see Figure~\ref{monopole bubbling}). From eq.~\eqref{condition for K}, $K$ is determined as
\begin{align}
K = \textrm{diag} (1)\;,
\end{align}
and thus $k=1$ (number of total boxes in Young diagram). Colored Young diagrams $\vec{Y}$ satisfying the condition \eqref{monopole bubbling : range 1} are 
\begin{align}
(Y_1  , Y_2 ) = (\;\Box \;, \;\cdot\;) \;, (\;\cdot \;, \;\Box\;) \;.
\end{align}
From these two colored Young diagrams, one gets (using formulas  in the previous section) 
\begin{align}
Z^{\mathbb{R}^3;U(2)}_{mono} (B,v)  =  \sum_{s =\pm 1} \frac{\Pi_\pm \sin [\pi (a_1 - a_2 \pm m + s \rho/2)]}{\sin [\pi (a_1  - a_2) ] \sin [ \pi (a_1 - a_2 + s  \rho) ]} \;, \; B=(2,0)\;, v=(1,1)\;. \label{Z^{R3}_{mono}(B=(2,0),v=(0,0))}
\end{align}
One can traslate this into monopole bubbling index on $S^3$ using eq.~\eqref{dictionaary} and \eqref{mono bubbling : S3 and N,S},
\begin{align}
Z^{S^3;U(2)}_{mono} (B,v)= \big{[}\frac{1-2x^2+x^4+  (\eta^{-1}+ \eta) (x+x^3) - 2x^2 (e^{-i (\lambda_1- \lambda_2)} +  e^{ i (\lambda_1 - \lambda_2)}) }{(1-e^{-i (\lambda_1 - \lambda_2)} x^2)(1-e^{ i (\lambda_1 - \lambda_2 )} x^2)} \big{]}^2\;, \label{Z_{mono}(B=(2,0),v=(0,0))}
\end{align}
for $B=(2,0)$ and $v=(1,1)$. Thus using \eqref{final index formula}, the index in the presence of  't Hooft line operator with $B=(2,0)$ can be written as
\begin{align}
I^{U(2)}_{B=(2,0)} (x, \eta)  
&=  \int_{0}^{2\pi}  \frac{d\lambda_1 d\lambda_2}{(2\pi)^2} (1-e^{i (\lambda_1 - \lambda_2)}x^2)(1-e^{-i (\lambda_1 - \lambda_2) }x^2)   \nn
\\
& \quad \quad \times \textrm{P.E}\big{[}\frac{ (\eta + \eta^{-1})x -2x^2}{1-x^2}(2+ e^{i (\lambda_1 -\lambda_2) }x^2+e^{-i (\lambda_1 -\lambda_2) }x^2)\big{]} \nn
\\
&\quad  + \frac{1}2\int_{0}^{2\pi}  \frac{d\lambda_1 d\lambda_2}{(2\pi)^2} Z_{mono}^{S^3;U(2)} (B,v)  (1-e^{i (\lambda_1 - \lambda_2)})(1-e^{-i (\lambda_1 - \lambda_2) }) \nn
\\
&\quad \quad \times \textrm{P.E}\big{[}\frac{ (\eta + \eta^{-1})x -2x^2}{1-x^2}(2+ e^{i (\lambda_1 -\lambda_2) }+e^{-i (\lambda_1 -\lambda_2) })\big{]} \;.
\end{align}
Here $Z^{S^3;U(2)}_{mono}(B,v)$ is given in eq~\eqref{Z_{mono}(B=(2,0),v=(0,0))}. On the other hand, for  the Wilson line operator in the tensor product of fundamental representations, the index is given by 
\begin{align}
I^{U(2)}_{R=(1,0)^2} & = \frac{1}2\int_{0}^{2\pi}  \frac{d\lambda_1 d\lambda_2}{(2\pi)^2}   (1-e^{i (\lambda_1 - \lambda_2)})(1-e^{-i (\lambda_1 - \lambda_2) }) (e^{i \lambda_1}+e^{i \lambda_2})^2 (e^{-i \lambda_1}+e^{-i \lambda_2})^2 \nn
\\
&\quad \quad \times \textrm{P.E}\big{[}\frac{ (\eta + \eta^{-1})x -2x^2}{1-x^2}(2+ e^{i (\lambda_1 -\lambda_2) }+e^{-i (\lambda_1 -\lambda_2) })\big{]} \;.
\end{align}
 Although two indices look  different, it can be shown that they are perturbatively same. Listing a few lowest orders in $x$, 
 \begin{align}
& I^{U(2)}_{B=(2,0)} ( x, \eta) = I^{U(2)}_{R=(1,0)^2}  ( x, \eta ) \nn
\\
&= 2 +5(\eta +\eta^{-1})x +(4+8 \eta^2 +8\eta^{-2})x^2 + (11\eta^3 +\eta^{-1}+\eta +11\eta^{-3}) x^3 \nn 
\\
&\quad + (4 + 14 \eta^{-4} +14 \eta^{4}) x^4 + (17 \eta^{-5} + 6 \eta^{-1}+ 6 \eta + 17 \eta^5)x^5 +\ldots  \end{align}
It is interesting to see how 't Hooft operators with non-minimal magnetic charge in $U(2)$ gauge group get translated under S-duality. The above result tells us that the quantum state of the 't Hooft operator with multiple charges  at the north pole and  its anti-object at the south pole would not be irreducible under the magnetic dual group. Rather than that, it is a sum of the irreducible ones which one could obtain by expanding the contribution for the corresponding Wilson line in irreducible representations.

For $SU(N)$ $\mathcal{N}=4$ SYM, the monopole bubbling index can be obtained from the index in $U(N)$  theory  simply by imposing traceless conditions on  $(a , B, v)$. Consider $SU(2)$ $\mathcal{N}=4$ SYM theory. Monopole charge $B$ in the theory can be labeled by a positive integer $p$,\footnote{In this case, half-integer entries in monopole charge $B$ do not violate the Dirac quantization conditions since all matters are in the adjoint representation and $\alpha(B) \in \mathbb{Z}$ with roots $\alpha$, weights of the adjoint. In the presence of fundamental matters, only even  $p$ is allowed since $\rho_{\textrm{fund}}(B) = \pm \frac{p}2$. }
\begin{align}
B = \frac{1}2 \textrm{diag} (p, -p)\;.
\end{align}
For $p>1$, the charge can be screened by monopole bubbling and the index can be written as $(Z^{\mathbb{R}^3}_{mono}(p,p)=1)$
\begin{align}
I^{SU(2)}_p &= \sum_{v=1,3,\ldots, p} \int [dU]_v [Z^{\mathbb{R}^3,SU(2)}_{mono}(p,v)]^2 Z^{1-loop}_{v} \;, \quad \textrm{odd $p$} \;, \nonumber
\\
&= \sum_{v=0,2,\ldots, p} \int [dU]_v  [Z^{\mathbb{R}^3,SU(2)}_{mono}(p,v)]^2 Z^{1-loop}_{v}\;,  \quad \textrm{even $p$}\;.
\end{align}
For $p=2$, the monopole bubbling index can be obtained from eq.~\eqref{Z^{R3}_{mono}(B=(2,0),v=(0,0))} by replacing $(a_1 , a_2)$  with $(a, -a)$.
\begin{align}
Z^{\mathbb{R}^3 ;SU(2) }_{mono} (2,0)  =  \sum_{s =\pm 1} \frac{\Pi_\pm \sin \pi (2a \pm m + s \rho/2)}{\sin (2 \pi a ) \sin (2 \pi a + s\pi  \rho) }.
\end{align} 
Using the dictionary \eqref{dictionaary}, it becomes
\begin{align}
Z^{N;SU(2)}_{mono} (2,0)= \frac{(1-2x^2+ x^4)+(\eta^{-1}+ \eta)(x+x^3) - 2x^2 (e^{-2i \lambda} +  e^{2 i \lambda}) }{(1-e^{-2 i \lambda} x^2)(1-e^{2 i \lambda} x^2)} \;. \label{Z20}
\end{align}
Following a similar procedure, one can obtain
\begin{align}
Z^{N;SU(2)}_{mono} (3, 1) = \frac{ 2 (1-x^2-x^4+ x^6)+ ( \eta^{-1}+ \eta) (x+x^3+ x^5)   - 3 x^3   (e^{ - 2 i \lambda}+e^{ 2 i \lambda}) }
{(1-e^{ - 2 i \lambda} x^3)( 1- e^{ 2 i \lambda} x^3)} . 
\label{Z31}
\end{align} 
Using these bubbling indices, one can calculate the indices of 't Hooft line operators with charge $p=2$ and $p=3$. In both cases, we can  check the exact agreements with the corresponding Wilson line operator indices, 
\begin{align}
&I^{SU(2)}_{B=  \textrm{diag}(1,-1)} (x, \eta) = I^{SU(2)}_{R=(1,0)^2} (x, \eta)\; \nn
\\
&= 2+ 3(\eta^{-1} +\eta) x + 3(\eta^{-2}+\eta^2) x^2 + (3 \eta^{-3}-2 \eta^{-1}-2\eta +3 \eta^3) x^3   \nn
\\
& \quad + (3 \eta^{-4}-\eta^{-2}+2 -\eta^{2}+3 \eta^4)x^4 +(3 \eta^{-5}+\eta^{-1}+\eta + 3\eta^5)x^5 +\ldots, \nn
\\
&I^{SU(2)}_{B=  \textrm{diag}(\frac{3}2,- \frac{3}2 )} (x, \eta) = I^{SU(2)}_{R=(1,0)^3} (x, \eta) \nn
\\
&=5+ 9(\eta^{-1}+\eta) x+(10 \eta^{-2}+1 +10\eta^2)x^2 +(10 \eta^{-3}-5\eta^{-1}- 5\eta +10 \eta^3)x^3 \nn
\\
& \quad +(10 \eta^{-4}-4\eta^{-2}+6-4\eta^{2}+10\eta^4) x^4 + (10\eta^{-5} -\eta^{-3} +3\eta^{-1} +3 \eta-\eta^3 +10 \eta^5)x^5\ldots \;. \nn
\end{align}
For the completeness, we  include the indices for  $B=(0,0)$ and $(\frac{1}2, \frac{1}2)$ 
where monopole bubbling effect is absent, 
\begin{align}
&I^{SU(2)}_{B=  \textrm{diag}(0,0)} (x, \eta)=I^{SU(2)}_{R= \textrm{trivial}} (x, \eta)
\nn
\\
&= 1+ (\eta^{-2}+1+\eta^2 ) x^2 -2 (\eta^{-1}+\eta) x^3 + (\eta^{-4} +2 \eta^{-2}+3 +2 \eta^2 + \eta^4) x^4  \nn
\\
&\quad - 2(\eta^{-3} +2 \eta^{-1}+ 2 \eta + \eta^3) x^5 + (\eta^{-6}+2 \eta^{-4}+5 \eta^{-2}+6+ 5 \eta^2 +2 \eta^4 + \eta^6) x^{6}+\ldots\; ,  \nn
\\
&I^{SU(2)}_{B=  \textrm{diag}(\frac{1}2,- \frac{1}2)} (x, \eta)=I^{SU(2)}_{R= \mathbf{2}} (x, \eta)
\nn
\\
&= 1+ (\eta^{-1}+\eta ) x + (\eta^{-2}+\eta^2) x^2 + (\eta^{-3} -\eta^{-1}-\eta + \eta^3 ) x^3 +(\eta^{-4} +1+ \eta^{4}) x^4 \nn
\\
&\quad   + (\eta^{-5}+ \eta^5) x^{5} +(\eta^{-6}-1 +\eta^6) x^6 +(\eta^{-7}-\eta^{-3}-\eta^{-1}-\eta - \eta^3 +\eta^7)x^7 +\ldots\; .  \nn
\end{align}
\\
\subsubsection{$\mathcal{N}=2$ $SU(2)$ theory with four flavors} \label{sec:4flavors}

Some ${\cal N}=2$ super conformal field theories have S-duality. One simple example is $\mathcal{N}=2$ $SU(2)$ theory with four fundamental hypermultiplets. In the case the minimal magnetic charge is $B=\textrm{diag}(1,-1)$ and it is not minuscule.  Therefore, we need the monopole bubbling index $Z_{mono}^{SU(2),N_f=4}(2,0)$ to calculate the correct index for the 't Hooft line operator. Unlike $U(N)$ cases,  a field theoretic algorithm for calculating a monopole bubbling index  is not yet developed  for general $\mathcal{N}=2$ $SU(N)$ theories.  Neverthless, a monopole bubbling index on $\mathbb{R}^3$ for the $SU(2)$ theory can be obtained using a 2d/4d correspondence \cite{Ito:2011ea}. For example,\footnote{It seems that an overall factor 4 is missing in the eq 8.27 in \cite{Ito:2011ea}.}
\begin{align}
Z^{\mathbb{R}^3;SU(2),N_f=4}_{mono}(2,0)  = -2 \cos \pi (\rho  - \sum_{i=1}^4 m_i) - 4\sum_{s=\pm 1} \frac{\prod_{i=1}^4 \sin \pi (sa - m_i + \frac{\rho}2)}{\sin (2\pi a) \sin \pi (s\rho +2a)}\;.
\end{align}
Translating this into the  bubbling index  on the north pole of $S^3$ using  \eqref{dictionaary}, one gets  ($i=1,2,3,4$.)
\begin{align}
Z^{N;SU(2),N_f =4}_{mono}(2,0) = - \frac{(x^2 + \prod_i \eta_i ) 
}{x \prod \eta_i^{1/2}} + \sum_{s=\pm 1}\frac{\prod_{i=1}{(x e^{ i s \lambda}-\eta_i})}{x(1-e^{2i s\lambda})(1-x^2 e^{2i s \lambda})\prod_i \eta_i^{1/2}}\;,
\end{align}
Using eq.~\eqref{final index formula} and \eqref{mono bubbling : S3 and N,S},  the index of 't Hooft line with $B=\textrm{diag}(1,-1)$ given by
\begin{align}
&I^{SU(2);N_f=4}_{B=(1,-1)} (x, \eta_i)  =\int  \frac{d\lambda}{2\pi} (1-e^{2i \lambda}x^2)(1-e^{-2i \lambda}x^2)  \nn
\\
&\quad \times  \textrm{P.E}\big{[}\frac{x(e^{i \lambda}x+e^{- i \lambda}x)}{1-x^2}\sum_i (\eta_i +\eta_i^{-1})- \frac{2x^2}{1-x^2} (e^{2i \lambda}x^2 + e^{-2i \lambda}x^2 +1)\big{]} \nonumber
\\
&+\frac{1}2 \int  \frac{d\lambda}{2\pi} (1-e^{2i \lambda})(1-e^{-2i \lambda}) \big{(}Z^{N;SU(2),N_f =4}_{mono}(2,0)\big{)}^2  \nn
\\
&\quad \times \textrm{P.E}\big{[} \frac{  x (e^{i \lambda}+e^{- i \lambda})}{1-x^2}\sum_i (\eta_i +\eta_i^{-1})- \frac{2x^2}{1-x^2} (e^{2i \lambda} + e^{-2i \lambda} +1)\big{]}\;. \label{SU(2) Nf=4 B=(1,-1)}
\end{align}
On the other hand, the index of the minimally charged Wilson line  is given by
\begin{align}
I^{SU(2);N_f=4}_{R=\mathbf{2}} (x, \eta_i)  &= \frac{1}2 \int  \frac{d\lambda}{2\pi} (1-e^{2i \lambda})(1-e^{-2i \lambda}) (e^{i \lambda}+e^{- i \lambda})^2  \nn
\\
&\quad \times \textrm{P.E}\big{[} \frac{x (e^{i \lambda}+e^{- i \lambda})}{1-x^2}\sum_i (\eta_i +\eta_i^{-1})- \frac{2x^2}{1-x^2} (e^{2i \lambda} + e^{-2i \lambda} +1)\big{]}\;.
\end{align}
Indices of two operators exactly match when $\eta_i=1$, 
\begin{align}
&I^{SU(2);N_f=4}_{B=(1,-1)} (x, \eta_i)|_{\eta_i=1}=I^{SU(2);N_f=4}_{R=\mathbf{2}} (x, \eta_i)|_{\eta_i=1}  \nn
\\
&= 1+ 62 x^2 + 896 x^4 + 7868 x^6 + 51856 x^8 +281836 x^{10} + 1328923 x^{12} \nn
\\
&\quad +5611146 x^{14}+ 21671145 x^{16}+77725908 x^{18} + 261809269 x^{20} +\ldots \;.
\label{4flavor-sdual}
\end{align}
Turning on the chemical potentials, two indices are related in the following way
\begin{align}
&I^{SU(2);N_f=4}_{B=(1,-1)} (x, \eta_i) =I^{SU(2);N_f=4}_{R=\mathbf{2}} (x, \tilde{\eta}_i )\;, \textrm{ where} \nn
\\
&
\tilde{\eta}_1 = \sqrt{\eta_1 \eta_2 \eta_3 \eta_4}\;, \;\tilde{\eta}_2= \frac{\sqrt{\eta_1 \eta_2}}{\sqrt{\eta_3 \eta_4}}\;, \; \tilde{\eta}_3 = \frac{\sqrt{\eta_2 \eta_4}}{\sqrt{\eta_1 \eta_3}}\;, \; \tilde{\eta}_4  = \frac{\sqrt{\eta_1 \eta_4}}{\sqrt{\eta_2 \eta_3}} \;.
\end{align}
Let $j_1  =  \eta_1  \eta_4, j_2 = \eta_1 \eta_4^{-1}, j_3 = \eta_2 \eta_3 , j_4 = \eta_2 \eta_3^{-1}$ and give similar relations for $\tilde{j}$s and $\tilde{\eta}$s. Then the above relations between $\eta$ and $\tilde{\eta}$ can be written as
\begin{align}
\tilde{j}_1 = j_1\; , \; \tilde{j}_2 = j_3 \; , \; \tilde{j}_3 = j_4\;, \;\tilde{j}_4 = j_2 \;.
\end{align}
These $j$s (or $\tilde{j}$s) are chemical potentials for Cartan generators of four $SU(2)$s in the $SO(8)$ global symmetry.  From this index computation,  we confirm the exchange of a (minimally charged) Wilson and a (minimally charged) 't Hooft line operator  and a permutation of four $SU(2)$s under S-duality in the $\mathcal{N}=2$ theory \cite{Gaiotto:2009we}.

\section{Index calculation using Verlinde loop operators}

In a recent paper \cite{Dimofte:2011py}, the authors consider a superconformal index in the presence of line operators. Rather than calculating the index directly from a field theory, they use the dictionary relating line operators in a $\mathcal{N}=2$ field theory and Verlinde loop operators in a 2d CFT.  In this section, we will review and extend their works to  compare with our field theory results. 

\subsection{${\cal N}=4$ $SU(2)$ theory}
For  $SU(2)$ $\mathcal{N}=4$  SYM theory, the superconformal index in the absence of line operators can be written as
\begin{align}
\langle \Pi_S | \Pi_N \rangle : = \sum_m \int_{0}^{2 \pi} \frac{ d \lambda}{2 \pi} \Delta_m (e^{i \lambda}, x ) \Pi_{m}^\dagger \Pi_{m}  \;, 
\label{DGG}
\end{align}
where the summation is over $m=0, \frac{1}2, 1, \frac{3}{2},\ldots,\infty$.  $\Pi_{m}(e^{i \lambda},x, \eta)$   denote a ``half-index", an index on southern (or northern) hemisphere.  Under the conjugation $^\dagger$,
\begin{align}
x^{\dagger} = x \; , \quad  \eta^{\dagger} = \eta^{-1}\;, \quad  (e^{i \lambda})^\dagger = e^{- i \lambda}\;.
\end{align}
 In the calculation in \cite{Dimofte:2011py}, they  introduce a mass parameter  for the adjoint hypermultiplet, which corresponds to the chemical potential $\eta$   in our index. Two half-indices are glued together with measure $\Delta_m$, which is the index for three dimensional theory on the boundary of hemispheres ($S^2 \times S^1$) where background gauge fields with magnetic flux $B$ are coupled to the global symmetry of the 3d theory \cite{Kapustin:2011jm}. When $B= \textrm{diag} (m ,-m )$, the 3d index $\Delta_m$ is given by 
\begin{align}
&\Delta_m (e^{i \lambda}, x) := (1-\frac{1}{2}\delta_{m,0})x^{-2m} (1-x^{2m} e^{2 i  \lambda})(1-x^{2m} e^{-2i \lambda}) \;. 
\end{align}
Factors $(1-\frac{1}{2}\delta_{m,0})$ and $x^{-2m}$ come from symmetry factor and Casimir energy in the 3d  index  respectively.  $x^{2m} \Delta_m$ is same with $[dU]_B$ defined in eq.~\eqref{1-loop for N=4 SYM}, the shifted Haar measure with magnetic charge $B= \textrm{diag} (m ,-m )$.  Half-index $\Pi_{m}$ is given as 
\begin{align}
&\Pi_{m} (e^{ i \lambda}, x, \eta )= \delta_{m,0}\textrm{P.E}\big{[}(\frac{x} {1-x^2} \eta- \frac{x^2}{1-x^2}) (e^{-2i \lambda} +1 + e^{2 i \lambda}) \big{]}  \;.
\end{align}
 P.E denotes the Plethystic exponential \eqref{P.E}. In the presence of line operators $\hat{O}_L$ at the north and south pole, the index becomes 
\begin{align}
I_L = \langle \hat{O}_L\cdot \Pi_S | \hat{O}_L\cdot \Pi_N \rangle =  \sum_m \int_{0}^{2 \pi} \frac{d \lambda}{2\pi} \Delta_m \big{(}\hat{O}_L\cdot  \Pi_{m}  \big{)}^\dagger \big{(}\hat{O}_L\cdot  \Pi_{m} \big{)}\;. 
\end{align}

 Now the question is how to identify the action of $\hat{O}$ on the half-index $\Pi_m$. In  \cite{Dimofte:2011py}, the authors proposed  a map between  Wilson-'t Hooft line operators $\hat{O}$ on $\Pi_m$ in $SU(2)$ $\mathcal{N}=4$ theory\footnote{Actually, they consider the $\mathcal{N}=2^*$ theory. But the theory is equivalent to  $\mathcal{N}=4$ theory twisted by  turning on chemical potential for $U(1)$  symmetry acting on a hypermultiplet. }  and Verlinde loop operators in a  2d Liouville theory. Such a relation originally  appears in the computation of $S^4$ partition function in the insertion of line operators located at great circle on $S^4$ via  AGT relation\cite{Alday:2009fs}\cite{Drukker:2009id}. One may expect that such a map also exists  for the index computation since the geometry near poles of $S^3 \times S^1$ is locally same with the geometry near the great circle of $S^4$, which are $\mathbb{R}^3 \times S^1$. 
According to the map in \cite{Dimofte:2011py}, the operator $\hat{O}_{1,0}$ corresponding to the line operator of the minimal magnetic charge, $B= \textrm{diag}(\frac{1}2 , - \frac{1}2 )$, is given as 
\begin{align}
\hat{O}_{1,0} =& \frac{  \left( \hat{x}( x^{\frac{1}{2}} \eta^{ \frac{1}{2}} )^{-1} - \hat{x}^{-1} (x^{\frac{1}{2}} \eta^{\frac{1}{2}} )  \right)  }
{ \hat{x} -\hat{x}^{-1} } \  \hat{p}^{ - 1/2}  \nonumber \\
& +  \frac{   \left( \hat{x} (x^{\frac{1}{2}} \eta^{ \frac{1}{2}} ) -\hat{x}^{-1} ( x^{ \frac{1}{2}} \eta^{ \frac{1}{2}} )^{-1} \right)}
{ \hat{x} - \hat{x}^{-1} } 
 \  \hat{p}^{1/2} \;.
\end{align}
where $\hat{x}, \hat{p}$ are defined as 
\footnote{%
Thus, for any function $f( e^{ i \lambda}, m)$, $ \hat{x}^{ \pm 1}, \hat{p}^{ \pm \frac{1}{2}}$ act as $\hat{x}^{ \pm 1} f( e^{i \lambda},m) = x^{\pm m} e^{  \pm i \lambda} f(e^{ i \lambda}, m) $,  $ \hat{p}^{ \pm \frac{1}{2} } f( e^{ i \lambda} , m) = f ( x^{ \pm\frac{1}2} e^{ i \lambda}, m \pm \frac{1}{2})$. For example, $\hat{p}^{\frac{1}2} \cdot \delta_{m,0} = \delta_{m+\frac{1}2 ,0}$. }
\begin{align} 
\hat{x} = x^m e^{ i \lambda}, \qquad \hat{p}= e^{ \partial_m } x^{ - i \frac{ \partial}{ \partial \lambda} } . \nonumber
\end{align}
Also, the map gives an operator $\hat{O}_{0,1}$ which corresponds to the fundamental Wilson loop operator  as
\begin{align} 
\hat{O}_{0,1} = \hat{x} + \hat{x}^{-1} .  \nonumber
\end{align}
$\hat{x}, \hat{p}$ satisfy a commutation relation $\hat{p} \hat{x} = x^2 \hat{x} \hat{p}$, 
 which is consistent with the OPE of line operators \cite{Dimofte:2011py}. Also note that these operators are hermitian w.r.t the measure $\sum_m \Delta_m$.  For instance, 
\begin{align}
\sum_m \int_{0}^{2 \pi} \frac{d \lambda}{2\pi} \Delta_m \big{(}\hat{O} \cdot  \Pi_{m}  \big{)}^\dagger\big{(}\hat{O} \cdot  \Pi_{m}  \big{)}
=\sum_m \int_{0}^{2 \pi} \frac{d \lambda}{2\pi} \Delta_m  \Pi_{m}^\dagger   ( \hat{O}^2 \cdot  \Pi_{m} ) \;. 
\nonumber 
 \end{align}
One can easily show that the expression of the index with a 't Hooft line of $B=\textrm{diag}(\frac{1}2, - \frac{1}2)$ obtained from this method is indeed identical to our result from the field theory computation for the $SU(2)$ ${\cal N}=4$ theory. 
\begin{align}
 &\sum_{m}  \int_0^{ 2 \pi } \frac{ d \lambda}{ 2 \pi}\Delta_m ( \hat{O}_{1,0} \Pi_{m} )^\dagger( \hat{O}_{1,0} \Pi_{m})  \; , \nonumber
 \\
 &= \int^{2\pi}_0 \frac{d\lambda}{2\pi} (1-e^{2 i \lambda}x)(1-e^{-2 i \lambda}x) \textrm{P.E}\big{[} \frac{(\eta+\eta^{-1}) x -2 x^2}{1-x^2} (1+e^{2i \lambda}x +e^{-2i \lambda}x)\big{]} \;.
 \end{align}  
 %
 %
The same holds true for the index in the presence of two fundamental Wilson loops. 
\begin{align}
&\sum_{m} \int_0^{ 2 \pi } \frac{ d \lambda}{ 2 \pi}  \Delta_m ( \hat{O}_{0,1} \Pi_{m} )^\dagger (\hat{O}_{0,1} \Pi_{m}) \nonumber
\\
&= \frac{1}2\int^{2\pi}_0 \frac{d\lambda}{2\pi} (1-e^{2 i \lambda})(1-e^{-2 i \lambda}) (e^{i \lambda}+e^{- i \lambda})^2\textrm{P.E}\big{[} \frac{(\eta+\eta^{-1}) x -2 x^2}{1-x^2} (1+e^{2i \lambda} +e^{-2i \lambda})\big{]} \;.
\end{align}
 
 The monopole bubbling effect, $Z^{SU(2)}_{mono} (B, v)$, can be obtained from this approach. 
For $B= \textrm{diag}(\frac{p}{2}, - \frac{p}{2})$ and $v= \textrm{diag}( \frac{s}{2}, - \frac{s}{2})$,  $Z^{SU(2),N}_{mono} (p,s)$ which is a function of $(e^{ i \lambda}, x, \eta)$ can be read from the following form 
\begin{align} 
\hat{O}_{1,0}^p \cdot \Pi_{m} =&  \sum_s x^{ \frac{ s}{2} } \eta^{ \frac{s}{2} } ( \delta_{m, \frac{s}{2} } + \delta_{m, - \frac{s}{2}} - \delta_{s, 0} \delta_{m,0} ) Z^{SU(2),N}_{mono} (p, s)
\nonumber \\
& \quad \times \textrm{P.E}\big{[}(\frac{x} {1-x^2} \eta^{-1} - \frac{x^2}{1-x^2}) (x^{s} e^{2i \lambda} +1+ x^s e^{-2 i \lambda}) \big{]},  
\label{AGT;bubble}
\end{align}
where the sum is over $s=0, 2, \cdots, p$ ($s=1, 3, \ldots, p$) if $p$ is even (odd). One can check that   $Z^{SU(2),N}_{mono} (2,0)$ and  $Z^{SU(2),N}_{mono} (3,1)$  obtained using the relation in eq~\eqref{AGT;bubble} are identical to those obtained from the field theory calculation,  eq.~\eqref{Z20} and \eqref{Z31}.  It is obvious that $Z_{mono}^{SU(2),N}(p,p)=1$ for any $p$. 
 
From eq.~\eqref{AGT;bubble}, one can get a recursion relation of the monopole bubbling effect as follows,
\begin{align}
& Z^{N;SU(2)}_{mono} (p, 0; e^{ i \lambda}, x, \eta) =  \frac{ ( 1 -  \eta e^{ - 2 i \lambda} x)( 1 -  \eta^{-1} e^{ - 2 i \lambda} x)  }{ ( 1 - e^{ - 2 i \lambda})(1-e^{ - 2 i \lambda} x^2 )} Z^{N;SU(2)}_{mono}(p-1, 1; x^{ - \frac{1}{2} } e^{ i \lambda}, x, \eta) \nonumber \\
&  \qquad \qquad \qquad \qquad \quad +  \frac{ ( 1 - \eta e^{ 2 i \lambda} x) ( 1 - \eta^{-1} e^{ 2 i \lambda} x)  }{ ( 1 - e^{  2 i \lambda})(1-e^{ 2 i \lambda} x^2 )} Z^{N;SU(2)}_{mono} (p-1, 1; x^{  \frac{1}{2} } e^{ i \lambda}, x, \eta) \;, \nn
\\ \nn
\nonumber \\
& Z^{N;SU(2)}_{mono}(p, s; e^{ i \lambda}, x , \eta) \stackrel{{\scriptsize \mbox{For }} s \geq 1}{=} Z^{N;SU(2)}_{mono}(p-1, s-1; x^{ - \frac{1}{2} } e^{ i \lambda}, x, \eta) \nonumber \\
&\qquad \qquad \qquad    + \frac{ (1 - \eta e^{ 2 i \lambda} x^{s+1} )(1 - \eta^{-1}  e^{ 2 i \lambda} x^{s+1} )  }{ ( 1- e^{ 2 i \lambda} x^s)(1- e^{ 2 i \lambda} x^{ 2+ s} )} Z^{N;SU(2)}_{mono} (p-1, s+1; x^{ \frac{1}{2} } e^{ i \lambda} , x, \eta)\; . 
\end{align}
This recursion relation is easier to use than the general formula in \cite{Ito:2011ea} in some respects, since the general formula becomes quickly cumbersome  as the involved instanton number $k$ (the total number of boxes in colored Young diagrams) increases as $ \frac{p^2-s^2}{4}$. For a special case that $s= p-2$,  one can solve the recursion relation 
\begin{align} 
& Z^{N;SU(2)}_{mono}(p, s)|_{s=p-2}  \nonumber \\
&= \frac{\left( -( 1 + x^2)\sum_{n=0}^{p-1} x^{2n}+ p( 1+ x^{2p})  \right)+ ( \eta^{-1}+ \eta)  \sum_{n=1}^{p} x^{ 2n-1} - p x^p ( e^{-2i \lambda}+  e^{2i \lambda})}
{(1 -  e^{-2i \lambda} x^p)(1-  e^{2i \lambda} x^p )}\;,
\label{rec:exact}
\end{align}
which includes the results for $Z^{N;SU(2)}_{mono}(2,0)$ and $Z^{N;SU(2)}_{mono}(3,1)$ given in eq.~\eqref{Z20}, \eqref{Z31}.  Let us first simplify the relation by turning off the chemical potential $x \to 1$, 
then the recursion relation can be solved for any $p, s$  as
\begin{align}
Z^{N;SU(2)}_{mono}(p, s)& =  \frac{ p!}{ ( \frac{ p+s}{2} )! ( \frac{ p-s}{2} )!}  \times  \frac{ ( 1 - \eta e^{ - 2 i \lambda} )^{ \frac{p-s}{2} } ( 1 - \eta^{-1} e^{ - 2 i \lambda})^{ \frac{ p-s}{2} } }
{ ( 1 - e^{ - 2 i \lambda})^{p-s} }, 
\label{mono-pestun}
\end{align}
which is consistent with eq~\eqref{rec:exact}. The result in eq~\eqref{mono-pestun} can be compared with eq (7.65) of \cite{Gomis:2011pf}, which summarizes monopole bubbling effects of the 't Hooft loop wrapping the great $S^1$ on $S^4$.  The results agree with each other once we map $( r \hat{a})_{theirs} =  ( ( \frac{ \lambda}{ 2 \pi} )_{ours} \pm \frac{s}{4} )$, $(r \hat{m})_{theirs}=  (\frac{ \ln \eta}{ 2 \pi i })_{ours}$, where $\pm$ denotes a sign ambiguity.

\begin{figure}[h!]
  \begin{center}
    \includegraphics[width=10cm]{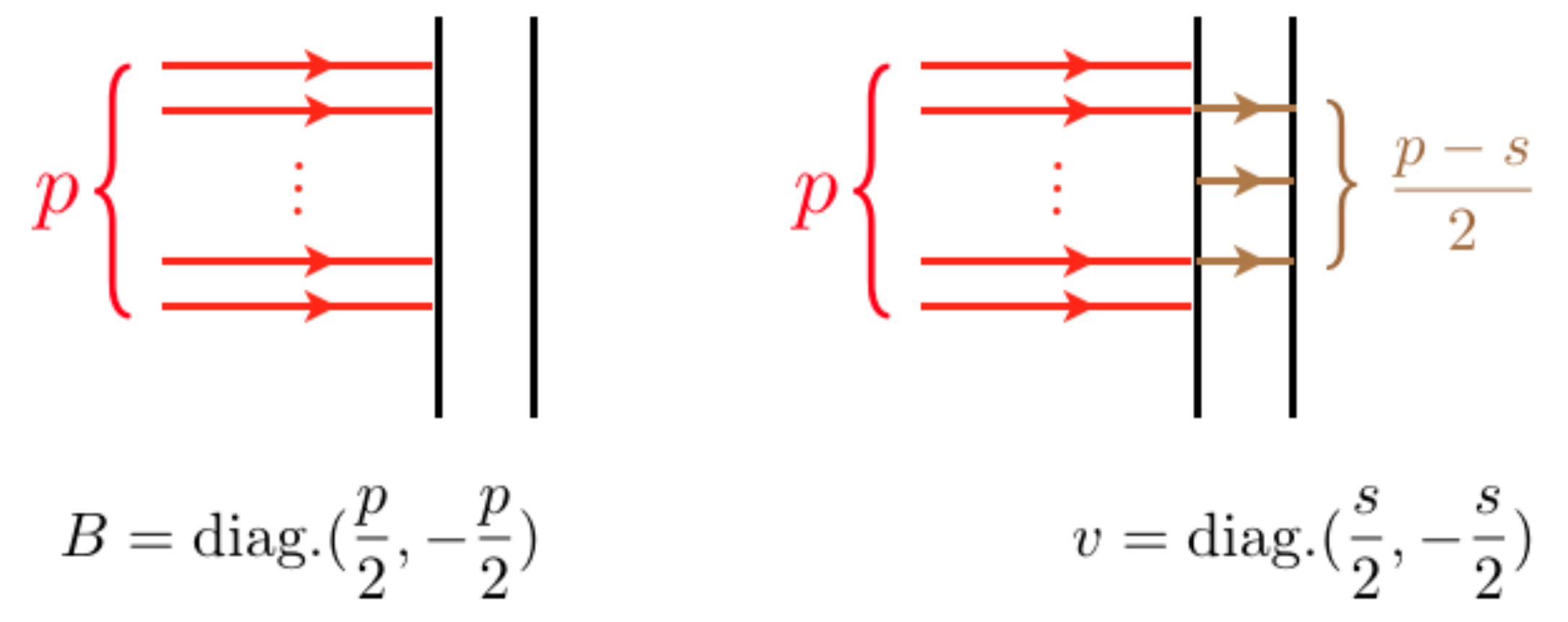}
\caption{$p$ infinitely stretched D1-branes (04, red lines) ending on two D3-branes (0123, black lines) correspond to a 't Hooft operator  with magnetic charge $\textrm{diag}. (p, 0)$, of which the traceless part is $B= \textrm{daig}. ( \frac{p}{2}, - \frac{p}{2})$ in the $SU(2)$ theory (left). Once $ \frac{p-s}{2}$ massless D1-branes (04, brown lines) end on the tops of  infinitely stretched D1-branes, the magnetic charge is reduced to $  \textrm{diag}.(p- \frac{p-s}{2}, \frac{p-s}{2})$, of which the traceless part is $v=\textrm{diag}.(\frac{s}{2}, - \frac{s}{2})$ (right).
}
\label{combinatorics}
  \end{center}
\end{figure}

If one turns off two chemical potentials $x \to 1, \eta \to1$, then the solution in eq~\eqref{mono-pestun} becomes 
\begin{align} 
Z^{N;SU(2)}_{mono} (p, s) = \frac{ p!}{ ( \frac{ p+s}{2} )! ( \frac{ p-s}{2} )!} . 
\label{com}
\end{align}
When all the chemical potentials are turned off, the superconformal index counts the number of vacua if all vacua are bosonic.  Eq.~\eqref{com} is the number of ways of picking $\frac{ p-s}{2}$ (or $\frac{ p+s}{2}$) ending sites for massless D1-branes from $p$ possibilities, which is the number of vacua in our picture of monopole bubbling (see Figure~\ref{combinatorics}).

\subsection{${\cal N}=4$ $SU(3)$ theory}

We can extend the methods in the previous section to $SU(3)$ theory. A generalization to $SU(N)$ theory will be discussed. For convenience, we will turn off mass parameters, setting $ \eta_i = 1$ hereafter. 

Let us first show the $SU(3)$ case explicitly. A representation of $SU(3)$ can be specified by $\vec{l}:=(l_1, l_2, l_3)$, for non-negative integers $l_i$ satisfying $l_1 \geq l_2 \geq l_3 = 0 $, which corresponds to the Young diagram with $l_i$ boxes in the $i$th row. The traceless part of $\textrm{diag} (l_1, l_2, l_3)$ can be defined as a diagonal matrix $\vec{m}$, i.e., 
\be
\vec{m}= \textrm{diag} ( \frac{2 l_1 - l_2}{3}, \frac{-l_1 +2l_2}{3}, - \frac{l_1+l_2}{3} ):=\textrm{diag} (m_1, m_2,m_3). 
\nn \ee
Note that  $m_3 =-(m_1+m_2)$. The root $ \alpha(\vec{m})$ are integers, since $ \alpha( \vec{m}) = \alpha( \vec{l})$. 
Since all matters in ${\cal N}=4$ theory are in the adjoint representation, a monopole charge given by $\vec{m}$ satisfies Dirac quantization condition. 

The index of ${\cal N}=4$ $SU(3)$ theory can be written in the following form 
\begin{align} 
I =\sum_{\vec m}  \int_{0}^{2 \pi} \left( \prod_{i=1}^{2} \frac{ d \lambda_i}{2 \pi} \right)  \Delta_{\vec{m}} (e^{ i \lambda}, x) | \Pi_{\vec{m}} (e^{ i \lambda}, x) |^2. 
\end{align}
The sum $\sum_{ \vec{m}}$ is over all possible sets of $\{ m_1, m_2 \}$ satisfying $l_1 \geq l_2 \geq 0$, i.e., $m_1 \geq m_2 \geq - \half m_1 $. 
The holonomy of $SU(3)$ is given as $\textrm{diag}(e^{i \lambda_1}, e^{i \lambda_2},e^{i \lambda_3})$, where $ \lambda_3 $ is understood to be $ \lambda_3= - ( \lambda_1+ \lambda_2)$. 
The half-index can be defined as 
\begin{align}
& \Pi_{(m_1, m_2)}= \delta_{\vec{m},(0,0)}  \hat{ \Pi}_{(0,0)}\;,
 \end{align}
where $\delta_{\vec{m},(s_1,s_2)}:= \delta_{m_1, s_1} \delta_{m_2, s_2}$, and $\hat{\Pi}_{\vec{s}}= \hat{ \Pi}_{(s_1, s_2)}$ is the part associated with monopole charge $ \textrm{diag}(s_1, s_2, -(s_1+s_2))$ given as 
\begin{align}
& \hat{\Pi}_{\vec{s}}   = \textrm{P.E}\big{[}\frac{x}{1+x}(2+\sum_{(i\neq j)=1}^3 x^{|s_{i}-s_j |})\big{]} \nn
\\
&=\prod_{n=0}^{ \infty}  \left(  \frac{(1-x^{2n+2})^2 }{(1-x^{2n+1})^{2}} \prod_{(i \neq j)=1}^3 \frac{  (1-x^{2n+2+|s_i - s_j|}e^{ i (\lambda_i - \lambda_j)}) }{(1-x^{2n+1+|s_i - s_j|}e^{  i(\lambda_i  - \lambda_j )})}  \right) \ . 
\end{align}
$\Delta_{\vec{m}}$ is a measure in the following form
\begin{align}
& \Delta_{(m_1, m_2)}  (e^{ i \lambda}, x) = \frac{1}{(\textrm{sym})}   x^{ - \frac{1}{2} \sum_{ \alpha}  | \alpha(\vec{m })| }  \prod_{ \alpha (\vec{m}) \neq 0} (1-x^{ | \alpha (\vec{m})|} e^{ i \alpha( \lambda)})\;,
\end{align}
where the symmetric factor denoted by $(\textrm{sym})$ is $3!=6$ for $l_1=l_2=0$, $2$ either for $l_1 > l_2=0$ or for $l_1 =l_2 >0$, $1$ for $l_1 >l_2 >0$.

To define line operators acting on the half-index, let us first define 
\begin{align}
\hat{p}_i  =x^{ - i  \p_{ \lambda_i} }  e^{ \p_{m_i}}. 
\nn \end{align}
Two minuscule representations of $SU(3)$ correspond to $\vec{l}=(1,0,0)$ and $ \vec{l}=(1,1,0)$ respectively. We denote line operators with magnetic charges $\textrm{diag}(1,0,0) \sim ( \frac{2}{3}, - \frac{1}{3}, - \frac{1}{3})$ and $\textrm{diag} (1,1,0)\sim (\frac{1}{3}, \frac{1}{3}, - \frac{2}{3})$ as $\hat{O}_{(1,0)}$ and $\hat{O}_{(1,1)}$.\footnote{The equivalence relation $B_1 \sim B_2$ means difference between $B_1$ and $B_2$ is proportional to identity matrix.} A line operator  corresponding to $\vec{l} = (l_1, l_2, 0)$ can be constructed by acting the following to the half-index, 
\begin{align}
(\hat{O}_{(1,1)} )^{l_2} (\hat{O}_{(1,0)} )^{l_1-l_2}.   \nn
\end{align}
Let us show the explicit form of $\hat{O}_{(1,0)}$, 
\begin{align}
\hat{O}_{(1,0)} &=x
\left( \frac{(1-x^{m_{12}-1} e^{i \lambda_{12} }) (1-x^{m_{13}-1} e^{i \lambda_{13}}) }
{(1-x^{m_{12}} e^{i \lambda_{12} }) (1-x^{m_{13}} e^{i \lambda_{13}}) } 
\right)
 \hat{p}^{ -\frac{2}{3}}_1 \hat{p}_2^{  \frac{1}{3}} \nn \\
& \quad +   \left(  \frac{(1- x^{m_{12} + 1} e^{ i \lambda_{12} })}{ ( 1- x^{m_{12} } e^{ i \lambda_{12} } )}
 \frac{ (1-x^{m_{23} -1} e^{ i \lambda_{23} })}{(1-x^{m_{23}} e^{ i \lambda_{23} } )} 
\right) \hat{p}^{ \frac{1}{3}}_1 \hat{p}_2^{ - \frac{2}{3}} \nn \\
& \quad  + x^{-1}  \left(
\frac{ (1- x^{m_{13}+1} e^{ i \lambda_{13} } )( 1 - x^{m_{23} +1} e^{ i \lambda_{23} } )}
{ (1- x^{m_{13}} e^{ i \lambda_{13} } )(1- x^{m_{23} } e^{ i \lambda_{23} })}
\right) \hat{p}_1^{ \frac{1}{3}} \hat{p}_2^{ \frac{1}{3}}
\nn 
\end{align}
where we defined $\lambda_{ij}:= \lambda_i- \lambda_j$, $m_{ij} := m_i - m_j$. The first line is to create magnetic charge $\textrm{diag}(\frac{2}{3}, - \frac{1}{3}, - \frac{1}{3}) \sim \textrm{diag}(1,0,0)$, while the second and third lines are associated with Weyl transformations of it. 

The index of the fundamental Wilson line can be obtained by
\begin{align} 
I^{SU(3)}_{R=\bf 3} &= \int_0^{ 2 \pi} \left( \prod_{i=1}^2 \frac{ d \lambda_i}{ 2 \pi} \right) \chi_{3} (e^{ i \lambda}) \chi_{ \bar{3}} ( e^{ i \lambda}) 
\Delta_{ (0,0)} (e^{ i \lambda}, x) |\hat{\Pi}_{(0,0)}(e^{ i \lambda}, x) |^2 \label{su(3)-wilson} \end{align}
where $ \chi_{\bf 3} $ is the character of the fundamental representation of $SU(3)$ given as
\begin{align}
& \chi_{{\bf 3}} (e^{ i \lambda})=  e^{ i \lambda_1} + e^{ i \lambda_2} + e^{  i \lambda_3} |_{ \lambda_3=-( \lambda_1+ \lambda_2)} , 
\nn 
\end{align}
and  $ \chi_{\bar{\bf 3}}(e^{i \lambda})= \chi_{{\bf 3}} (e^{ - i \lambda})$. 
The index of 't Hooft line with magnetic charge $B= \textrm{diag}( \frac{2}{3},- \frac{1}{3}, - \frac{1}{3})$ is given as 
\begin{align} 
 I^{SU(3)}_{B= ( \frac{2}{3},- \frac{1}{3}, - \frac{1}{3})} =\sum_{\vec m}  \int_{0}^{2 \pi} \left( \prod_{i=1}^{2} \frac{ d \lambda_i}{2 \pi} \right)  \Delta_{\vec{m}} (e^{ i \lambda}, x) | \hat{O}_{(1,0)} \Pi_{\vec{m}} (e^{ i \lambda}, x) |^2
\nn 
\end{align}
which becomes 
\begin{align}
 I^{SU(3)}_{B= ( \frac{2}{3},- \frac{1}{3}, - \frac{1}{3})} &= \int_0^{ 2 \pi} \left( \prod_{i=1}^2 \frac{ d \lambda_i}{ 2 \pi} \right)  x^2 \Delta_{( \frac{2}{3}, - \frac{1}{3}) }  (e^{ i \lambda}, x) | \hat{ \Pi}_{ ( \frac{2}{3} , - \frac{1}{3}) } (e^{ i \lambda}, x) |^2 , 
\label{su(3)-tHooft} 
\end{align}
since $\sum_{\vec{m}}$ is over $\{m_1, m_2 \}$ such that $m_1 \geq m_2 \geq - \frac{1}{2}m_1$. 
The fundamental Wilson line index \eqref{su(3)-wilson} matches the 't Hooft line index \eqref{su(3)-tHooft},  as expected from S-duality
\begin{align} 
& I^{SU(3)}_{R={\bf 3}} = I^{SU(3)}_{B= ( \frac{2}{3},- \frac{1}{3}, - \frac{1}{3})}
\nn \\
& = 1+ 2 x+ 5 x^2 + 4 x^3 + 6 x^4 + 6 x^5 + 13 x^6 + 8 x^7 + 8 x^8 + 10 x^9 +O(x^{10}). 
\nn 
\end{align}

Let us now consider the next simplest example: the index of 't Hooft line with magnetic charge $B=\textrm{diag}(\frac{4}{3}, - \frac{2}{3}, - \frac{2}{3})$ and the index of a product of two fundamental Wilson lines.  The 't Hooft line operator index can be obtained by
\begin{align} 
 I^{SU(3)}_{B= ( \frac{4}{3},- \frac{2}{3}, - \frac{2}{3})} =\sum_{\vec m}  \int_{0}^{2 \pi} \left( \prod_{i=1}^{2} \frac{ d \lambda_i}{2 \pi} \right)  \Delta_{\vec{m}} (e^{ i \lambda}, x) | \hat{O}^2_{(1,0)} \Pi_{\vec{m}} (e^{ i \lambda}, x) |^2.
\nn 
\end{align}
The relevant part of $ \hat{O}_{(1,0)}^2 \Pi_{m}$ is 
\begin{align} 
& \hat{O}^2_{(1,0)} \Pi_m \supseteq x^2 \delta_{\vec{m},(\frac{4}{3}, - \frac{2}{3})}  \hat{ \Pi}_{ ( \frac{4}{3}, - \frac{2}{3}) } (e^{ i \lambda},x)+ x  \delta_{\vec{m}, (\frac{1}{3}, \frac{1}{3})} Z_{mono}^{N;SU(3)}(e^{i\lambda},x)  \hat{ \Pi}_{ ( \frac{1}{3}, \frac{1}{3}) } (e^{i \lambda},x),
\nn 
\end{align}
where $Z^{N;SU(3)}_{mono}(e^{i \lambda},x)$ is the monopole bubbling index on the northern hemisphere (which is equivalent to the corresponding index on $\mathbb{R}^3$) with reduced charge $v=\textrm{diag}(\frac{1}{3}, \frac{1}{3}, -  \frac{2}{3})$,
\begin{align}
Z^{N;SU(3)}_{mono}(e^{i \lambda},x)=  \frac{(1-x e^{- i \lambda_{12}})^2}{(1-e^{-i \lambda_{12}})(1-x^2 e^{- i \lambda_{12}}) }+ \frac{(1-x e^{i \lambda_{12}})^2}{(1-e^{i \lambda_{12}})(1-x^2 e^{ i \lambda_{12}}) }  
\label{m-bub} \ .
\end{align}
 When $x=1$,  eq~\eqref{m-bub} becomes $2$, reflecting two possible choices of the position of a massless D1-brane. 
The index of 't Hooft line  can be written as
\begin{align}
 I^{SU(3)}_{B= ( \frac{4}{3},- \frac{2}{3}, - \frac{2}{3})} &= \int_0^{ 2 \pi} \left( \prod_{i=1}^2 \frac{ d \lambda_i}{ 2 \pi}\right)  \left( x^4 \Delta_{(\frac{4}{3}, - \frac{2}{3}) } | \hat{ \Pi}_{ (\frac{4}{3} , - \frac{2}{3} )}|^2 
 + x^2 \Delta_{(\frac{1}{3}, \frac{1}{3})} |Z^{N;SU(3)}_{mono} (e^{i\lambda},x)\hat{\Pi}_{(\frac{1}{3}, \frac{1}{3})}|^2
 \right).
 \end{align}
On the other hand, the index of the product of two fundamental Wilson lines can be obtained by
\begin{align} 
I^{SU(3)}_{(W_{\bf 3})^2} &= \int_0^{ 2 \pi} ( \frac{ d \lambda_i}{ 2 \pi})^2 
\left( \chi_{3} (e^{ i \lambda}) \chi_{ \bar{3}} ( e^{ i \lambda}) \right)^2
\Delta_{ (0,0)} (e^{ i \lambda}, x) |\hat{\Pi}_{(0,0)}(e^{ i \lambda}, x)|^2.  
\end{align}
Again, we can check that they match,
\begin{align} 
& I_{(W_{\bf 3})^2 }^{SU(3)} =  I^{SU(3)}_{ B= (\frac{4}{3}, - \frac{2}{3}, - \frac{2}{3} )} \nn \\
&=2+ 8 x+ 19 x^2 + 24 x^3 + 25 x^4 + 32 x^5 + 53 x^6 + 48 x^7 + 33 x^8 + 56 x^9 +O(x^{10}). 
\end{align}

The other operator with magnetic charge $ \textrm{diag}(1,1,0) \sim \textrm{diag}(\frac{1}{3}, \frac{1}{3}, - \frac{2}{3})$ can be constructed similarly. It has the following form 
\begin{align}
\hat{O}_{(1,1)} &= ( \cdots) \hat{p}_1^{ \frac{2}{3} } \hat{p}_2^{ - \frac{1}{3}}+ ( \ldots) \hat{p}_1^{ - \frac{1}{3}} \hat{p}_2^{ \frac{2}{3}}   +  ( \cdots) \hat{p}^{ - \frac{1}{3}}_1 \hat{p}^{- \frac{1}{3}}_2 \nn
\end{align}
where $(\cdots)$ denote functions of $x, \vec{m}, e^{ i \lambda}$.  
We expect that the operator is S-dual to the anti-fundamental Wilson line. The index of line operator associated with $\hat{O}_{1,1}^2$ also shall be same with the index of the product of two anti-fundamental Wilson lines. In the case, the reduced charge of monopole bubbling will be $v= \textrm{diag}(\frac{2}{3}, - \frac{1}{3}, - \frac{1}{3})$, rather than $\textrm{diag}( \frac{1}{3}, \frac{1}{3} , - \frac{2}{3})$ in the previous case.

  The generalization to $SU(N)$ for $N \geq 4$ is now obvious.  The line operator with magnetic charge $\textrm{diag}(1, 0, \ldots, 0) \sim (\frac{N-1}{N}, - \frac{1}{N}, \ldots, - \frac{1}{N})$ contains creation operators in the form of $\hat{p}_1^{ \frac{N-1}{N}} \hat{p}_2^{ - \frac{1}{N}} \ldots \hat{p}_{N-1}^{- \frac{1}{N}}$ and the other operators  corresponding to its Weyl transformations. Coefficients can be determined from the constraint that the monopole bubbling effect of the highest weight is 1.  It would be interesting to explicitly work it out and compare the result with \cite{Gomis:2010kv}, but we do not pursuit it here.

\subsection{${\cal N}=2$ $SU(2)$ theory with four flavors}

The index formula for the $SU(2)$ ${\cal N}=2$ theory can be written in the form of eq.~\eqref{DGG} with a following half-index,
\begin{align}
\Pi_{m,S} ( e^{ i \lambda}, x, \eta_i) & = \delta_{m,0} \textrm{P.E} [  ( \sum_{i=1}^4 \frac{x}{1-x^2} \eta_i) (e^{ i \lambda} + e^{ - i \lambda}) - \frac{x^2}{1-x^2} 
 (e^{- 2 i \lambda}+1+ e^{ 2 i \lambda}) ]  \;.
\end{align}
Hereafter we set $ \eta_i=1$ for convenience.

As discussed in section \ref{sec:4flavors}, the minimal charge of 't Hooft line in this theory is $B ={\textrm{diag}}(1, -1)$, which is not minuscule. Thus the line operator with the minimal magnetic charge  contains  contributions from monopole bubbling effect, 
\begin{align}
\hat{O}_{1,0} = H_{+} ( e^{ i \lambda}, x)  \hat{p} + h_{m} (e^{ i \lambda}, x  ) + H_{-} (e^{ i \lambda}, x)  \hat{p}^{-1}, \nn
\end{align}
where 
\begin{align}
 H_{ \pm} ( e^{ i \lambda} ,  x) & =  \frac{ x  ( 1 -  e^{\mp i \lambda  } x^{ \mp m - 1 } )^4 }{ ( 1- e^{\mp 2 i \lambda } x^{\mp 2m-2})( 1- e^{\mp 2 i \lambda} x^{ \mp 2m})} \ ,  \nn 
\\
h_{m} ( e^{ i \lambda}, x)  &  -  \frac{8 x^{ m} e^{ i \lambda} }{ (1+ x^{m-1} e^{ i \lambda} )(1+ x^{m+1} e^{ i \lambda})  } \;. 
\label{hm}
\end{align}
Especially, the function $h_{m}(e^{i \lambda}, x)$ can be read from the result of the 't Hooft line operator in AGT context given in eq.5.39 of \cite{Alday:2009fs}, which becomes
\begin{align}
h_{m} = - \frac{4}{ \cos \pi b^2 - \cos b \pi ( 2a - Q)}  \;,
\end{align}
when hypermultiplets are massless.
 Applying  a map   $ e^{ i \pi b ( 2a - Q)} \to - x^m e^{ i \lambda}$, $e^{  \pi i b^2} \to x $ similarly to \cite{Dimofte:2011py}, 
 one gets the result in eq~\eqref{hm}
  \footnote{ 
  The sign flip of the first map is determined from a condition that the  function $h_{0} (e^{ i \lambda}, x)$ should coincide with our previous result of monopole bubbling, i.e., $h_{0}  = Z_{mono}^{N;SU(2), N_f=4} (2,0)$. }. 
From the construction, $\hat{O}_{1,0} \Pi_m$ corresponds to the half-index of the 't Hooft operator with magnetic charge $B=\textrm{diag}(1,-1)$, 
\begin{align}
\hat{O}_{1,0} \Pi_m (e^{i \lambda}, x) &= ( \delta_{m, -1} + \delta_{m, 1} ) x  \hat{ \Pi }_{1} (e^{ i \lambda}, x)+  \delta_{m, 0} Z_{mono}^{N;SU(2), N_F=4} (2,0) \  \hat{ \Pi}_0(e^{ i \lambda} , x) \;,
\end{align}
where $\hat{\Pi}_s (e^{ i \lambda},x)$ are the part associated with the monopole charge $\textrm{diag}(s,-s)|_{s\geq 0}$
\begin{align} 
\hat{\Pi}_s ( e^{ i \lambda}, x ) &= \textrm{P.E} \big{[} \frac{x^{1+s}}{1-x^2} ( e^{i \lambda}+ e^{- i \lambda}) - \frac{x^2}{1-x^2} (1+ x^{2s} x^2 e^{2i \lambda}+x^{2s}x^{2}e^{-2i \lambda})\big{]} \nn
\\
=&\prod_{n=0}^{ \infty} 
\frac{ (1- x^{2n+2})(1- e^{-2i \lambda} x^{2n+2s+2} )(1- e^{2i \lambda} x^{2n+2s+2} )}
{  (1-  e^{-i \lambda} x^{1+s + 2n} )^4( 1 -  e^{ i \lambda} x^{ 1+s+2n} )^4 } \ .
\label{Pis}
\end{align}
The index of 't Hooft line with $B= \textrm{diag}(1,-1)$ calculated in this way give the same expression  in eq.~\eqref{SU(2) Nf=4 B=(1,-1)}. In section \ref{sec:4flavors}, we check that the  't Hooft line index  is same with the index of the fundamental Wilson line (see eq.~\eqref{4flavor-sdual}).

Let us check the S-duality for the next simplest case. The index of 't Hooft line operator with $B={\textrm{diag}}(2,-2)$ can be written as
\begin{align}
I^{SU(2); N_f=4}_{B=(2,-2)} (x)= \sum_{m=0}^{ \infty}  \int_{0}^{  2\pi } \frac{ d \lambda}{ 2 \pi}  \Delta_m |\hat{O}_{1,0}^2 \Pi_m |^2, \nn 
\end{align}
where explicitly, 
\begin{align}
& \hat{O}^2_{1,0}  \Pi_m ( e^{ i \lambda}, x) \nn \\
&  = ( \delta_{m, -2} + \delta_{m, 2} ) x^2 \hat{ \Pi}_{2} +  \left( \delta_{m, -1} + \delta_{m, 1} \right)  \left(h_{-1} ( e^{ i \lambda}, x) + h_0 ( x e^{ i \lambda}, x) \right) 
 x  \hat{ \Pi}_1 \nn \\
& \quad + \delta_{m,0} \left( \left(h_0 ( e^{ i \lambda}, x)  \right)^2 +  \sum_{ \pm}  \frac{ ( 1 - e^{ \pm i \lambda})^8}{(1 - e^{  \pm 2 i \lambda})( 1- e^{ \pm 2 i \lambda} x^2)^2 ( 1- e^{ \pm 2 i \lambda} x^4)}  \right)
 \hat{ \Pi}_{0}.
\end{align}
In the second line,  we used that $ h_{-1} (e^{ i \lambda},x) + h_0 (x e^{ i \lambda}, x ) = h_{1} ( e^{ i \lambda}, x) + h_0 (x^{ - 1} e^{ i \lambda}, x)$ for $h_m(e^{ i \lambda}, x)$ in eq.~\eqref{hm}. 
On the other hand, the index of the product of two fundamental Wilson lines can be written as 
\begin{align}
I_{R=(1,-1)^2}^{SU(2); N_F=4} (x) &= \int_{0}^{ 2 \pi} \frac{ d \lambda}{2 \pi} \Delta_0 (e^{ i \lambda}, x)  (e^{ i \lambda}+ e^{ - i \lambda})^4 | \hat{\Pi}_0 (e^{ i \lambda}, x) |^2 \;. 
\end{align}
Indices of two operators exactly match again, 
\begin{align}
& I^{SU(2); N_f=4}_{B=(2,-2)} (x)= I_{R=(1,-1)^2}^{SU(2); N_F=4} (x) \nn \\
& \quad = 2+158 x^2 +2618 x^4+ 24606 x^6 + 169300 x^8 + 947738 x^{10} + O(x^{12}). 
\end{align}

\section{Holography}

\subsection{Fundamental string/fundamental Wilson line}
Another important feature of  $U(N)$ $\mathcal{N}=4$ SYM is that it admits a well-established gravity dual description.   The gravity dual of Wilson line operators (in the fundamental  at the north pole and  anti-fundamental at the south pole) in the field theory on $\mathbb{R}\times S^3$ is a fundamental string wrapping  $AdS_2$ in global $AdS_5$. In global coordinates of $AdS_5$, 
\begin{align}
ds^2_{AdS_5} =- \cosh^2 \rho dt^2 + d \rho^2 +\sinh^2 \rho d\Omega_3^2, 
\end{align}
the string  world-volume is given by (located at a point on $S^5$)
\begin{align}
0 \leq \rho \leq \infty, \quad -\infty \leq t \leq \infty, \quad  \vec{\Omega}_3 = (1,0,0,0),(-1,0,0,0) \;.
\end{align}
\begin{figure}[h!]
  \begin{center}
    \includegraphics[width=4.5cm]{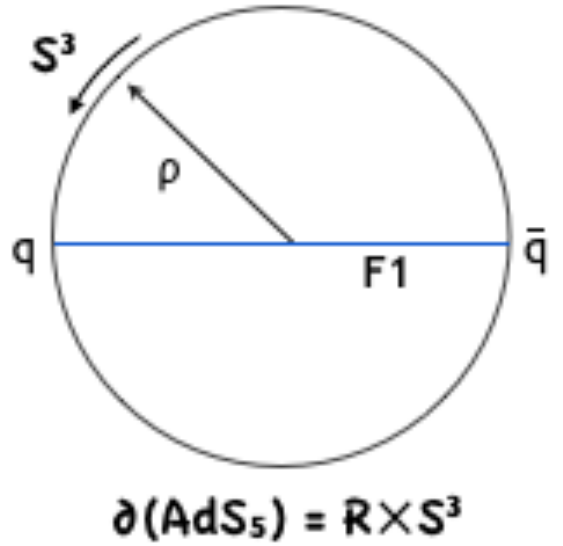}
\caption{Holographic  description of the quark-antiquark ($q\bar{q}$) system in $U(N)$ ${\cal N}=4$ SYM. The figure describes  a constant (global) time slice  of $AdS_5$.  A fundamental string (F1)  is stretched along $\rho$ direction (radial direction in the figure) and meets  the boundary (located at $\rho = \infty$) of $AdS_5$ at two points, the south and north poles on $S^3$. }
  \end{center}
\end{figure}
They preserve bosonic $SO(1,2)\times SO(3)\times SO(5)$ symmetries, which coincide with the symmetries preserved by the Wilson lines. In \cite{Faraggi:2011bb}, the fluctuation modes  of  the string wrapping  $AdS_2$ are analyzed. In terms
of $SO(1,2)\times SO(3)\times SO(5)$ representations, the complete spectrum is given by
\begin{align}
(1, 0, \mathbf{5})\oplus (\frac{3}2,  \frac{1}2 , \mathbf{4})\oplus (2,  1, 1) \;. 
\end{align}
For $SO(1,2)$, the representation is labelled by the conformal dimension $\epsilon$ of the highest weight state. The Cartan $j_z$ of  the SO(3) is equal to $j_L +j_R$  in the definition of our index. Note that $r_1$- and $r_3$-charges (where $r_3$ is conjugate to the chemical potential $\eta$) for  $\mathbf{5}$ and $\mathbf{4}$ of $SO(5)$ are given by
\begin{align}
&\{(r_1,r_3)\} (\mathbf{5}) = \{ (1,1),(1,-1),(-1,-1),(-1,1),(0,0) \}\; , \nonumber
\\
&\{ ( r_1, r_3)\} (\mathbf{4}) =\{ (1,0),(-1,0),(0,1),(0,-1) \} \;. 
\end{align}
Thus two states in $\mathbf{5}$ and one state in $\mathbf{4}$ saturate the BPS bound $\Delta = \epsilon- j_L - j_R -r_1=0$. The single particle index from the fluctuations of the string is given by
\begin{align}
I_{sp;F1} (x) = (\eta +\eta^{-1})x - x^2 \;. 
\end{align} 
Taking the large $N$ limit of SCI in the presence of fundamental Wilson line in $\mathcal{N}=4$ $U(N)$ theory,  we find the following factorization property  (ignoring overall factors independent of $x,\eta$ and $e^{i \lambda_i}$)
\begin{align}
&I^{N\rightarrow \infty}_{R=\mathbf{N}} (x,\eta)  \nonumber
\\
&=\frac{1}{N!} \int \frac{d^N \lambda_i}{(2\pi)^N} \big{[} \prod_{i \neq j}^N (1-e^{i (\lambda_i - \lambda_j)}) \big{]} (\sum_i e^{i \lambda_i})(\sum_i e^{- i\lambda_i})\exp \big{[}\sum_{n=1}^\infty \sum_{i,j}^N  \frac{1}n  f(x^n, \eta^n) e^{i n (\lambda_i - \lambda_j)}\big{]} \;, \nonumber
\\
&\propto \int \prod_{n=1}^\infty d\rho_n d\rho_{-n} (\rho_1  \rho_{-1}) \exp \big{[} N^2 \sum_{n=1}^\infty -\frac{1 -f(x^n, \eta^n)}{n}\rho_n \rho_{-n} \big{]} \; , \nonumber
\\
&   \propto I_{0} (x,\eta) I_{W} (x,\eta) \;.
\end{align}
Here we define the function $f(x, \eta)$ as
\begin{align}
f(x, \eta) = (\eta +\eta^{-1})\frac{x}{1-x^2} - \frac{2x^2}{1-x^2} \;.
\label{f}
\end{align}
To take the large $N$ limit of the index, we introduced Fourier transformation coefficients $\rho_n$ of the density function $\rho (\theta) : = \frac{1}N \sum_i^N \delta(\theta - \lambda_i) $. 
\begin{align}
\rho_n = \frac1 {2 \pi }\int d \theta \rho(\theta ) e^{i n\theta}= \frac{1}{N}\sum_i e^{i n \lambda_i} \;.
\end{align}
In the large $N$ limit,
\begin{align}
 &\int \frac{d^N \lambda_i}{(2\pi)^N} \; \rightarrow \; \int D\rho = \int \prod_{n=-\infty}^\infty d \rho_n\;, \nonumber
 \\
 & \prod_{i \neq j}^N (1-e^{i (\lambda_i - \lambda_j)})= \exp \big{[}-\sum_{n=1}^\infty \frac{1}n (N^2 \rho_n \rho_{-n} -N)\big{]}\;.
\end{align}
$I_0 (x,\eta)$ denotes the large $N$ index in the absence of  line operators, which is same with the gravity index from supergravity spectrum on $AdS_5 \times S^5$ \cite{hep-th/0510251}. 
\begin{align}
I_{0}(x, \eta) = \prod_{n=1}^\infty \frac{1}{1-f(x^n, \eta^n)}\;.
\label{sugra}
\end{align}
$I_W$ can be thought as  the index contribution due to the insertion of  fundamental Wilson line operator
\begin{align}
I_W (x) = \frac{1}{1-f(x, \eta)}  = \textrm{P.E} [ I_{sp;W} (x,\eta)], \;\textrm{with } I_{sp;W}(x,\eta) = (\eta+ \eta^{-1})x -x^2 \;. 
\end{align}
As  expected, the index  from fundamental Wilson line operator (at large $N$) is same with the index from  fluctuation modes of fundamental string wrapping  $AdS_2$ in $AdS_5$,   $I_{sp;W} = I_{sp;F1}$.

\subsection{D5-brane/anti-symmetric Wilson line}
Let us consider the $k$-th anti-symmetric Wilson line in the $U(N)$ ${\cal N}=4$ theory. When $k, N$ are both large while $k/N$ is fixed, the holography dual of the Wilson line is known to be a D5-brane wrapping $AdS_2 \times S^4 \subset AdS_5 \times S^5$  \cite{Yamaguchi:2006tq}. Since taking this limit of the anti-symmetric Wilson line index  in eq.~\eqref{Wilson in A_k} seems to be rather difficult,  let us consider the limit of the index of  't Hooft line operators with charge $B = (1^k, 0^{N-k})$, which is S-dual to the anti-symmetric Wilson lines. The 't Hooft line index is given in eq.~\eqref{tHooft in A_k}.
To take both of large $N$ and large $k$ limit, one needs to introduce two density functions 
\begin{align} 
\rho_n =  \frac{1}{k} \sum_{i=1}^{k} e^{ i n \lambda_i} , \qquad \tilde{ \rho}_n = \frac{1}{N-k} \sum_{i=k+1}^N e^{ n i \lambda_i} . \nonumber 
\end{align}
At the large $N$ and $k$ limit, following the same procedure with the fundamental Wilson line, the index becomes 
\begin{align}
&I^{N,k\rightarrow \infty}_{B=(1^k, 0^{N-k})} ( x, \eta ) \nn
\\
&\propto \int \prod_{n=1}^{ \infty} ( \prod_{ \pm} d \rho_{ \pm n}  d \tilde{ \rho}_{ \pm n}  )
\exp \big[ -\sum_{n=1}^{ \infty}\frac{ 1 - f(x^n , \eta^n)}{ n} \big{(} \rho_n \rho_{-n} + \tilde{ \rho}_n \tilde{\rho}_{-n} + x^n \sum_{ \pm} ( \rho_{ \pm n} \tilde{ \rho}_{ \mp n} ) \big{)} \big] \;,  \nonumber 
\end{align}
where $f(x, \eta)$ is defined in eq~\eqref{f}.
Doing the Gaussian integral results in
\begin{align}
I^{N,k\rightarrow \infty}_{B=(1^k, 0^{N-k})}(x, \eta) & \propto I_0 ( x, \eta) I_B ( x, \eta) \;, 
\nonumber 
\end{align}
where $I_0(x, \eta)$ is  the gravity index given in eq~\eqref{sugra}, and $I_B (x)$ is given as 
\begin{align} 
I_B (x) &= \prod_{n=1}^{ \infty} \frac{1}{1- f(x^n, \eta^n)}  \frac{1}{1- x^{2n}} = {\rm P.E} [ \frac{ \eta x}{ 1- \eta x} + \frac{ \eta^{-1} x}{1- \eta^{-1} x} ] . 
\label{i_b;sp}
\end{align}

The fluctuations around a D5-brane wrapping $AdS_2 \times S^4 \subset AdS_5 \times S^5$ are summarized in table 1 of \cite{Faraggi:2011ge}(see also \cite{Faraggi:2011bb}). We reproduce the table in our convention here for the reader's convenience (see table \ref{tab:d3}). Their quantum number $l$ corresponds to $|q_1|+|q_2|+ |q_3|$ here, where $q_i$'s are 3 Cartans of $SO(6)$,  related with the three Cartans of $SU(4)$ by $r_1 = q_2 + q_3, r_2 = q_1 - q_2, r_3 = q_2 - q_3$ (see eq. (C.4) of \cite{hep-th/0510251}). For instance, for the field $a_l$, the highest $r_1$-charge of ${\bf 5}$ of $so(5)$ representation is $1$, and the highest $r_1$ charge of $|q_1| + |q_2| +| q_3|=l$ is $l$, thus the highest $r_1$ of the representation becomes $l+1$. 
\begin{table}[htbp]
   \centering
   \begin{tabular}{@{} l|c|c|c|c|c|c @{}} 
      \toprule

       & field  & $\epsilon$ & $j_L + j_R$ &$r_1$  &  $so(5)$ &$|q_1|+|q_2|+|q_3|$ \\
      \midrule
      bosons & $\eta_l$ ($ l \geq 1$) & $l$ & $ 0$ & $l$ & 1&  $ l $  \\
                & $ \zeta_l$   & $l+4$ & 0 & $l$ &1& $l$  \\
            & $ a_l$ & $ l+3$ & 0  &  $(l+1)_2$ &${\bf 5}$ & $ l $\\
          & $ \chi_l^{ \underbar{i}}$  & $l+2$  & $  1$ & $l$ &1& $ l $ \\
          \midrule
     fermions & $\psi_{l+}$  & $ l + \frac{3}{2}$ & $   \frac{1}{2} $ &$l+1$ &$ { \bf 4}$  & $l$ \\
     & $ \psi_{l+}$ & $ l+ \frac{7}{2}$ & $  \frac{1}{2}$ &$l+1$ & ${\bf 4}$ &$ l$ \\
      \bottomrule
   \end{tabular}
   \caption{Fluctuations around the D5-brane wrapping $AdS_2 \times S^4 \subset AdS_5 \times S^5$ (cf. table 1 of \cite{Faraggi:2011ge}). The numbers in the columns under $j_L+j_R$ and $r_1$  denote the highest weight (charge) of the representation. The subscript of $(l+1)_2$ denotes degeneracy. $l \geq 0$ except for the $ \eta_l$. }
   \label{tab:d3}
\end{table}

One can see that the field $ \eta_l$ and $ \psi_{l+}$ for $(\epsilon , j_L + j_R, r_1) =( l,  0, l)$ and $( l+ \frac{3}{2},  \frac{1}{2}, l+1)$ satisfy the BPS bound $\epsilon= (j_L+j_R)+r_1$ respectively. The BPS bound is saturated when $q_1=0,  q_2+q_3=l$ in both cases, and $\{(r_1, r_3) \} ({ \bf 4})= (1, 0)$ for $\psi_{l+}$. For $q_2+q_3=l$, $(q_2, q_3)$-charges can be $\{ (0, l), (1, l-1), \ldots, (l, 0)\}$. Thus the  $r_3$-charge can run from $-l$ to $l$ by 2 in both cases. In sum, the contributions to the index $\Tr [ (-1)^F \eta^{r_3} x^{ \epsilon+j_L+j_R} ]$ are 
\begin{align} 
I_{sp;D5} ( x, \eta)  & =  \sum_{l=1}^{ \infty} ( \eta^{ - l} + \eta^{- l+2} + \ldots + \eta^l) x^{l}+(-1)  \sum_{l=0}^{ \infty} ( \eta^{-l} + \eta^{-l+2} + \ldots + \eta^l ) x^{ l+ 2} , \nonumber
\end{align}
where the first (second) term comes from the field $ \eta_l$ ($ \psi_{l+}$) which saturates the BPS bound. The result can be rewritten as 
\begin{align}
I_{sp;D5} (x, \eta)  &=  \frac{ \eta x}{1- \eta x} + \frac{ \eta^{-1} x}{ 1- \eta^{-1} x} 
 \end{align}
 which exactly agrees with the single particle index obtained in \eqref{i_b;sp}. 

 For Wilson line operators in the $k$-th symmetric representation with large $k$,  the holographic dual object is a D3-brane wrapping $AdS_2 \times S^2$ in $AdS_5$. Spectrum of fluctuations around the brane is  analyzed in \cite{Faraggi:2011bb}.  However, in this case, the large N calculation of the field theory index seems  difficult since rewriting the character  in terms of the Fourier coefficients, $\rho_n$, becomes quickly messy as $k$ increses. This difficulty also exists for Wilson line in  anti-symmetric representation. However, in this case, we could circumvent the difficulty using S-duality. 
\\
\\
\\
\noindent{\bf Acknowledgements}
We are grateful to Seok Kim, Hiroaki Nakajima, Hee-Cheol Kim, Sung-soo Kim, and Takuya Okuda 
for helpful discussions. This work is supported by  the National Research Foundation of Korea Grants    2006-0093850 (KL), 2009-0084601 (KL), 2010-0007512 (DG) and 2005-0049409 through the Center for Quantum Spacetime(CQUeST) of Sogang University (KL).

\newpage

\centerline{\large \bf Appendix}

\appendix
\section{$\mathcal{N}=4$ SYM on $\mathbb{R}\times S^3$}In this section, we review the action for $\mathcal{N}=4$ SYM theories on $\mathbb{R}\times S^3$  \cite{Okuyama:2002zn},\cite{hep-th/0605163}. We follow the notation used in \cite{hep-th/0605163}.
The action (using 10d spinors) is
\begin{align}
S= \frac{1}{g_{YM}^2} \int d^4 x \sqrt{-g} \textrm{Tr} \big{(} - \frac{1}4 F_{a b} F^{a b} - \frac{1}2 D_a \phi_m D_a \phi^m - \frac{1}2 \phi_m^2  \nonumber
\\
- \frac{i}2 \bar{\lambda}\Gamma^a D_a \lambda - \frac{1}2 \bar{\lambda}\Gamma^m [\phi_m , \lambda] + \frac{1}4 [\phi_m, \phi_n]^2\big{)} \; ,
\end{align}
where $a,b$ are local Lorentz indices and run from 0 to 3, and $m$  from 4 to 9. $\Gamma^m$ are the 10-dimensional gamma matrices.
\\
\\
Relations between $\mathbf{6}$ of $SO(6)$  and $\Box \Box$  of $SU(4)$ are, 
\begin{align}
&X_{i4}  = \frac{1}2 (X_{i+3} +i X_{i+6}) \; (i=1,2,3)\; , \nonumber
\\
&X_{AB} = - X_{BA}, \; X^{AB} = - X^{BA} = X_{AB}^\dagger,\; X^{AB} = \frac{1}2 \epsilon^{ABCD}X_{CD} \;. \label{SO(6) and SU(4)}
\end{align}
There are similar relations between $\Gamma_m$s and $\Gamma_{AB}$s.
\\
\\
The 10d Majorana-Weyl spinor can be decomposed into to 4 d Weyl spinors. 
\\
We use a decomposition of 10d Gamma matrices given as 
\begin{align}
\Gamma^a = \gamma^a \otimes \mathbb{I} \; ,  \quad \Gamma^{AB} = \gamma_5 \otimes \left(\begin{array}{cc}0 & - \tilde{\rho}^{AB} \\ \rho^{AB} & 0\end{array}\right) = - \Gamma^{BA} \; ,
\end{align}
where $\gamma^a$ are the 4-dimensional gamma matrices ($a=0, 1,2,3$) and $\gamma^5 \equiv -i \gamma^0 \gamma^1 \gamma^2 \gamma^3$. $\Gamma^{AB}$ satisfy $\{\Gamma^{AB}, \Gamma^{CD} \} = \epsilon^{ABCD}$, and $2\times 2$ matrices $\rho^{AB}$ and $\tilde{\rho}^{AB}$ are defined by
\begin{align}
(\rho^{AB})_{CD} = \delta^A_C \delta^B_D - \delta^A_D \delta^B_C, \quad (\tilde{\rho}^{AB})^{CD} = \epsilon^{ABCD} \;.
\end{align}
This is compatible with \eqref{SO(6) and SU(4)}. The charge conjugation matrix and chirality matrix are given by
\begin{align}
C_{10}  = C_4 \otimes \left(\begin{array}{cc} 0 & \mathbb{I}_4 \\ \mathbb{I}_4 & 0\end{array}\right) \; , \quad \Gamma^{11} = \Gamma^0 \ldots \Gamma^9 = \gamma_5 \otimes \left(\begin{array}{cc} \mathbb{I}_4 & 0\\ 0  & -\mathbb{I}_4\end{array}\right) \; .
\end{align}
A 10d Majorana-Weyl spinors $ \lambda$ can be decomposed into
\begin{align}
\lambda = \Gamma_{11} \lambda = \left(\begin{array}{c}\lambda^A_+ \\ \lambda_{-A}\end{array}\right) \; , 
\end{align}
where $\lambda_{-A}$ is the charge conjugation of $\lambda_+^A$ :
\begin{align}
\lambda_{-A} = C_4 (\bar{\lambda}_{+A})^T \; , \quad \gamma_5 \lambda_\pm = \pm \lambda_\pm \;.
\end{align}
The action can be rewritten in the $SU(4)$ covariant form as follows :
\begin{align}
S = \frac{1}{g_{YM}^2} \int d^4  x \sqrt{-g} \textrm{Tr} \big{(}- \frac{1}4 F_{ab}F^{ab} - \frac{1}2 D_a X_{AB} D^a X^{AB} - \frac{1}2 X_{AB}X^{AB} - i \bar{\lambda}_{+A} \gamma^a D_a \lambda^A_+ \nonumber
\\
- \bar{\lambda}_{+A}[X^{AB}, \lambda_{-B}] - \bar{\lambda}^A_- [X_{AB}, \lambda_+^B] + \frac{1}4 [X_{AB}, X_{CD}][X^{AB}, X^{CD}] \big{)} \; .
\end{align}
We choose the 4d gamma matrices $\gamma^a$ to be
\begin{align}
\gamma^0 = i \mathbb{I}_2 \otimes \mathbb{\sigma}_1 \; , \quad \gamma^i = \sigma^i \otimes \sigma_2\;.
\end{align}
It leads to
%
%
\begin{align}
\gamma_5 =  \mathbb{I}_2 \otimes \sigma_3 \; , \quad C_4 = \gamma^0 \gamma^3 = -\sigma_2\otimes \sigma_3 \;. 
\end{align}
We introduce two-component spinors $ \psi^A$:
\begin{align}
\lambda^A_+ = \psi^A \otimes \left(\begin{array}{c}1 \\0\end{array}\right) \; . 
\end{align}
The $\psi^A$ can be thought as spinors  on $S^3$. Using the two-component spinor, we can express the action as follows :
\begin{align}
S = \frac{1}{g_{YM}^2}  \int d^4 x \big{(} - \frac{1}4 F_{ab}F^{ab} - \frac{1}2 D_aX_{AB}D^a X^{AB} - \frac{1}2 X_{AB}X^{AB} + i \psi_A^\dagger D_0 \psi^A + i \psi_A^\dagger \sigma^i D_i \psi^A \nonumber
\\
+ \psi_A^\dagger \sigma^2 [X^{AB}, (\psi^\dagger_B)^T] - \psi^{AT}\sigma^2 [X_{AB}, \psi^B] + \frac{1}4[X_{AB}, X_{CD}][X^{AB},X^{CD}] \big{)} \; .
\end{align}
$D_i$ are covariant derivatives on $S^3$ ($i=1,2,3$ are vielbein indices).
\begin{align}
D_i \psi = \partial_i \psi + \frac{1}8 \omega^{j}_{i,k} [\eta^j, \eta^k] \psi - i [A_i ,\psi ] \; .
\end{align}
\section{Spectrum on $S^3$}
\subsection{Scalar modes}
We are considering spectrum of the following operator
\begin{align}
M_q^2 =- D_{q}^2 +1 +\frac{q^2}{\sin^2 \chi} = - \partial^2_\chi -2 \frac{\cos \chi}{\sin\chi} \partial_\chi + \frac{-D^2_{S^2:q}+ q^2}{\sin^2 \chi}+1 \ .
\end{align}
Following \cite{Maor:2007xr}, first expand $Y(\chi, \Omega_2)$ in the monopole harmonics $W_{J,m}(\Omega_2)$ on $S^2$
\begin{align}
Y (\chi, \Omega_2) = \zeta ( \chi) W_{J,m} (\Omega_2) \; . 
\end{align}
Scalar monopole harmonics on $S^2$ are denoted as $W^q_{J,m}$, given as
\begin{align}
& D_{S^2:q}^2 W^q_{J,m} = - [J(J+1)-q^2] W^q_{J,m},  
\end{align}
where $ J=|q|, |q|+1 ,\ldots $ and  $m=-J, \ldots, J $. Then the eigenvalue problem is simplified as
\begin{align}
- ( \partial_\chi^2   +2 \frac{\cos \chi}{\sin \chi} \partial_\chi  - \frac{J(J+1)}{\sin^2 \chi} +1)\zeta  = \omega^2 \zeta
\end{align}
Making the substitutions $x = \cos \chi$ and $\zeta = (1-x^2)^{J/2} \eta$, the equation become
\begin{align}
(\partial_x^2 - (2J+3)\frac{x}{1-x^2}\partial_x  + \frac{(\omega^2 -1)- J(J+2)}{1-x^2}) \eta =0
\end{align}
This is Gegenbauer’s equation which has solutions regular at $x = \pm 1$ only for quantized values $\omega^2 = (n+1)^2$ with $n= J , J+1 , \ldots$. The solutions are
\begin{align}
Y^q_{n,J,m} (\chi , \Omega_2)  =  \sin^J(\chi)G^{J+ (1/2)}_{n-J}  (\cos \chi) W^q_{J,m}(\Omega_2) \; .
\end{align}
Here $G^{\alpha +\frac{1}2}_{n-J}$ is Gegenbauer polynomial.
Summarizing the analysis in the section, the eigenfunctions of $M_q^2$ are $\{ Y^q_{n,J,m} \}$ with eigenvalue $(n+1)^2$. Here, $J= |q|,|q|+1, \ldots, m =- J, \ldots, J $ and $n = J, J+1, \ldots$.

\subsection{fermionic modes}
A metric on $ S^3$ is 
\begin{align}
 d \chi^2 + \sin^2 \chi (d \theta^2+ \sin^2 \theta d \varphi^2) \; , 
\end{align}
where Vielbeins are
\begin{align}
 \theta^1 = d \chi, \; \theta^2 = \sin \chi d \theta,\; \theta^3 = \sin \chi \sin \theta d \varphi \; .
\end{align}
The spin connection form $ \omega$ can be obtained as follows
\begin{align}
&d \theta^1=0  \; , \; \nonumber
\\
&d \theta^2 = \frac{\cos \chi}{\sin \chi} \theta^1  \wedge \theta^2\; , \quad {\omega_{\theta}^{2}}_1 = \cos \chi \;, \nonumber
\\
& d \theta^3 = \frac{\cos \chi}{\sin \chi} \theta^1 \wedge \theta^3 + \frac{1}{\sin \chi}\frac{\cos \theta}{\sin \theta} \theta^2 \wedge \theta^3 \; ,  \quad {\omega_{\varphi}^{3}}_2 = \cos \theta, \;  {\omega_{\varphi}^{3}}_1 = \cos \chi \sin \theta
\end{align}
Thus the Dirac operator on $S^3$ is given as
\begin{align}
&i \gamma^i \nabla_i  \nonumber
\\
&=i \big{(} \sigma^1 \partial_\chi + \frac{\sigma^2}{\sin \chi} ( \partial_\theta + \frac{1}4 \cos \chi [\sigma^2, \sigma^1]) + \frac{\sigma^3}{\sin \chi \sin \theta} (\partial_\varphi + \frac{1}4 \cos \theta [\sigma^3, \sigma^2] + \frac{1}4 \cos \chi \sin \theta [\sigma^3, \sigma^1]) \big{)} \nonumber
\\
&= i \big{(} \sigma^1 \partial_\chi + \frac{\cos \chi}{\sin \chi} \sigma^1 + \frac{1}{\sin \chi}(\sigma^2 \partial_\theta + \frac{\sigma^3}{\sin \theta} \partial_\varphi + \frac{1}2 \frac{\cos \theta}{\sin \theta} \sigma^2) \big{)}  =  i \sigma^1 (\partial_\chi + \cot \chi ) +\frac{1}{\sin \chi} \slashed \nabla_{S^2}\; . 
\end{align}
When turning on magnetic fluxes with charge $q$ along $S^2$, the derivative $\slashed\nabla$ is modified into a covariant derivative $\slashed D_q$
 w.r.t the monopole. Monopole spinor harmonics on $S^2$ are (see Appendix C in \cite{Benna:2009xd})
 \begin{align}
 &i \slashed D_{S^2} \psi_{qJm}^\pm  =\pm \mu_{Jq} \psi_{q Jm}^\pm \; \quad J=|q|+\frac{1}2, |q|+ \frac{3}2, \ldots, \quad m= -J, -J+1, \ldots, J\;.  \nonumber
 \\
 & i \slashed D_{S^2} \psi^0_{qJm} = 0 \; , \quad J = |q|- \frac{1}2 \;. 
 \end{align}
 Here $\mu_{Jq} = \sqrt{(J+\frac{1}2)^2 - q^2}$. Note that $\psi^0_{qJm}$  exist only for $|q| \geq \frac{1}2$.  One important property for the spinor harmonics is
 \begin{align}
 &\sigma_1 \psi^{\pm}_{qJm} = \psi^{\mp}_{qJm} \; , \; \sigma_1  \psi^0_{qJm} =\textrm{ sgn} (q) \psi^0_{qJm} \;.  \nonumber
 \end{align}
Consider the following eigenvalue problem for a fermionic operator $\slashed M_q$
 \begin{align}
\left(\begin{array}{cc} i \slashed  D_{q} &   \frac{ q}{\sin \chi}  \\    \frac{ q}{\sin \chi}   &- i \slashed  D_{q}\end{array}\right)\left(\begin{array}{c} \chi_1 \\  \chi_2\end{array}\right) = \lambda \left(\begin{array}{c} \chi_1 \\  \chi_2\end{array}\right) \; . 
 \end{align}
 Expanding $S^3$ spinor $\Psi$ in  $S^2$ monopole harmonics, (for an exceptional case  $J = |q|- \frac{1}2$, we set  $\chi_1 = f_1 ( \chi) \psi^0_{qJm} ( \theta, \varphi), \chi_2 = f_3 ( \chi) \psi^0_{qJm} ( \theta, \varphi)$ ), 
 one gets
 \begin{align}
 &\chi_1   =   f_1 (\chi) \psi^+_{qJm}(\theta, \varphi)+ f_2 (\chi) \psi^-_{qJm}(\theta, \varphi) \; ,  \nonumber
 \\
 &\chi_2 =   f_3  (\chi )\psi^+_{qJm} (\theta , \varphi) + f_4 (\chi) \psi^-_{qJm} (\theta, \varphi) \;.
 \end{align}
 The spectral problem becomes ($\mu := \mu_{Jq}$)
 \begin{align}
 &
 \left(\begin{array}{cc}i (\partial_\chi  + \cot \chi) \sigma^1 + \frac{\mu}{\sin \chi} \sigma^3  & \frac{q}{\sin \chi} \mathbb{I}_2 \\ \frac{q}{\sin \chi} \mathbb{I}_2  & -i (\partial_\chi  + \cot \chi) \sigma^1 - \frac{\mu}{\sin \chi} \sigma^3 \end{array}\right) \left(\begin{array}{c}f_1 \\ f_2 \\ f_3 \\ f_4\end{array}\right)  = \lambda  \left(\begin{array}{c}f_1 \\ f_2 \\ f_3 \\ f_4\end{array}\right) \;.  \nonumber 
 \end{align}
More compactly, the $4\times 4$ matrix can be written as
\begin{align}
\slashed M_q &= i (\partial_\chi  + \cot \chi) \sigma^1 \otimes \sigma^3 + \frac{\mu}{\sin \chi} \sigma^3 \otimes \sigma^3 + \frac{q}{\sin \chi} \mathbb{I}_2 \otimes \sigma_1 \; \nonumber
\\
& :=  i (\partial_\chi  + \cot \chi)  T_0  + \frac{\mu}{\sin \chi} T_x + \frac{q}{\sin \chi} T_y \; . 
\end{align}
Note that $\{ T_x, T_y, T_z \} := \{ \sigma^3 \otimes \sigma^3 , \mathbb{I}_2 \otimes \sigma^1, \sigma^3 \otimes \sigma^2 \} $ form the $SU(2)$-algebra and $T_z$ commutes with $T_0$. Thus with a proper choice of $\epsilon$, one can see that
\begin{align}
\tilde{\slashed M_q} & =e^{- i \epsilon T_z} (\slashed M_q) e^{i \epsilon T_z} = i( \partial + \cot \chi) T_0 + \frac{\sqrt{\mu^2 + q^2}}{\sin \chi} T_x  \;  \nonumber
\\
&=  i( \partial + \cot \chi) \sigma^1 \otimes \sigma^3 + \frac{(J+\frac{1}2)}{\sin \chi} \sigma^3 \otimes \sigma^3 \; , 
\end{align}
The eigenvalue problem for $\slashed M_q$ is equivalent to the problem for $\tilde{\slashed M_q}$. 
 If we set the eigen-spinor $\Psi = \left(\begin{array}{c}g^+_1 \\ g^+_2\end{array}\right) \otimes \left(\begin{array}{c}1 \\0\end{array}\right) +\left(\begin{array}{c}g^-_1 \\ g^-_2\end{array}\right) \otimes \left(\begin{array}{c}0 \\1\end{array}\right)$, the eigenvalue problem for $\tilde{\slashed M_q}$ can be decomposed into two independent equations, 
\begin{align}
[i (\partial_\chi  + \cot \chi ) \sigma^1  + \frac{(J+1/2)}{\sin \chi} \sigma^3 ]\left(\begin{array}{c}g^\pm_1 \\ g^\pm_2\end{array}\right) = \pm \lambda \left(\begin{array}{c}g^\pm_1 \\ g^\pm_2\end{array}\right) \;.
\label{eigen value problem for gpm}
\end{align}
Since the two equations for $g_i^+$ and $g_i^-$ are identical up to the sign of $\lambda$, we will concentrate on the eigenvalue problem for $(g_1^+ , g^+_2)$ and let $g_1^+ = g_1 , g_2^+ = g_2$ for simplicity. The eigenvalue equation is ($g_s := g_1 + g_2, g_a := g_1 - g_2$)
\begin{align}
&\big{[} i (\partial_\chi  + \cot \chi ) - \lambda \big{]} g_s (\chi) + \frac{(J+1/2)}{\sin \chi} g_a (\chi) =0 \; , \nonumber
\\
&\big{[} i (\partial_\chi  + \cot \chi ) + \lambda \big{]} g_a (\chi) - \frac{(J+1/2)}{\sin \chi} g_s(\chi) =0 \; .
\label{eigen value problem for g}
\end{align}
Two solutions $g^{(1),(2)}$ for these coupled linear differential equations can be expressed in terms of hypergeometric functions, 
\begin{align}
&g^{(1)}_s (\chi)  = u^{2 + \lambda}(1-u^2)^{J-1/2} {}_2F_1 (J+\frac{3} 2 , J+ \lambda +1 , \lambda+ \frac{3}2 ; u^2) \; \nonumber
\\
&g^{(1)}_a (\chi) =\frac{i(\lambda + 1/2)}{J+\frac{1}2} u^{\lambda+1}(1-u^2)^{J -\frac{1}2}{}_2 F_1(J + \frac{1}2  , J+\lambda+ 1 , \lambda + \frac{1}2;u^2)\;, \quad u:=e^{i \chi}\;. \nonumber
\\
&g^{(2)}_s (\chi) = u^{1- \lambda} (1-u^2)^{J-\frac{1}2} (2\lambda+1) {}_2 F_1 (J+ \frac{1}2 , J- \lambda +1, -\lambda+ \frac{1}2 ; u^2) \;, \nonumber
\\
&g^{(2)}_a (\chi) = -2i u^{2- \lambda}(1-u^2)^{J-\frac{1}2} {}_2 F_1 (J+\frac{3}2 , J - \lambda +1 , \frac{3}2- \lambda ;u^2) \;. 
\end{align}
The hypergeometric function ${}_2F_1 (a,b,c;z)$ is characterized by the ratio of successive coefficients in the power expansion in $z$
\begin{align}
{}_2F_1 (a,b,c;z) = \sum_{n=0}^\infty c_n z^n\; , \quad \frac{c_{k+1}}{c_k} = \frac{(k+a)(k+b)}{(k+c)(k+1)} \;. 
\end{align}  
Since our expansion parameter $|u^2| = |e^{2  i \chi}| =1$, the series  should terminate at a finite order to give a convergent expression. From the condition,
\begin{align}
&J+ \lambda +1  = (\textrm{non-positive integer}), \quad  \textrm{for the first solution of $g^{(1)}_{s,a}$} \; , \nonumber
\\
&J - \lambda +1  = (\textrm{non-positive integer}), \quad  \textrm{for the second solution of $g^{(2)}_{s,a}$} \;.
\end{align}
Thus the spectrum for eigenvalue problem in \eqref{eigen value problem for g} is
\begin{align}
&\lambda^{q,\pm}_{n,J,m} = \pm (n+1), \quad n=J,J+1,J+2,\ldots\; \nonumber
\\
& J= |q|- \frac{1}2 \; (\textrm{ exist for $|q|\neq 0$ })  , |q| + \frac{1}{2} , |q|+ \frac{3}2 \dots, \nonumber
\\
& m = - J, -J+1, \ldots , J \; .
\end{align}
Let the eigen-spinor with eigenvaule $\lambda^{q}_{n,J,m}$ be $\Psi^{q,\pm}_{n,J,m}$. Taking into account the spectrum of $g^{-}_i$ in eq. \eqref{eigen value problem for gpm}, the spectrum of $\slashed M_q$ is the double copy the above spectrum, $\textrm{Spec}( \slashed M_q) =  \{ \lambda^{q;\pm,\kappa}_{n,J,m} \}|_{\kappa = 1,2}$. For the exceptional case ($J=|q|-\frac{1}2 $), the eigen-spinors $\Psi^{\kappa=1}$ and $\Psi^{\kappa=2}$ are not independent  and we  abandon the second one.

\end{document}